\documentclass[modern]{aastex62}
\usepackage{amsmath, amsthm, amssymb, amsfonts}
\usepackage{chngcntr}
\usepackage{etoolbox}
\usepackage{color}
\pretocmd{\abstractname}{\newpage}{}{}

\newcommand{\e}{\textnormal{e}^{-1}}
\newcommand{\Rp}{R_p}
\newcommand{\Rs}{R_{\star}}
\newcommand{\RpRs}{\Rp/\Rs}
\newcommand{\Rjup}{R_{\textnormal{J}}}
\newcommand{\Mjup}{M_{\textnormal{J}}}
\newcommand{\Rsun}{R_{\odot}}

\newcommand{\aRs}{a/\Rs}
\newcommand{\Tmid}{T_{\textnormal{mid}}}
\newcommand{\Ts}{T_{\star}}
\newcommand{\mb}{\mathbf}
\newcommand{\gb}{\boldsymbol}

\newcommand{\um}{\mu\textnormal{m}}
\newcommand{\FeH}{\textnormal{Fe}/\textnormal{H}}
\newcommand{\iLphi}{L_{\phi}^{-1}}

\newcommand{\iLx}{L_{x}^{-1}}
\newcommand{\iLy}{L_{y}^{-1}}

\shorttitle{WASP-121b optical transmission spectrum}
\shortauthors{Evans et al.}

\begin{document}

\title{An optical transmission spectrum for the ultra-hot Jupiter WASP-121b measured with the Hubble Space Telescope}

\correspondingauthor{Thomas M.\ Evans}
\email{tmevans@mit.edu}

\author[0000-0001-5442-1300]{Thomas M.\ Evans}
\affil{Physics and Astronomy, Stocker Road, University of Exeter, Exeter, EX4 3RF, UK}
\affil{Kavli Institute for Astrophysics and Space Research, Massachusetts Institute of Technology, 77 Massachusetts Avenue, 37-241, Cambridge, MA 02139, USA}

\author{David K.\ Sing}
\affil{Physics and Astronomy, Stocker Road, University of Exeter, Exeter, EX4 3RF, UK}
\affil{Department of Earth and Planetary Sciences, Johns Hopkins University, Baltimore, MD, USA}

\author{Jayesh Goyal}
\affil{Physics and Astronomy, Stocker Road, University of Exeter, Exeter, EX4 3RF, UK}

\author{Nikolay Nikolov}
\affil{Physics and Astronomy, Stocker Road, University of Exeter, Exeter, EX4 3RF, UK}

\author{Mark S.\ Marley}
\affil{NASA Ames Research Center, Moffett Field, California, USA}

\author{Kevin Zahnle}
\affil{NASA Ames Research Center, Moffett Field, California, USA}

\author{Gregory W.\ Henry}
\affil{Center of Excellence in Information Systems, Tennessee State University, Nashville, TN 37209, USA}

\author{Joanna K.\ Barstow}
\affil{Department of Physics and Astronomy, University College London, Gower Street, London WC1E 6BT, UK}

\author{Munazza K.\ Alam}
\affil{Harvard-Smithsonian Center for Astrophysics, 60 Garden Street, Cambridge, MA 02138, USA}

\author{Jorge Sanz-Forcada}
\affil{Centro de Astrobiologı\'{i}a (CSIC-INTA), ESAC Campus, Camino Bajo del Castillo, E-28692 Villanueva de la Canada, Madrid, Spain}

\author{Tiffany Kataria}
\affil{NASA Jet Propulsion Laboratory, 4800 Oak Grove Drive, Pasadena, CA 91109, USA}

\author{Nikole K.\ Lewis}
\affil{Department of Astronomy and Carl Sagan Institute, Cornell University, 122 Sciences Drive, 14853, Ithaca, NY, USA}

\author{Panayotis Lavvas}
\affil{Groupe de Spectroscopie Mol\'{e}culaire et Atmosph\'{e}rique, Universit\'{e} de Reims, Champagne-Ardenne, CNRS UMR F-7331, France}

\author{Gilda E.\ Ballester}
\affil{Lunar and Planetary Laboratory, University of Arizona, Tucson, AZ 85721, USA}

\author{Lotfi Ben-Jaffel}
\affil{Sorbonne Universit\'{e}s, UPMC Universit\'{e} Paris 6 and CNRS, UMR 7095, Institut d'Astrophysique de Paris, 98 bis boulevard Arago, F-75014 Paris, France}

\author{Sarah D.\ Blumenthal}
\affil{Physics and Astronomy, Stocker Road, University of Exeter, Exeter, EX4 3RF, UK}

\author{Vincent Bourrier}
\affil{Observatoire de l'Universit\'{e} de Gen\`{e}ve, 51 chemim des Maillettes, 1290 Sauverny, Switzerland}

\author{Benjamin Drummond}
\affil{Physics and Astronomy, Stocker Road, University of Exeter, Exeter, EX4 3RF, UK}

\author{Antonio Garc\'{i}a Mu\~{n}oz}
\affil{Zentrum f\"{u}r Astronomie und Astrophysik, Technische Universit\"{a}t Berlin, Hardenbergstrasse 36, D-10623 Berlin, Germany}

\author{Mercedes L\'{o}pez-Morales}
\affil{Harvard-Smithsonian Center for Astrophysics, 60 Garden Street, Cambridge, MA 02138, USA}

\author{Pascal Tremblin}
\affil{Maison de la Simulation, CEA, CNRS, Univ\ Paris-Sud, UVSQ, Universit\'{e} Paris-Saclay, F-91191 Gif-sur-Yvette, France}

\author{David Ehrenreich}
\affil{Observatoire de l'Universit\'{e} de Gen\`{e}ve, 51 chemim des Maillettes, 1290 Sauverny, Switzerland}

\author{Hannah R.\ Wakeford}
\affil{Space Telescope Science Institute, 3700 San Martin Drive, Baltimore, Maryland 21218, USA}

\author{Lars A.\ Buchhave}
\affil{DTU Space, National Space Institute, Technical University of Denmark, Elektrovej 328, DK-2800 Kgs. Lyngby, Denmark}

\author{Alain Lecavelier des Etangs}
\affil{Sorbonne Universit\'{e}s, UPMC Universit\'{e} Paris 6 and CNRS, UMR 7095, Institut d'Astrophysique de Paris, 98 bis boulevard Arago, F-75014 Paris, France}

\author{\'{E}ric H\'{e}brard}
\affil{Physics and Astronomy, Stocker Road, University of Exeter, Exeter, EX4 3RF, UK}

\author{Michael H.\ Williamson}
\affil{Center of Excellence in Information Systems, Tennessee State University, Nashville, TN 37209, USA}

\begin{abstract}

We present an atmospheric transmission spectrum for the ultra-hot Jupiter WASP-121b, measured using the Space Telescope Imaging Spectrograph (STIS) onboard the Hubble Space Telescope (HST). Across the $0.47$--$1\,\um$ wavelength range, the data imply an atmospheric opacity comparable to -- and in some spectroscopic channels exceeding -- that previously measured at near-infrared wavelengths ($1.15$--$1.65\,\um$). Wavelength-dependent variations in the opacity rule out a gray cloud deck at a confidence level of $3.8\sigma$ and may instead be explained by VO spectral bands. We find a cloud-free model assuming chemical equilibrium for a temperature of $1500$\,K and metal enrichment of $10$--$30\times$ solar matches these data well. Using a free-chemistry retrieval analysis, we estimate a VO abundance of $-6.6_{-0.3}^{+0.2}$\,dex. We find no evidence for TiO and place a $3\sigma$ upper limit of $-7.9$\,dex on its abundance, suggesting TiO may have condensed from the gas phase at the day-night limb. The opacity rises steeply at the shortest wavelengths, increasing by approximately five pressure scale heights from $0.47$ to $0.3\,\um$ in wavelength. If this feature is caused by Rayleigh scattering due to uniformly-distributed aerosols, it would imply an unphysically high temperature of $6810 \pm 1530$\,K. One alternative explanation for the short-wavelength rise is absorption due to SH (mercapto radical), which has been predicted as an important product of non-equilibrium chemistry in hot Jupiter atmospheres. Irrespective of the identity of the NUV absorber, it likely captures a significant amount of incident stellar radiation at low pressures, thus playing a significant role in the overall energy budget, thermal structure, and circulation of the atmosphere.

\end{abstract}

%---------------------------------------------------------------
\section{Introduction} \label{sec:intro}

Spectroscopic observations made during the primary transit of an exoplanet allow the atmospheric transmission spectrum of the day-night boundary region to be probed \citep{2000ApJ...537..916S}, while the same type of observation made during secondary eclipse provides the emission spectrum of the dayside hemisphere \citep{1998ApJ...502L.157S}. Much of the transmission and emission spectroscopy work published to date has employed the Hubble Space Telescope (HST), primarily with the Space Telescope Imaging Spectrograph (STIS), covering the $0.1$--$1\,\um$ UV-optical wavelength range, and Wide Field Camera 3 (WFC3), covering the $0.8$--$1.65\,\um$ near-IR wavelength range.

A non-exhaustive list of HST transmission spectroscopy highlights at optical through IR wavelengths include:\ the detection of Na on HD\,209458b \citep{2002ApJ...568..377C}; multiple detections of H$_2$O \citep[e.g.][]{2013ApJ...774...95D,2016ApJ...822L...4E,2014Natur.513..526F,2013MNRAS.434.3252H,2015ApJ...814...66K,2018AJ....155..156T,2017Sci...356..628W,2018AJ....155...29W}; widespread evidence for aerosols \citep[e.g.][]{2014Natur.505...69K,2014MNRAS.437...46N,2015MNRAS.447..463N,2008MNRAS.385..109P,2015MNRAS.446.2428S,2016Natur.529...59S}; and a detection of He in the extended atmosphere of WASP-107b \citep{2018Natur.557...68S}. At UV wavelengths, transit observations made with STIS have probed the hydrogen exospheres of hot Jupiters \citep[e.g.][]{2003Natur.422..143V} and warm Neptunes \citep[e.g.][]{2015Natur.522..459E}, while heavier elements such as oxygen have been detected using the HST Cosmic Origins Spectrograph \citep[e.g.][]{2010ApJ...714L.222F,2013A&A...553A..52B}. For emission, a similar list includes: detections of H$_2$O absorption \citep{2017AJ....154..158B,2014Sci...346..838S}; evidence for H$_2$O emission \citep{2017Natur.548...58E}; evidence for TiO emission \citep{2015ApJ...806..146H}; constraints on optical reflection spectra \citep{2013ApJ...772L..16E,2017ApJ...847L...2B}; and multiple featureless thermal spectra \citep[e.g.][]{2018MNRAS.474.1705N,2018AJ....156...10M}. 

This paper reports a transmission spectrum measured for the ultra-hot ($T_{\textnormal{eq}} \gtrsim 2500$\,K) Jupiter WASP-121b across the $0.3$--$1\,\um$ wavelength range using STIS. Discovered by \cite{2016MNRAS.tmp..312D}, WASP-121b orbits a moderately bright ($V = 10.5 $) F6V host star, which has an estimated radius of $1.458 \pm 0.030\,\Rsun$ \citep{2016MNRAS.tmp..312D} and measured parallax of $3.676 \pm 0.021$\,mas \citep{2018A&A...616A...1G}, corresponding to a system distance of $272.0 \pm 1.6$\,parsec. WASP-121b itself has a mass of $1.18 \pm 0.06\,\Mjup$, an inflated radius of $\sim 1.7\Rjup$, and a dayside equilibrium temperature above $2400$\,K. Together, these properties make WASP-121b an excellent target for atmospheric characterization \citep{2016MNRAS.tmp..312D,2016ApJ...822L...4E,2017Natur.548...58E}. 

We previously published the near-IR $1.15$--$1.65\,\um$ transmission spectrum for WASP-121b measured using WFC3 in \cite{2016ApJ...822L...4E}. Those data revealed absorption due to the H$_2$O band centered at $1.4\,\um$, along with a second bump across the $1.15$--$1.3\,\um$ wavelength range, which we suggested could be a signature of FeH or VO. Analyzing the same dataset, \cite{2018AJ....155..156T} reproduced the $1.15$--$1.3\,\um$ feature and presented a best-fit model including absorption by TiO and VO, although they did not discuss FeH. In \cite{2016ApJ...822L...4E}, we also compared the WFC3 transmission spectrum with transits measured at optical wavelengths by \cite{2016MNRAS.tmp..312D} using ground-based photometry. This comparison implied significantly deeper transits at optical wavelengths relative to the near-IR, which we speculated could be evidence for a strong opacity source such as TiO and/or VO. Subsequent modeling of these data confirmed such an interpretation to be plausible \citep[e.g.][]{2017ApJ...845L..20K,2018A&A...617A.110P}.

In \cite{2017Natur.548...58E}, we presented a secondary eclipse observation for WASP-121b, also made with WFC3 at near-IR wavelengths. The measured spectrum indicates a mean photosphere temperature of approximately $2700$\,K and shows the $1.4\,\um$ H$_2$O band in emission, rather than absorption, implying the dayside hemisphere has a vertical thermal inversion. As for the transmission spectrum, the emission data exhibit a second bump across the $1.15$--$1.3\,\um$ wavelength range, which can be fit with VO in emission. To do so, however, requires assuming a VO abundance over $1000 \times$ higher than expected for solar elemental composition in chemical equilibrium, casting doubt on this interpretation. Models assuming chemical equilibrium and abundances closer to solar do not reproduce the $1.15$--$1.3\,\um$ bump \citep[e.g.][]{2018A&A...617A.110P}. For now, we do not have a satisfying explanation for this feature, but the fact that it has been observed in both the transmission spectrum and emission spectrum is intriguing.

Our understanding of the atmosphere of WASP-121b remains a work in progress. For instance, the thermal inversion measured for the dayside hemisphere implies significant heating at low pressures ($\lesssim 100$\,mbar), though it is unclear what causes this. One possibility is absorption of incident stellar radiation at optical wavelengths by TiO and VO \citep[e.g.][]{2003ApJ...594.1011H,2008ApJ...678.1419F}. However, neither of these species have yet been definitively detected in the atmosphere of WASP-121b, despite the hints described above. Furthermore, it has been pointed out that TiO and VO could be removed from the upper atmospheres of even very hot planets by cold-trapping \citep[e.g.][]{2009ApJ...699.1487S,2009ApJ...699..564S,2017AJ....154..158B}. Additionally, the dayside temperatures of ultra-hot Jupiters such as WASP-121b are likely high enough for significant thermal dissociation of TiO and VO, along with other molecules such as H$_2$O, to occur \citep{2018ApJ...855L..30A,2018AJ....156...17K,2018ApJ...866...27L,2018A&A...617A.110P}. Nonetheless, evidence for TiO has been detected on the dayside of WASP-33b \citep{2015ApJ...806..146H,2017AJ....154..221N}, which has a mean photosphere temperature of around $3000$\,K at near-IR wavelengths, making it even hotter than WASP-121b. An optical transmission spectrum measured for another ultra-hot Jupiter, WASP-19b, also exhibits a prominent TiO band \citep{2017Natur.549..238S}, although this may have been the signature of unocculted star spots \citep{2018MNRAS.tmp.2569E}. Despite the picture remaining unclear, observations such as these imply TiO, and presumably VO, can perhaps persist at low pressures in ultra-hot Jupiter atmospheres. As will be described in the following sections, the STIS transmission spectrum for WASP-121b provides new evidence for VO absorption at optical wavelengths.

Absorption at UV wavelengths may also play a significant role in heating the upper atmospheres of strongly-irradiated planets such as WASP-121b. For instance, \cite{2009ApJ...701L..20Z} examined non-equilibrium sulfur chemistry in the context of hot Jupiter atmospheres and concluded that SH and S$_2$ could be important absorbers across the $0.24$--$0.4\,\um$ wavelength range. These species may be driven to higher-than-equilibrium abundances via reactions involving the photolytic and photochemical destruction of H$_2$S. As will be reported below, the measured transmission spectrum for WASP-121b exhibits a strong signal at wavelengths shortward of $\sim 0.47\,\um$ and absorption by SH appears to provide a viable explanation.

We begin, however, by describing our observations and the steps taken to extract the spectra from the raw data frames in Section \ref{sec:observations_datared}. We present analyses of the white lightcurves in Section \ref{sec:whitelc} and spectroscopic lightcurves in Section \ref{sec:speclcs}. The results are discussed in Section \ref{sec:discussion}, including the implications of the measured transmission for the planetary atmosphere. Our conclusions are given in Section \ref{sec:conclusion}.

%---------------------------------------------------------------
\clearpage
\section{Observations and data reduction} \label{sec:observations_datared}

We observed three primary transits of WASP-121b using HST/STIS as part of the Panchromatic Comparative Exoplanet Treasury (PanCET) survey (Program 14767; P.I.s Sing and L\'{o}pez-Morales). This was comprised of two visits made on 2016 Oct 24 and 2016 Nov 6 with the G430L grating, and one visit made on 2016 Nov 12 with the G750L grating. In what follows, we shall refer to the first and second G430L visits as the G430Lv1 and G430Lv2 datasets, respectively. For all three STIS visits, the target was observed for 6.8 hours, covering five consecutive HST orbits. Observations were made using the widest available slit ($52 \times 2$ arcsec) to minimize slit losses and the detector gain was set to 4 $\e$/DN. Overheads were reduced by only reading out a $1024 \times 128$ pixel subarray containing the target spectrum. Exposure times of 253\,s and 161\,s were used for the G430L and G750L observations, respectively. We also took a short 1\,s exposure at the start of each HST orbit for both gratings, but discarded these exposures in the subsequent analysis. This was done because STIS observations typically suffer from a systematic in which the first exposure of each HST orbit has anomalously lower counts relative to the immediately-following exposures \citep[e.g.][]{2013ApJ...772L..16E,2014MNRAS.437...46N,2015MNRAS.447..463N,2015MNRAS.446.2428S} and we wanted to minimize the integration time lost to this effect. With this observing setup, we acquired a total of 48 science exposures for each G430L visit and 70 science exposures for the G750L visit.

\begin{figure}
\centering  % this centres figure in column
\includegraphics[width=0.62\columnwidth]{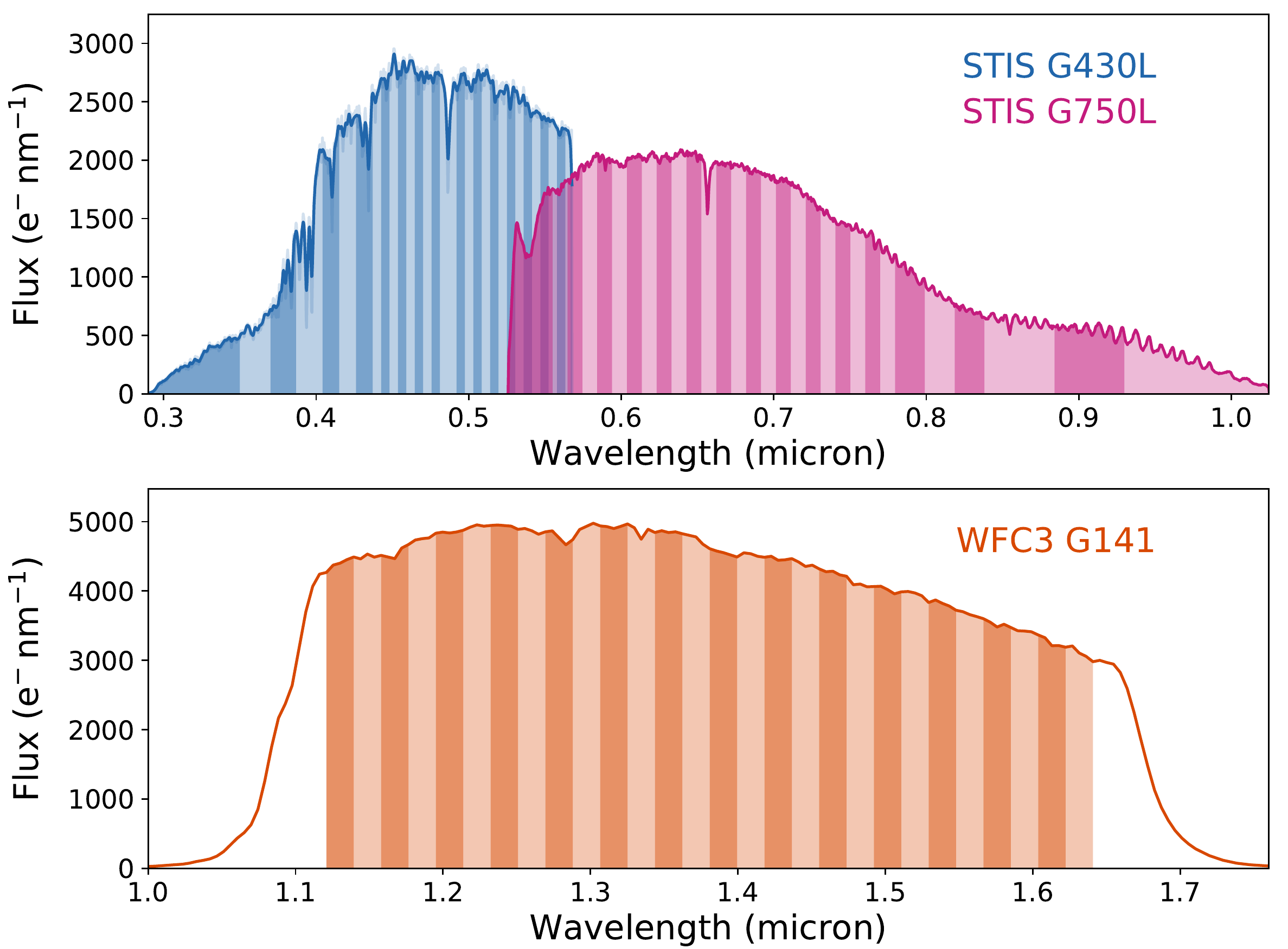}
\caption{Example spectra for the G430L and G750L gratings \textit{(top panel)}, and the G141 grism \textit{(bottom panel)}. Dark and light vertical bands indicate the wavelength channels adopted for the spectroscopic lightcurves.}
\label{fig:example_spectra}
\end{figure}

The STIS datasets were reduced following the methodology described in \cite{2014MNRAS.437...46N,2015MNRAS.447..463N}. Raw data frames were bias-, dark-, and flat-corrected using the CALSTIS pipeline (v3.4) with relevant calibration frames. Cosmic ray events and pixels flagged as `bad' by CALSTIS were removed and interpolated over. Overall, we found $\sim 4$\% of pixels were affected by cosmic rays for all visits with a further $\sim 5$\% flagged as bad by CALSTIS. To extract spectra from the cleaned 2D frames, we used the IRAF procedure \texttt{apall} with aperture radii of 4.5, 6.5, 8.5, and 10.5 pixels for both the G430L and G750L datasets. The dispersion axis was mapped to a wavelength solution using the x1d files produced by CALSTIS. 

In addition to the STIS data, a single primary transit of WASP-121b was observed on 2016 Feb 6 with the G141 grism (Program 14468; P.I. Evans). This dataset was originally published in \cite{2016ApJ...822L...4E}, to which the reader is referred for further details.

Example G430L, G750L, and G141 spectra are shown in Figure \ref{fig:example_spectra}.

\begin{figure}
\centering  % this centres figure in column
\includegraphics[width=\columnwidth]{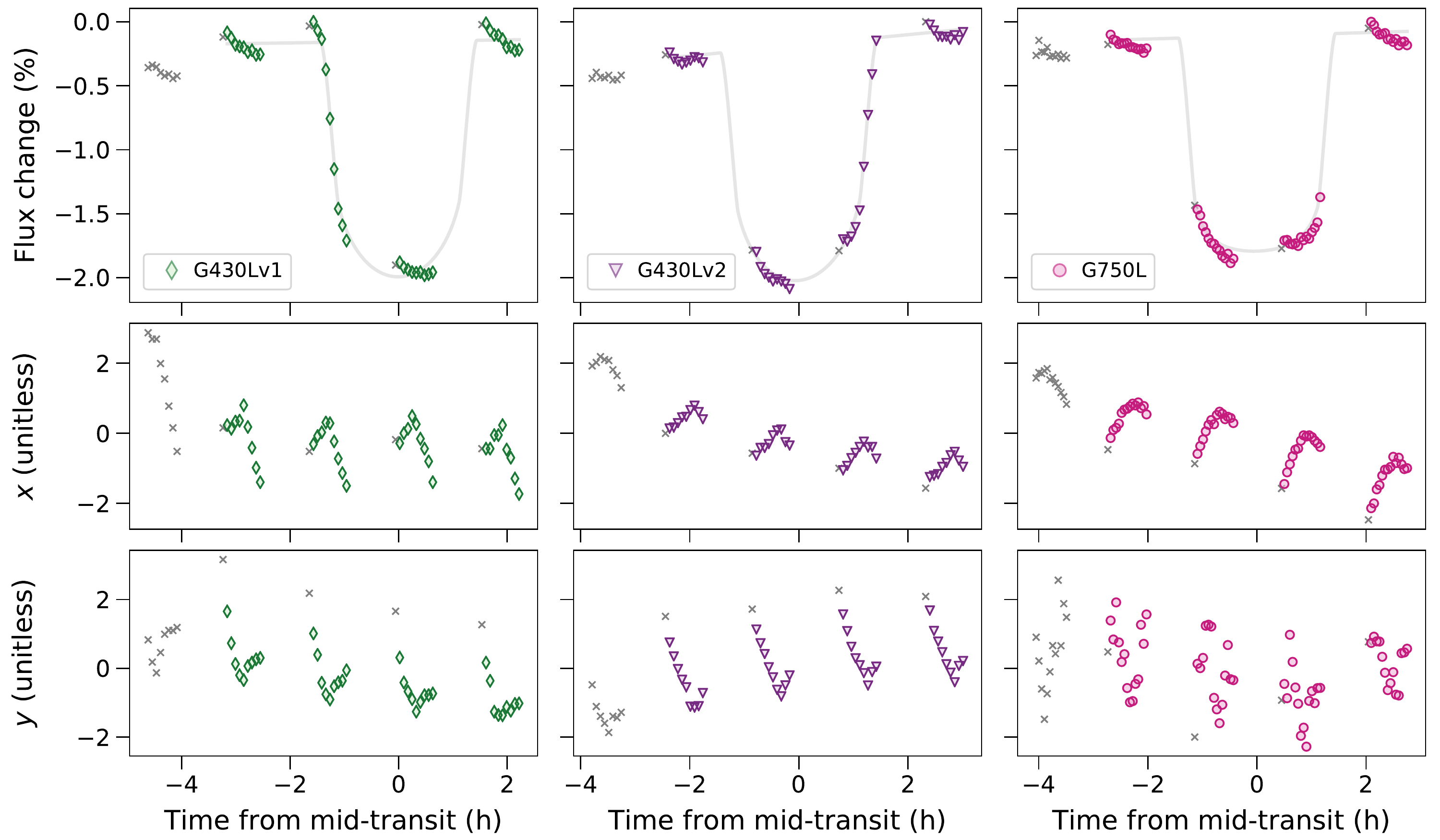}
\caption{ \textit{(Top row)} Raw white lightcurves for the G430Lv1, G430Lv2, and G750L datasets. Gray lines show the best-fit transit signals with linear baseline trends. \textit{(Middle row)} Dispersion drift variable for each dataset. \textit{(Bottom row)} Cross-dispersion drift variable for each dataset. In all panels, colored symbols indicate data points that were included in the analysis and gray crosses indicate those that were excluded for reasons explained in the main text. The two drift variables are unitless as they have been standardized, i.e.\ mean subtracted and normalized by their standard deviations.}
\label{fig:stis_timeseries}
\end{figure}

%---------------------------------------------------------------
\section{White Lightcurve Analyses} \label{sec:whitelc}

White lightcurves were constructed for each dataset by summing the flux of each spectrum across the full dispersion axis. The resulting lightcurves are shown in the top row of Figure \ref{fig:stis_timeseries}. As in our previous work \citep{2013ApJ...772L..16E,2016ApJ...822L...4E,2017Natur.548...58E}, we followed the methodology outlined by \cite{2012MNRAS.419.2683G} and treated each lightcurve as a Gaussian process (GP). Under this approach, the posterior likelihood is described by a multivariate normal distribution of the form $\mathcal{N}\left( \mb{d}\,|\,\gb{\mu}, \mb{K}+\gb{\Sigma} \right)$, where: $\mb{d}$ is an $N$-length vector containing the flux measurements; $\gb{\mu}$ is a vector containing the deterministic mean function; $\mb{K}$ is an $N \times N$ matrix describing the correlations between data points; and $\gb{\Sigma}$ is an $N \times N$ diagonal matrix containing the squared white noise uncertainties, $\sigma_j^2$, for each data point $j = 1, \ldots, N$.

For the mean function, we adopted a \cite{2002ApJ...580L.171M} transit model multiplied by a linear trend in time ($t$) of the form $c_0+c_1\,t$. We assumed a circular orbit with a period ($P$) of 1.2749255\,days \citep{2016MNRAS.tmp..312D}. We allowed the normalized planet radius ($\RpRs$) and transit mid-time ($\Tmid$) to vary as free parameters with uniform priors. As described in Section \ref{sec:whitelc:orbparsfree}, we first performed fits with the normalized semimajor axis ($\aRs$) and impact parameter ($b$) allowed to vary as free parameters, both with uniform priors. Then, as described in Section \ref{sec:whitelc:orbparsfixed}, we fixed $\aRs$ and $b$ to their weighted-mean values and repeated the fitting.

In all fits, we assumed a quadratic limb darkening law and treated both coefficients ($u_1$, $u_2$) as free parameters. We first estimated values for $u_1$ and $u_2$ by fitting to the limb darkening profile of a stellar model over the appropriate bandpass. Specificially, we used a 3D stellar model from the STAGGER grid \citep{2013A&A...557A..26M} with $\Ts = 6500$\,K, $\log_{10} g = 4$\,cgs, and $\left[\FeH\right] = 0$\,dex, as this was the grid point closest to the properties of the WASP-121 host star \citep[{$\Ts = 6460 \pm 140$\,K, $\log_{10} g = 4.242 \pm 0.2$\,cgs, $\left[\FeH\right] = +0.13 \pm 0.09$\,dex;}][]{2016MNRAS.tmp..312D}. We then applied broad normal priors to $u_1$ and $u_2$ in the model fitting, with means set to these estimated values and standard deviations of 0.6, providing plenty of flexibility for the model to be optimized.

For the GP covariance matrix $\mb{K}$, we adopted a squared-exponential kernel\footnote{We refer the reader to previous studies such as \cite{2012MNRAS.419.2683G}, \cite{2013ApJ...772L..16E}, and \cite{2014MNRAS.445.3401G} for further details of the squared-exponential kernel.} with three input variables that it is reasonable to assume could correlate with the instrumental systematics: namely, HST orbital phase ($\phi$), dispersion drift ($x$), and cross-dispersion drift ($y$). This resulted in four free parameters for each dataset: namely, the covariance amplitude ($A$) and correlation length-scales ($L_k$) for each input variable, $k = \{ \phi, x, y \}$. For the white noise matrix, $\gb{\Sigma}$, we adopted the formal photon noise values $\sigma_j$ multiplied by a rescaling factor ($\beta$) which was allowed to vary as a free parameter. The latter affords some flexibility to handle high-frequency systematics that are pseudo-white-noise in nature, which would otherwise bias the model toward impractically small $L_k$ values.

For the GP covariance amplitude $A$, we adopted Gamma priors of the form $p(A) \propto e^{-100A}$, to favor smaller correlation amplitudes. This can help prevent a small number of outliers having a disproportionate influence on the inferred covariance amplitude. For the correlation length scales $L_k$, we followed previous studies \citep[e.g.][]{2017Natur.548...58E,2017MNRAS.467.4591G} and fit for the natural logarithm of the inverse correlation length scales $\ln\eta_k = \ln L_k^{-1}$, adopting uniform priors for each. In practice, this favors longer correlation length scales, with the intention of capturing the lower-frequency systematics present in the data, as these are most degenerate with the planet signal. Higher-frequency systematics can be accounted for through the $\beta$ parameter, for which we adopted a normal prior with mean of 1 and standard deviation of 0.2, to favor values close to the formal photon noise.

\begin{figure}[t!]
\centering  % this centres figure in column
\includegraphics[width=0.75\columnwidth]{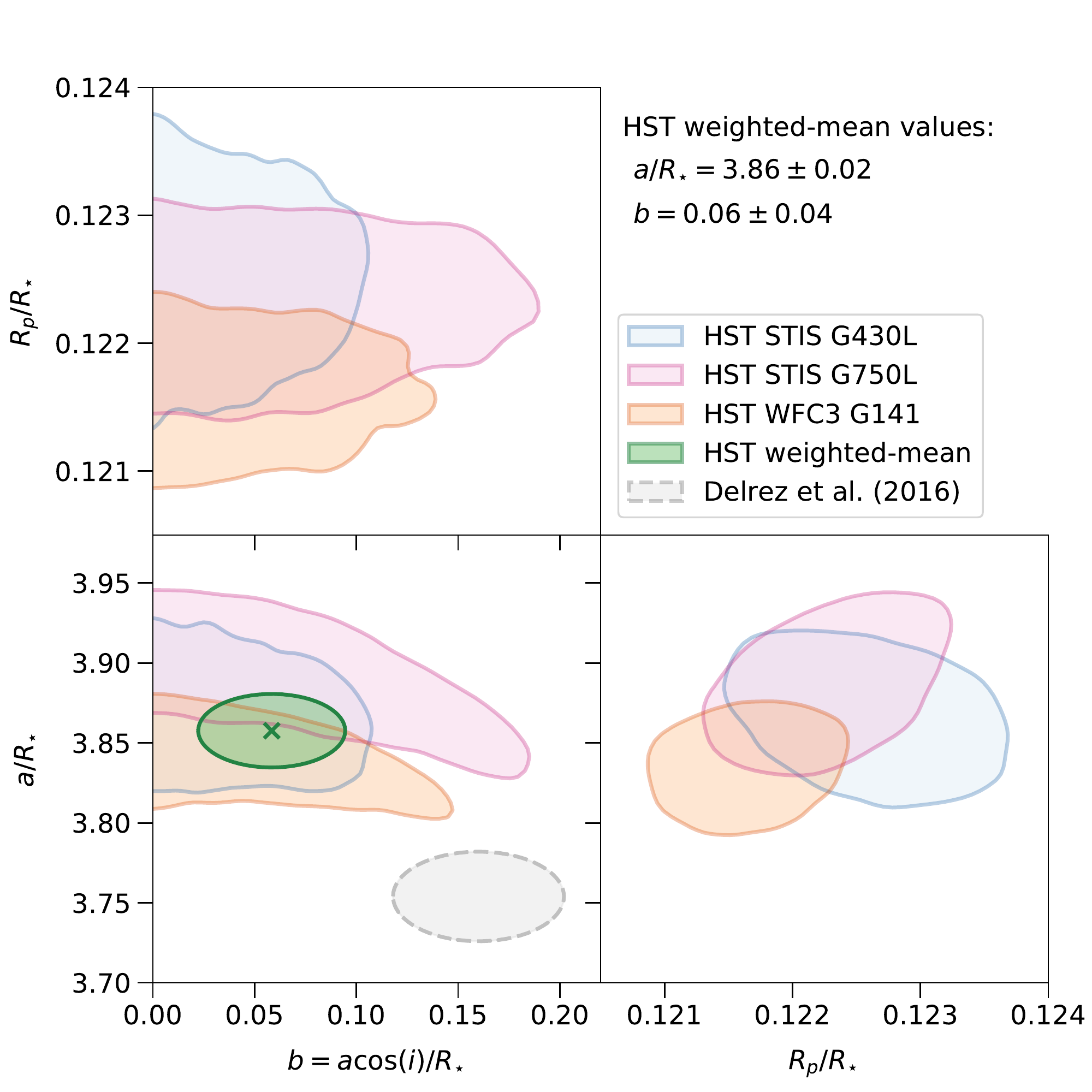}
\caption{Posterior distributions for $\RpRs$, $\aRs$, and $b$ obtained from the white lightcurve analyses described in Section \ref{sec:whitelc:orbparsfree}. Blue, pink, and orange regions indicate smoothed contours containing 68\% of MCMC samples for the G430L, G750L, and G141 analyses, respectively. Green region indicates the weighted-mean of the HST posterior distributions and gray region indicates the $1\sigma$ range reported by \cite{2016MNRAS.tmp..312D}.}
\label{fig:orbpars}
\end{figure}

We modeled the white lightcurves for G430L, G750L, and G141 separately. For the G430L lightcurves, we assumed $\RpRs$, $\aRs$, and $b$ were the same for both visits, while allowing $\Tmid$, $\beta$, $A$, $\ln\eta_{\phi}$, $\ln\eta_x$, and $\ln\eta_y$ to vary separately for each visit. The posterior distributions were marginalized using affine-invariant Markov chain Monte Carlo (MCMC), as implemented by the \texttt{emcee} Python package \citep{2013PASP..125..306F}. In all fits, we randomly distributed five groups of 150 walkers throughout the parameter space and allowed them to run for 100 steps to locate the peak of the posterior distribution. We then re-initialized the five groups of 150 walkers in a tighter ball around this peak and allowed them to run for 500 steps, of which we discarded the first 250 steps as burn-in and combined the remaining 250 steps into a single chain for each walker group. At this point, a comparison of the chains from each walker group confirmed that they appeared well-mixed and converged, with Gelman-Rubin statistic values within 2\% of unity for each free parameter \citep{GelmanRubin92}. Table \ref{table:whitefit} summarizes the resulting posterior distributions. For each of the STIS lightcurves produced using the different trial apertures (see Section \ref{sec:observations_datared}), we obtained results consistent to within $1\sigma$ for the planet parameters (e.g.\ $\RpRs$) and report only those for the $8.5$\,pixel aperture. 

\subsection{$a/\Rs$ and $b$ allowed to vary} \label{sec:whitelc:orbparsfree}

The purpose of the model fits in which $\aRs$ and $b$ were allowed to vary as free parameters was to use the HST data to refine our estimates of these system properties. Previously, the only published measurements were those provided in the original discovery paper by \cite{2016MNRAS.tmp..312D}, which reported $\aRs = 3.754_{-0.028}^{+0.023}$ and $b = 0.160_{-0.042}^{+0.040}$. Figure \ref{fig:orbpars} shows the posterior distributions obtained from our analyses for comparison, with values reported in Table \ref{table:whitefit}. We find good agreement for both $\aRs$ and $b$ across our fits to the G430L, G750L, and G141 white lightcurve datasets. Taking the arithmetic weighted-mean of these results, we estimate $\aRs = 3.86 \pm 0.02$ and $b = 0.06 \pm 0.04$, implying $i = 89.1 \pm 0.5 $\,deg. We note that our HST results differ from those of Delrez et al.\ by $3.5\sigma$ for $\aRs$ and $2\sigma$ for $b$. The reason for this disagreement is unclear and will likely be resolved by additional transit observations that are currently planned or in the process of being analyzed (Evans et al., in prep.). For the present study, we note that the primary consequence of assuming slightly different values for $\aRs$ and $b$ will be to perturb the inferred values for $\RpRs$. Importantly, this will be a wavelength-independent effect and thus should not affect our interpretation of the atmospheric transmission spectrum. For this reason, and given the mutual agreement between the G430L, G750L, and G141 datasets, we adopt the HST weighted-mean values for $\aRs$ and $b$ in all subsequent lightcurve fits.

\subsection{$a/\Rs$ and $b$ held fixed} \label{sec:whitelc:orbparsfixed}

\begin{figure}
\centering  % this centres figure in column
\includegraphics[width=\columnwidth]{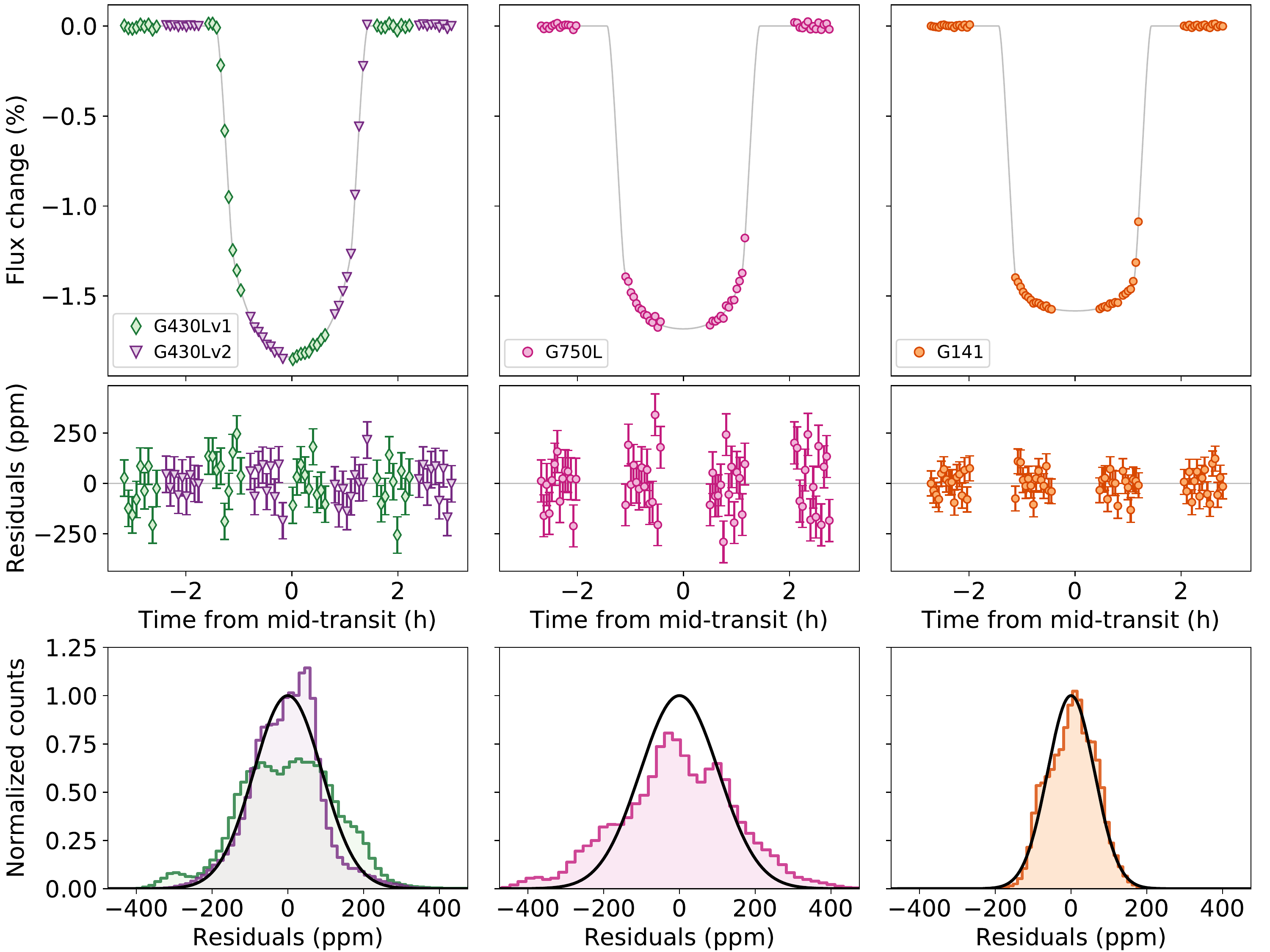}
\caption{White lightcurves for G430L, G750L, and G141 datasets analyzed in this study. \textit{(Top row)} Relative flux variation after removing the systematics contribution inferred from the GP analyses (see Figure \ref{fig:whitefit_gps}), with best-fit transit signals plotted as solid lines. \textit{(Middle row)} Corresponding model residuals, with photon noise errorbars. \textit{(Bottom row)} Normalized histograms of residuals obtained by subtracting from the data a random subset of GP mean functions obtained in the MCMC sampling. Solid black lines correspond to normal distributions with standard deviations equal to photon noise (i.e.\ prior to rescaling by the $\beta$ factor described in the main text).}
\label{fig:whitefit_lcs}
\end{figure}

Inferred values for $\RpRs$ can be biased by differences in the assumed values for $\aRs$ and $b$ across datasets. For this reason, we held the latter parameters fixed to the HST weighted-mean values determined in the previous section and repeated the white lightcurve analyses. This is physically motivated by the fact that the true values of $\aRs$ and $b$ should be constant across our datasets, and we are primarily interested in wavelength-dependent variations of $\RpRs$ arising due to the planetary atmosphere.

\begin{figure}
\centering  % this centres figure in column
\includegraphics[width=0.8\columnwidth]{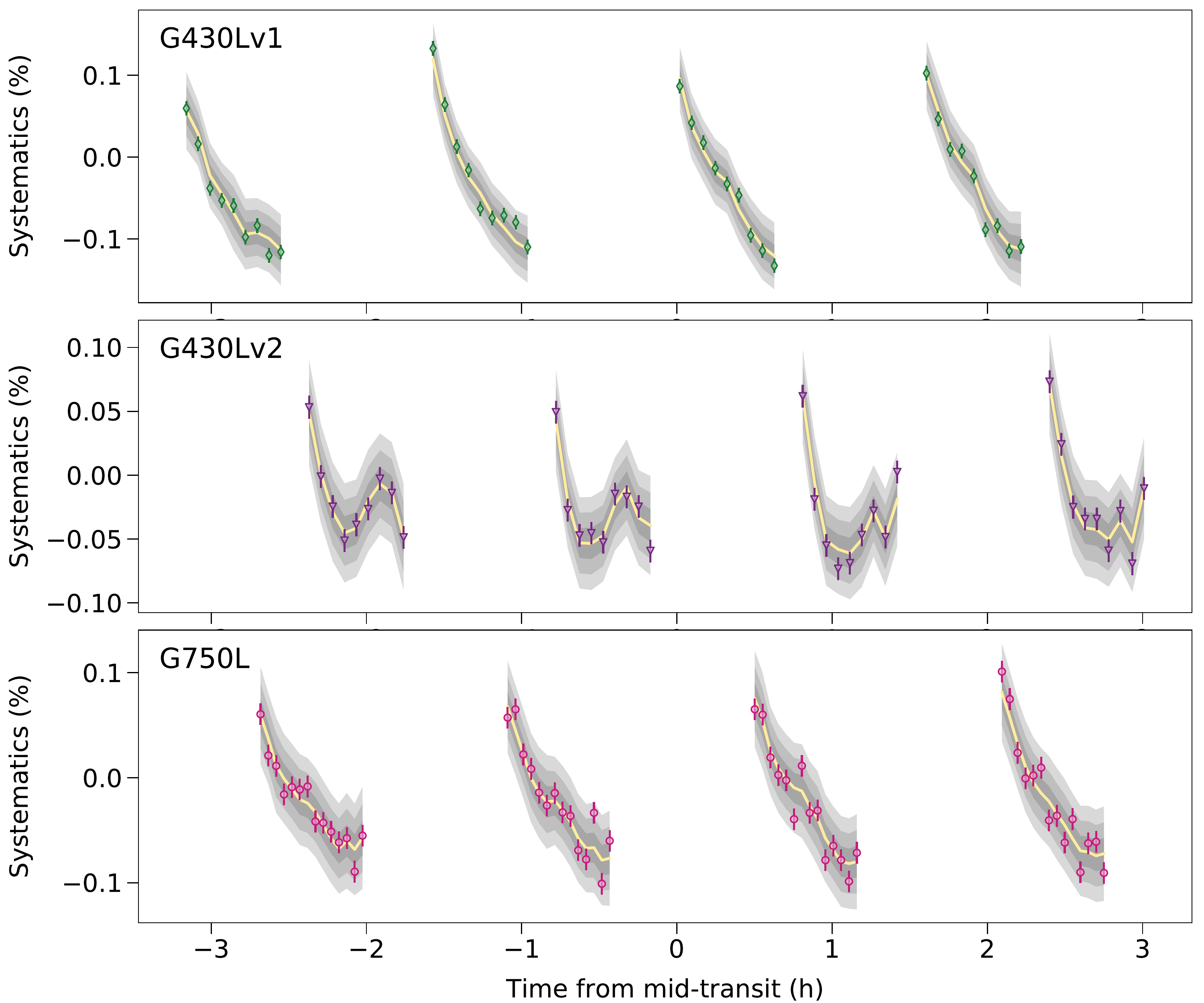}
\caption{Systematics in the white lightcurves for G430L and G750L datasets. Effectively, these are the residuals after dividing the raw flux time series by the transit signals with linear baseline trends shown in Figure \ref{fig:stis_timeseries}. Yellow lines and gray shaded regions, respectively, show the means and $1\sigma$, $2\sigma$, and $3\sigma$ ranges of the best-fit GP distributions. Note that in practice the transit signal, linear baseline trend, and GP are fit simultaneously. The purpose of this figure is only to highlight the structure of the systematics.}
\label{fig:whitefit_gps}
\end{figure}

Figure \ref{fig:whitefit_lcs} shows the best-fit transit models compared with the data after removing the systematics contribution inferred by the GP. The latter are shown separately in Figure \ref{fig:whitefit_gps} and Table \ref{table:whitefit} summarizes the posterior distributions. The resulting estimates for $\RpRs$, $u_1$, and $u_2$ are all within $1\sigma$ of those obtained for the fits in which $\aRs$ and $b$ were allowed to vary. Unsurprisingly, we obtain similar estimates for $\beta$, as this parameter is sensitive to high-frequency noise in the data that is unlikely to be significantly correlated with $\aRs$ and $b$. The inferred $\beta$ values imply scatters that are $\sim 20$--$40$\% and $\sim 10$\% above the photon noise floor for the STIS and WFC3 datasets, respectively. This is illustrated in Figure \ref{fig:whitefit_lcs}, which shows the model residuals. For $\Tmid$, we find the inferred values shift by $\sim 5$--$20$\,sec, but remain within $\sim 1\sigma$ of those obtained for the fits in which $\aRs$ and $b$ were allowed to vary.

\begin{figure}
\centering  % this centres figure in column
\includegraphics[width=\columnwidth]{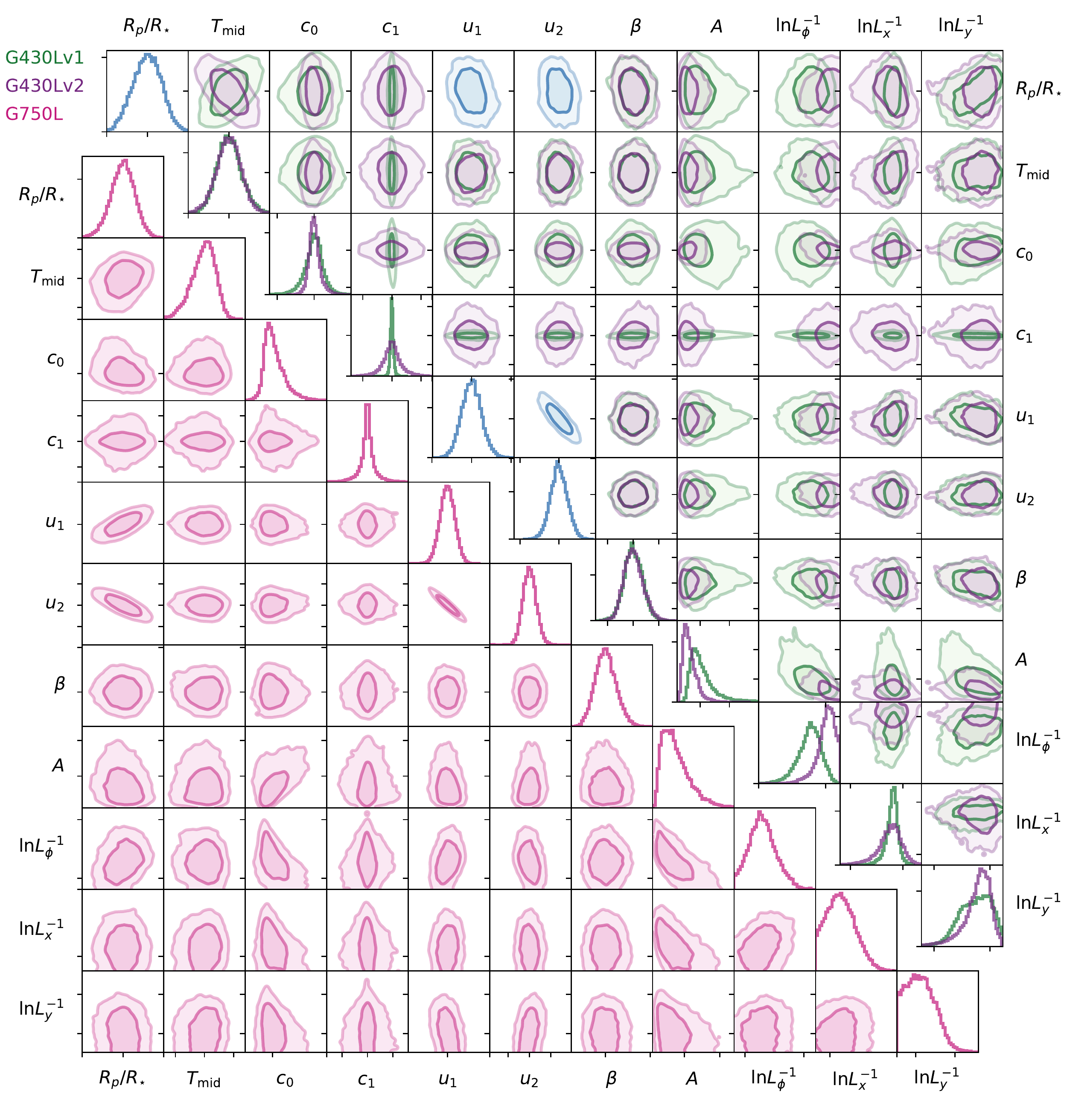}
\caption{Posterior distributions obtained from the white lightcurve analyses described in Section \ref{sec:whitelc:orbparsfixed}. Top right panels show results for the joint analysis of both G430L visits and bottom left panels show results for the G750L analysis. Plotted contours contain 68\% and 95\% of the MCMC samples. Panels along the diagonal show marginalized posterior distributions. Note that $\Tmid$, $c_0$, $c_1$, and $\beta$ have been median-subtracted to allow both G430L visits to be plotted on the same axes. The purpose of this figure is to visually illustrate correlations between model parameters. Numerical values for all parameter distributions are summarized in Table \ref{table:whitefit}.}
\label{fig:whitefit_post}
\end{figure}

%---------------------------------------------------------------
\section{Spectroscopic Lightcurve Analyses} \label{sec:speclcs}

Spectroscopic lightcurves were constructed by first summing the spectra of each dataset within the wavelength channels shown in Figure \ref{fig:example_spectra}. Median channel widths were: 20 pixels ($\sim 55$\,\AA) for both G430L datasets; 20 pixels ($\sim 98$\,\AA) for the G750L dataset; and 4 pixels ($\sim 186$\,\AA) for the G141 dataset. Care was taken to avoid the edges of prominent stellar lines and to maintain similar levels of flux within each channel. Thus, subsets of the G430L and G750L channels were broader than these nominal widths. The resulting raw lightcurves for the STIS datasets are shown in Figures \ref{fig:speclcs_raw_g430lv1}-\ref{fig:speclcs_raw_g750l}.

We next generated common-mode (i.e.\ wavelength-independent) signals for each dataset by dividing the raw white lightcurves by the corresponding best-fit transit signals obtained in Section \ref{sec:whitelc} and shown in Figure \ref{fig:whitefit_lcs}. Each of the raw spectroscopic lightcurves were then divided by the resulting common-mode signals. Note that in addition to removing common-mode systematics, this latter step also has the effect of dividing each spectroscopic lightcurve by the intrinsic scatter of the white lightcurve. However, this is acceptable, as the spectroscopic lightcurves have a larger intrinsic scatter than the white lightcurves: dividing white noise by lower-amplitude white noise should on average have zero net effect on the scatter of the resulting corrected lightcurves. Meanwhile, applying a common-mode correction of this nature -- as opposed to dividing through by the best-fit systematics model from the white lightcurve fits -- has the potential advantage of removing systematics in the spectroscopic lightcurves that may not be captured by our white lightcurve systematics model. The common-mode corrected lightcurves for the STIS datasets are shown in Figures \ref{fig:speclcs_cmcorr_g430lv1}-\ref{fig:speclcs_cmcorr_g750l}.

\begin{figure}
\centering  % this centres figure in column
\includegraphics[width=0.97\columnwidth]{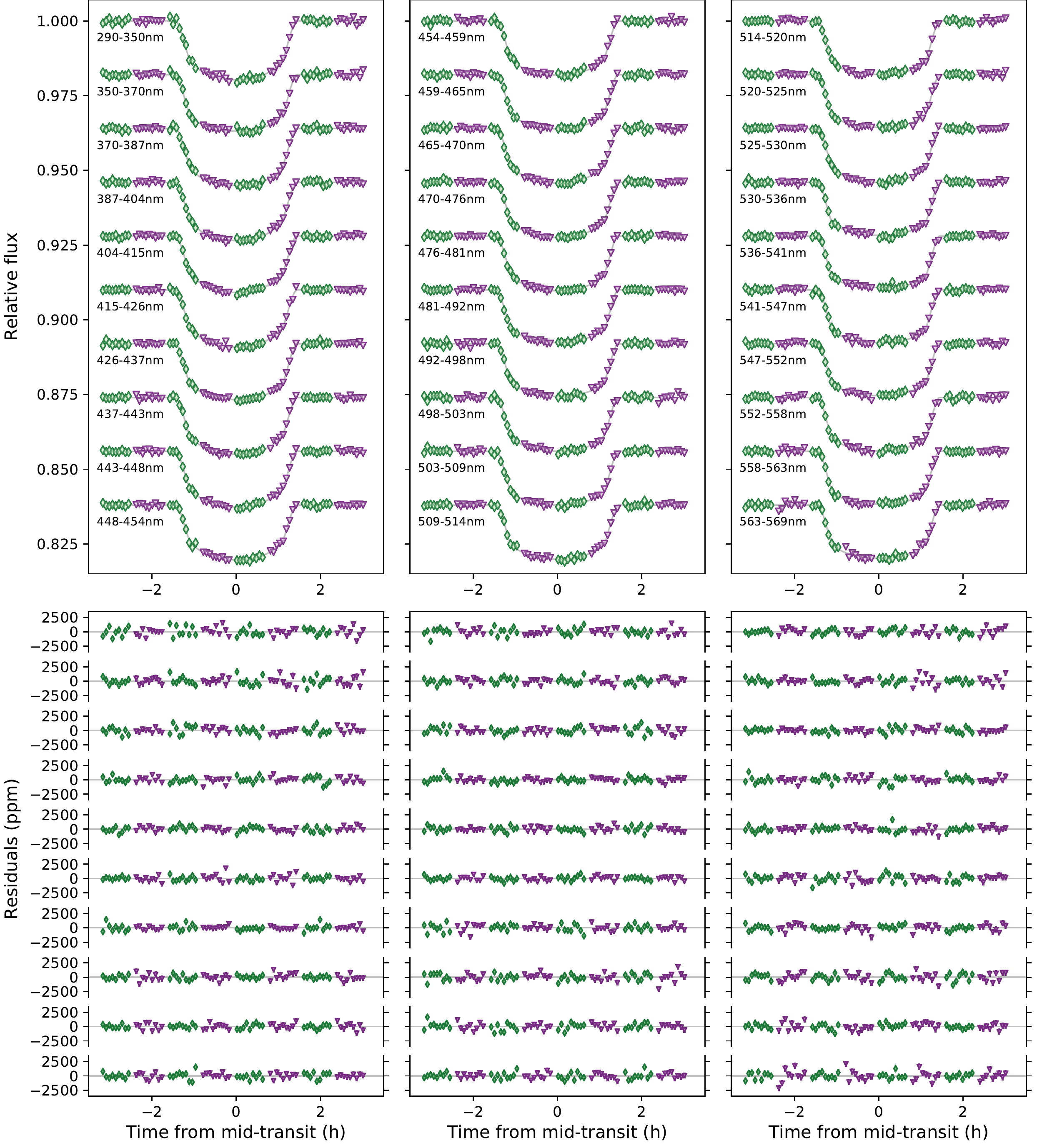}
\caption{Spectroscopic lightcurves for the G430Lv1 and G430Lv2 datasets after removing the systematics contributions inferred from the GP analyses, with best-fit transit signals plotted as solid lines. Green triangles and purple diamonds correspond to the G430Lv1 and G430Lv2 datasets, respectively.}
\label{fig:specfits_g430l}
\end{figure}

To fit the spectroscopic lightcurves, we used the same approach as described in Section \ref{sec:whitelc}. The only exception was that we fixed $\Tmid$ to the best-fit values listed in Table \ref{table:whitefit}. Thus, for the spectroscopic transit signals, the free paramaters were the radius ratio ($\RpRs$) and quadratic limb darkening coefficients ($u_1$, $u_2$). For the G430L analysis, we fit both visits jointly with shared values for $\RpRs$, $u_1$, and $u_2$, as was done for the white lightcurve analysis. In all fits, we again accounted for systematics by fitting for a linear trend in $t$ and a GP with $\{ \phi, x, y \}$ as inputs to a squared-exponential covariance kernel. White noise levels were allowed to vary for each individual lightcurve via $\beta$ rescaling parameters. Marginalization of the posterior distributions was performed in the manner described above, using affine-invariant MCMC.

\begin{figure}
\centering  % this centres figure in column
\includegraphics[width=0.97\columnwidth]{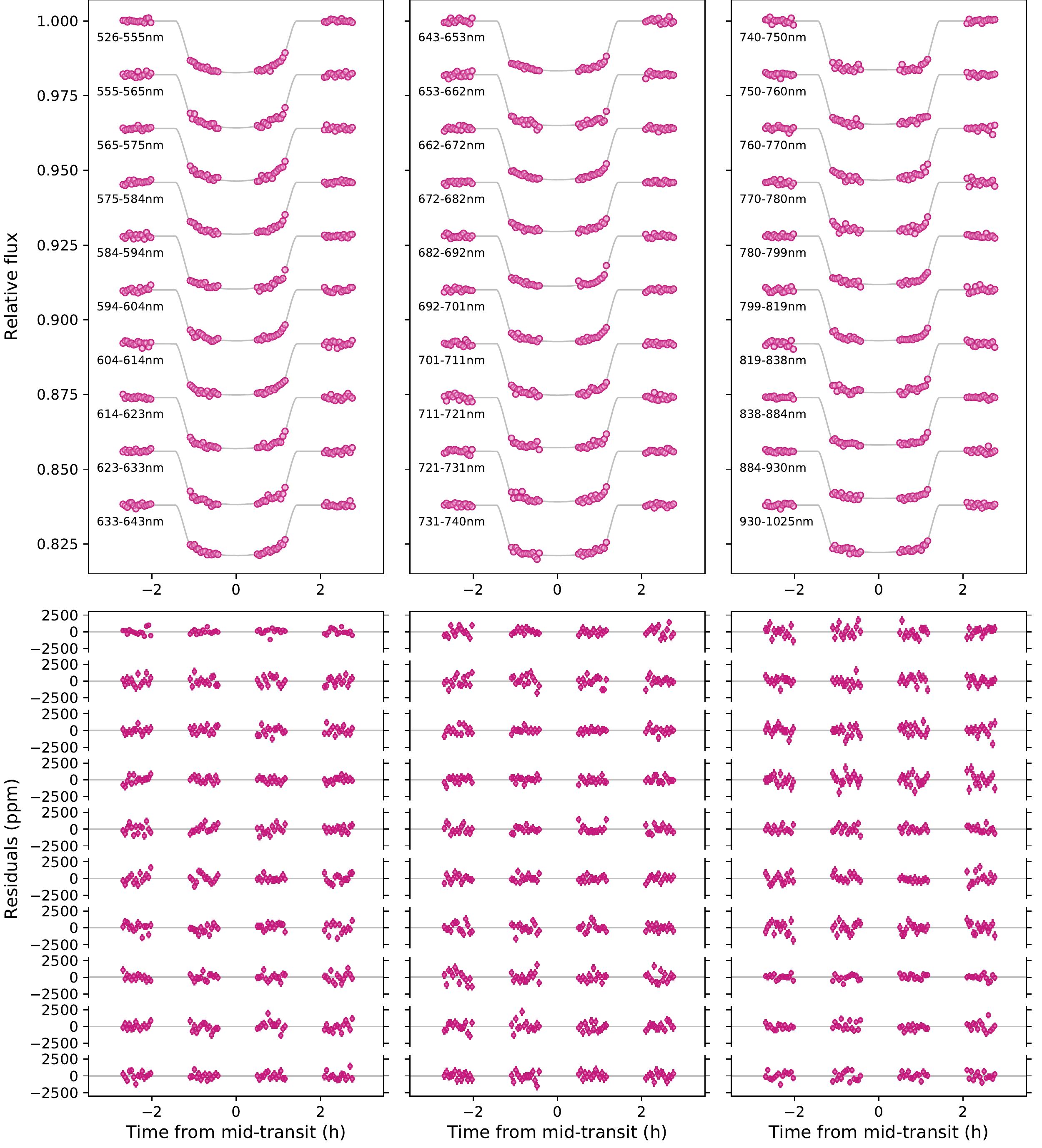}
\caption{Similar to Figure \ref{fig:specfits_g430l}, but for the G750L spectroscopic lightcurves. }
\label{fig:specfits_g750l}
\end{figure}

\begin{figure}[b!]
\centering  % this centres figure in column
\includegraphics[width=\columnwidth]{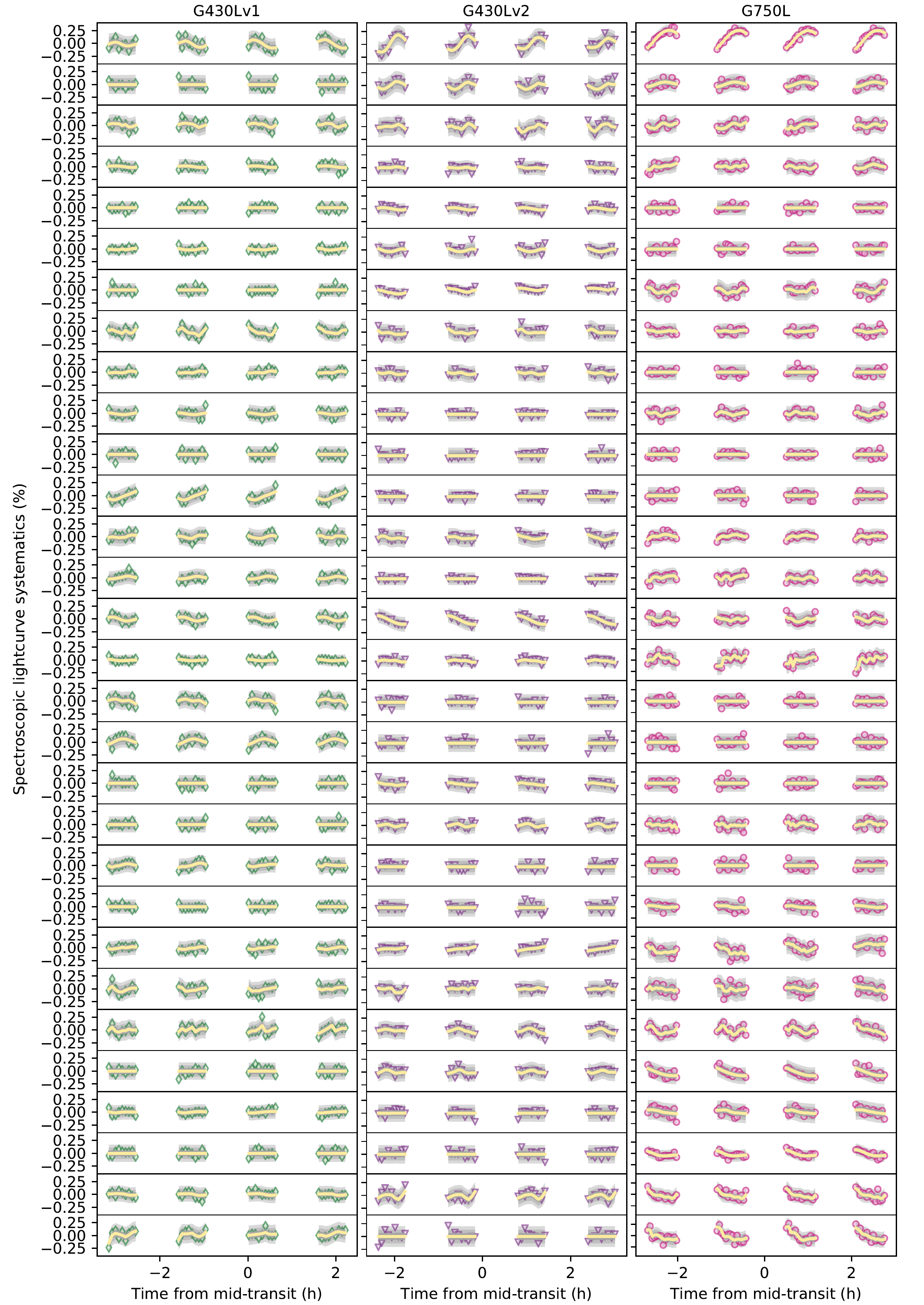}
\caption{Similar to Figure \ref{fig:whitefit_gps}, but showing the systematics and GP fits for the spectroscopic lightcurves. In all columns, wavelength increases from top to bottom.}
\label{fig:specfits_gps}
\end{figure}

The best-fit transit signals and model residuals are shown in Figure \ref{fig:specfits_g430l} for G430L and Figure \ref{fig:specfits_g750l} for G750L. Figure \ref{fig:specfits_gps} shows the systematics and GP fits for each spectroscopic lightcurve. Histograms of residuals are shown in Figures \ref{fig:speclcs_cmcorr_g430lv1}-\ref{fig:speclcs_cmcorr_g750l}. For the G141 spectroscopic lightcurve fits, the results were essentially identical to those presented in \cite{2016ApJ...822L...4E}, so we do not duplicate them here. The only difference for the latter is a wavelength-uniform shift of $\RpRs$ by $0.0007$, in line with the revised white lightcurve analysis which gives $\RpRs = 0.1218 \pm 0.0004$ (Table \ref{table:whitefit}), compared with the previous estimate of $\RpRs = 0.1211 \pm 0.0003$ \citep{2016ApJ...822L...4E}.\footnote{The revised value for $\RpRs$ within the G141 bandpass can be attribued to the updated values for $\aRs$ and $b$ adopted in the present study (Section \ref{sec:whitelc:orbparsfree}).}

\begin{figure}
\centering  % this centres figure in column
\includegraphics[width=\columnwidth]{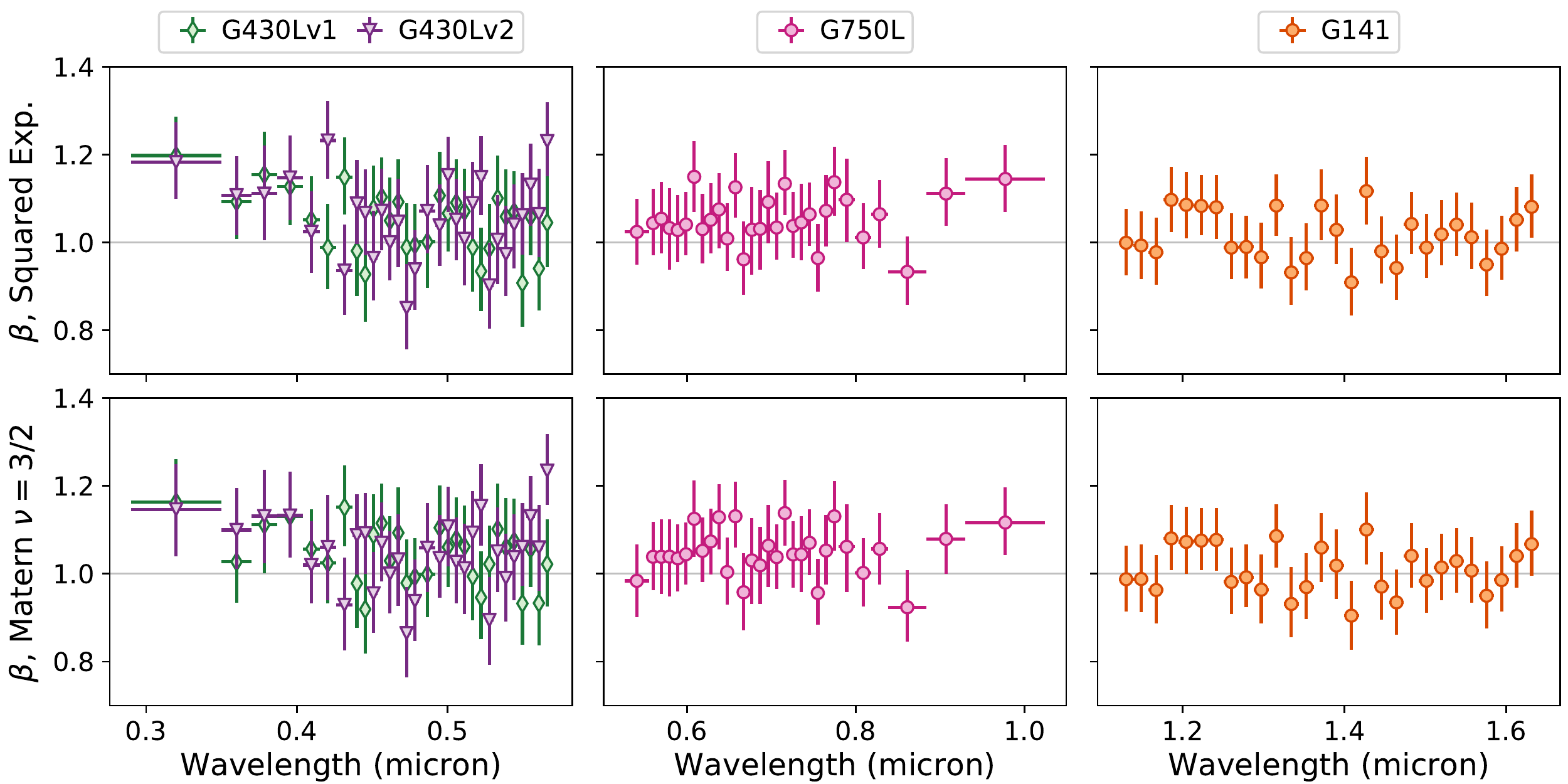}
\caption{\textit{(Top row)} Inferred white noise rescaling parameters $\beta$ for the GP analyses adopting a squared-exponential covariance kernel. \textit{(Bottom row)} The same, but for the GP analyses adopting a Mat\'{e}rn $\nu=3/2$ covariance kernel.}
\label{fig:speclc_betapars}
\end{figure}

As shown in Figure \ref{fig:speclc_betapars}, we obtain means and standard deviations for the inferred $\beta$ values across spectroscopic channels of: $1.05 \pm 0.07$ for the G430lv1 dataset; $1.06 \pm 0.09$ for the G430Lv2 dataset; $1.05 \pm 0.05$ for the G750L dataset; and $1.02 \pm 0.06$ for the G141 dataset. The consistency of these results with $\beta = 1$ indicate that the GP models are broadly successful at marginalizing over the correlations in the lightcurves, implying in turn that degeneracies between the systematics and planet signal are properly accounted for in our estimates of parameters such as $\RpRs$, which we are primarily interested in.

To investigate the sensitivity of our results to the choice of covariance kernel, we repeated the spectroscopic lightcurve fitting using the Mat\'{e}rn $\nu=3/2$ kernel, which can be more suitable for modeling high-frequency signals than the squared-exponential kernel \citep[e.g.\ see][]{2012MNRAS.419.2683G}. For all channels, we found the inferred $\RpRs$ values remained unchanged to well-within $1\sigma$, regardless of which covariance kernel was used. However, the $\beta$ values inferred using the Mat\'{e}rn $\nu=3/2$ kernel were on average slightly closer to unity, as illustrated in Figure \ref{fig:speclc_betapars}. This suggests some of the channels may contain high-frequency noise that can be suitably accounted for either by inflating the white-noise level above the photon noise floor via $\beta>1$ or by employing a covariance kernel with enough flexibility to marginalize over signals of this nature, such as the Mat\'{e}rn $\nu=3/2$. Given the results for $\RpRs$ are found to be insensitive to the choice of covariance kernel, we adopt those obtained using the squared-exponential for the remainder of this paper.

The corresponding posterior distributions for $\RpRs$, $u_1$, and $u_2$ are summarized for each STIS dataset in Tables \ref{table:specfit_g430l} and \ref{table:specfit_g750l}. The median uncertainties on $\RpRs$ are 800\,ppm for G430L, 900\,ppm for G750L, and 500\,ppm for G141, which translates to uncertainties on the transit depth $(\RpRs)^2$ of approximately 200\,ppm, 220\,ppm and 125\,ppm, respectively. For comparison, a change in the effective planetary radius of one atmospheric pressure scale height $H$ corresponds to a transit depth variation of $\sim 150$--$200$\,ppm for WASP-121b, assuming average limb temperatures in the range of $1500$--$2000$\,K, a planetary surface gravity of 940\,cm\,s$^{-2}$, and an atmospheric mean molecular weight of $\mu = 2.22$ atomic mass units (i.e.\ equal to that of Jupiter).

%---------------------------------------------------------------
%\clearpage
\section{Discussion} \label{sec:discussion}

\begin{figure}[t!]
\centering  % this centres figure in column
\includegraphics[width=\columnwidth]{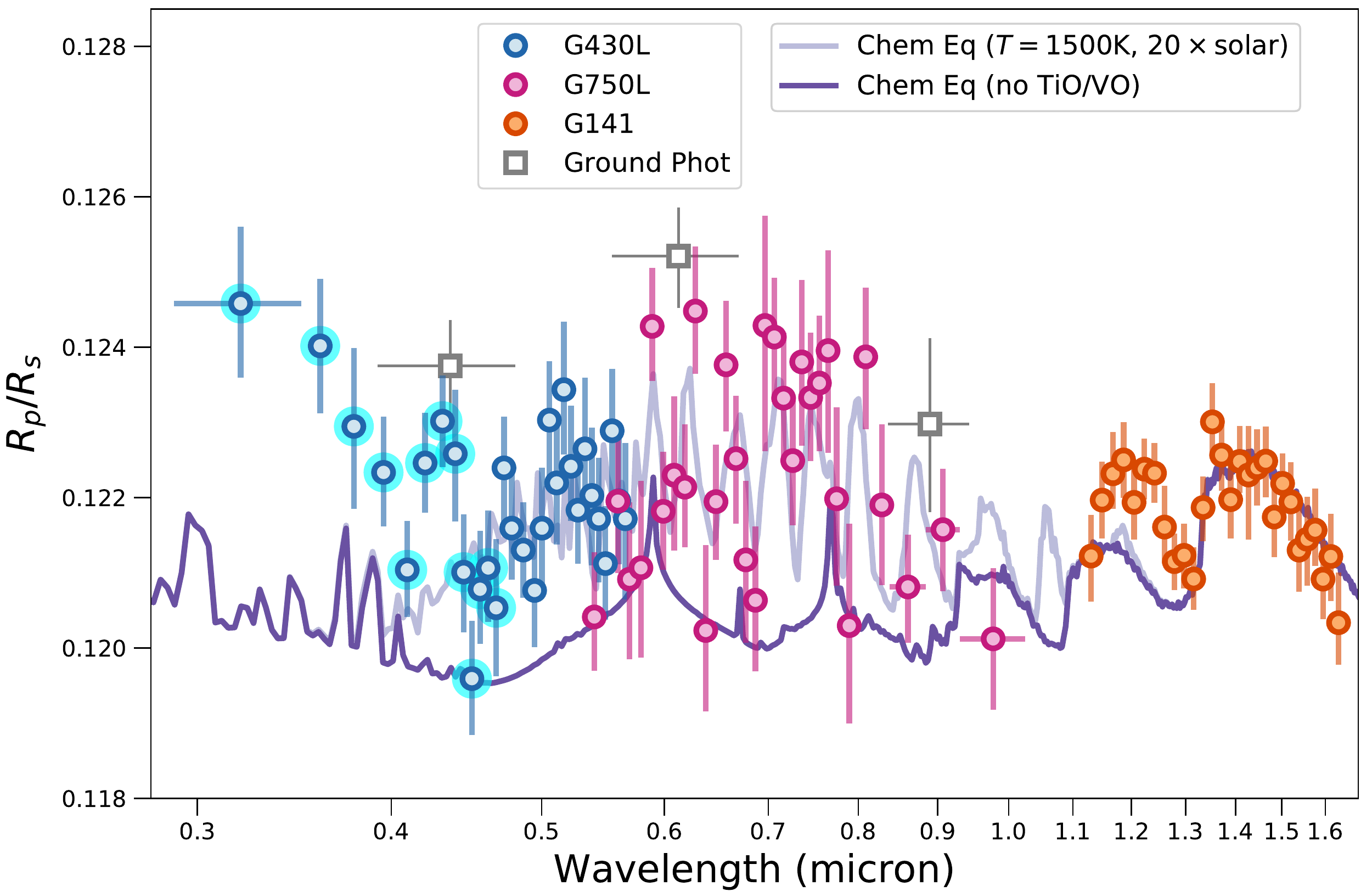}
\caption{Transmission spectrum for WASP-121b obtained using STIS and WFC3 (colored circles) and ground-based photometry from \cite{2016MNRAS.tmp..312D} (unfilled squares). Note that the latter are taken from the re-analysis of \cite{2016ApJ...822L...4E}, although very similar results were obtained by Delrez et al. Light blue halos indicate the subset of G430L data that we refer to as the blue data in the main text. Two forward models assuming chemical equilibrium are also shown, both with a temperature of $1500$\,K and $20\times$ solar metallicity. One model includes TiO/VO opacity (light purple line) and the other does not include TiO/VO opacity (dark purple line).}
\label{fig:trspec_stiswfc3}
\end{figure}

The measured transmission spectrum is shown in Figure \ref{fig:trspec_stiswfc3} and has a number of notable features. In particular, the G430L data exhibit a steep rise toward shorter wavelengths from $\sim 0.47\,\um$, where $\RpRs \sim 0.121$, to $\sim 0.28\,\um$, where $\RpRs \sim 0.125$. This corresponds to a change in effective planetary radius of approximately five pressure scale heights. At longer optical wavelengths covered by the G430L and G750L gratings ($\sim 0.47$--$1\,\um$), $\RpRs$ is measured to vary across spectroscopic channels, implying a wavelength-dependent atmospheric opacity. Within some of these optical channels, the atmospheric opacity is found to be even higher than that measured within the H$_2$O band at $1.4\,\um$, which is detected in the G141 bandpass \citep{2016ApJ...822L...4E}.

Figure \ref{fig:trspec_stiswfc3} also shows the $\RpRs$ values measured using ground-based photometry in the $B$, $r^\prime$, and $z^\prime$ bandpasses. The latter were originally reported by \cite{2016MNRAS.tmp..312D} and an independent analysis of the same lightcurves was presented by \cite{2016ApJ...822L...4E}. Both studies obtained similar estimates for $\RpRs$ in each bandpass that are larger than those obtained from the HST data. It is unclear what is responsible for this tension. As noted above, updated values for $\aRs$ and $b$ were used in the lightcurve fits of the present study. However, for the G141 dataset, this had the effect of shifting the mean $\RpRs$ value to a higher value from $0.1211 \pm 0.0003$ \citep{2016ApJ...822L...4E} to $0.1218 \pm 0.0004$ (Table \ref{table:whitefit}). A similar upward shift for the ground-based photometry would make those data more discrepant relative to the HST data. Alternatively, during the photometry data reduction, effects such as aperture light-losses or an over-estimated background may have artificially deepened the transit signals, resulting in $\RpRs$ estimates above the true values. Another more speculative possibility is intrinsic variability of the atmosphere from epoch to epoch. For example, \cite{2013A&A...558A..91P} report a 3D GCM study showing that significant variations in passive tracer abundances over $\sim 100$ day timescales are possible at the planetary limb of hot Jupiters. As those authors note, if this occurs for strongly-absorbing species such as TiO and VO, it could have significant implications for transmission spectra measured at different epochs. Indeed, the ground-based photometry and HST STIS observations were separated by over $100$ days. However, the current data are insufficient to test this theory, and we consider it more likely that the difference is due to some unaccounted-for systematic in the ground-based photometry.

To evaluate the robustness of the HST transmission spectrum, we performed a number of tests, full details of which are reported in Appendix \ref{app:robustness}. First, we find that the measured transmission spectrum is insensitive to our treatment of limb darkening. Second, we investigated the inclusion of time $t$ as an additional GP input variable in the lightcurve fits and obtain very similar results to those reported here. Third, for the G430L data, we find that the measured transmission spectrum is repeatable when each of the two visits are analyzed separately. Fourth, we conclude that stellar activity is unlikely to have significantly affected the measured transmission spectrum, based on: (1) the lack of photometric variability and modest X-ray flux of the WASP-121 host star; (2) the epoch-to-epoch repeatability of the G430L datasets; (3) the good level of agreement obtained across the overlapping wavelength range of the G430L and G750L datasets; and (4) the inability of unocculted spots to explain the shape of the measured spectrum under reasonable assumptions. In the following sections, we therefore seek to interpret the measurements shown in Figure \ref{fig:trspec_stiswfc3} as the signal of the planetary atmosphere.

\subsection{Rayleigh scattering and a gray cloud-deck} \label{sec:discussion:scattering}

The signature of aerosol scattering is ubiquitous in observations of exoplanet atmospheres \citep[e.g.][]{2008MNRAS.385..109P,2014Natur.505...69K,2014MNRAS.437...46N,2015MNRAS.447..463N,2015MNRAS.446.2428S}. For hot Jupiter transmission spectra, this is unsurprising given the large number of refractory species expected to condense at the temperatures and pressures characteristic of these atmospheres \citep[e.g.][]{2018A&A...614A...1W}, as well as the highly-sensitive nature of the grazing geometry to even trace opacity sources \citep{2005MNRAS.364..649F}. Indeed, the rise in opacity toward shorter wavelengths that we measure for WASP-121b is somewhat reminiscent of transmission spectra previously obtained for other hot Jupiters, which can be explained by Rayleigh scattering due to high-altitude layers of submicron aerosols \citep{2016Natur.529...59S}. In addition, an optically-thick cloud deck could act as a gray opacity source, if present at low pressures.

We investigated how well the WASP-121b transmission spectrum can be explained by aerosols by first fitting simple Rayleigh scattering and cloud deck models to the STIS data spanning the G430L and G750L gratings. For the Rayleigh component, we followed the methodology outlined by \cite{2008A&A...481L..83L} (L08), who provide relations between the slope of the transmission spectrum and the atmospheric temperature, under the assumption of scattering particles distributed uniformly with pressure. For the cloud deck component we assumed a wavelength-independent opacity, implemented as a horizontal flat line in $\RpRs$ that was allowed to float vertically relative to other spectral features in the transmission spectrum. For this initial analysis, we excluded the G141 dataset, as it exhibits a clear spectral feature due to H$_2$O, which would add additional complexity to the model. This is addressed in Section \ref{sec:discussion:retrieval}, where we perform a free-chemistry fit to the combined STIS+WFC3 dataset that includes opacity due to both gas-phase species and aerosols.

\begin{figure}
\centering  % this centres figure in column
\includegraphics[width=\columnwidth]{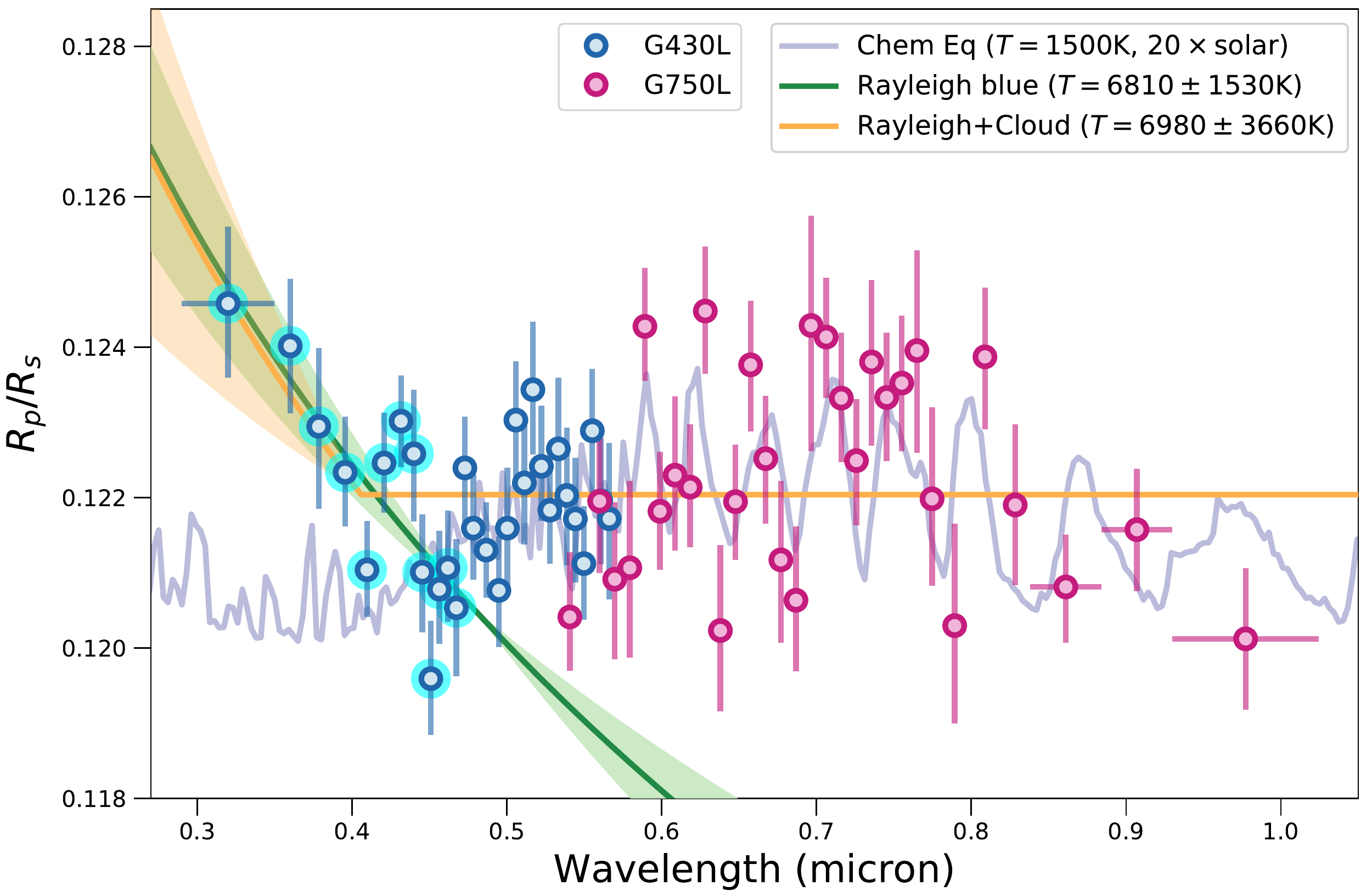}
\includegraphics[width=\columnwidth]{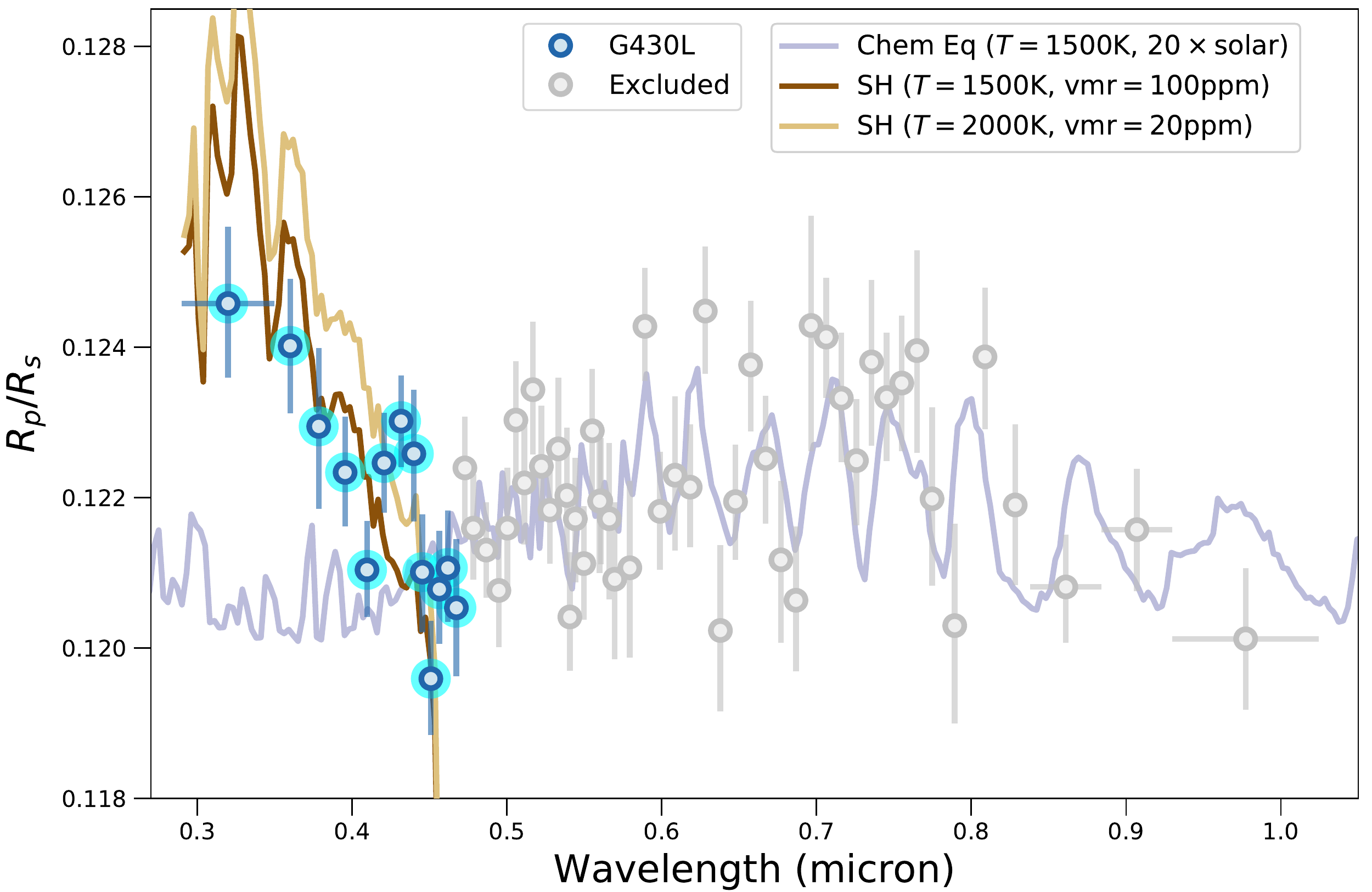}
\caption{Similar to Figure \ref{fig:trspec_stiswfc3}, but showing only the STIS data. \textit{(Top panel)} Rayleigh scattering fit to the NUV data only (green line) and a hybrid Rayleigh+cloud model fit to the complete STIS dataset (yellow line). Although Rayleigh scattering gives a good fit to the NUV data, it requires invoking an unphysically high temperature. The Rayleigh+cloud model is ruled out at $3.7\sigma$ confidence, due to the opacity variations measured across optical wavelengths. \textit{(Bottom panel)} Models illustrating the expected opacity contribution due to SH for temperatures of 1500\,K and 2000\,K with volume mixing ratios $100$\,ppm and $20$\,ppm, respectively (brown lines).}
\label{fig:trspec_nuvopt}
\end{figure}

Our best-fit model combining a Rayleigh slope with cloud-deck is shown in Figure \ref{fig:trspec_nuvopt}. It provides a poor fit to the data, with a reduced $\chi^2$ of 1.8 for 57 degrees of freedom, allowing us to exclude it at $3.7\sigma$ confidence. This is due to the inability of a featureless cloud deck to explain the optical data across the $0.47$--$1\,\um$ wavelength range. Furthermore, the temperature inferred from the Rayleigh slope is $6980 \pm 3660$\,K, which is improbably high for the atmospheric pressures probed in transmission. For instance, if WASP-121b absorbs all incident radiation on its dayside hemisphere (i.e.\ the Bond albedo is zero), then the substellar point would have a temperature of $T_\star/\sqrt{a/R_\star} \sim 3280$\,K and the day-night boundary probed by the transmission spectrum should be considerably cooler. Furthermore, at such high temperatures, no condensates are expected to exist and molecules should be thermally dissociated, including H$_2$.

To be conservative, we also tried dividing the NUV-optical data into different wavelength sections and fitting them one at a time. For convenience, we will refer to these subsets as the blue ($0.3$--$0.47\,\um$) and red ($0.47$--$1\,\um$) data. In principle, a good fit to one or both of these datasets separately should be easier to achieve than a good joint fit, as the models need not be self-consistent.

First, we fit a Rayleigh profile to the blue data, as this is where the transmission spectrum exhibits a strong slope. Although we obtain a better statistical fit with a reduced $\chi^2$ of 1.6 for 10 degrees of freedom (Figure \ref{fig:trspec_nuvopt}), the inferred temperature remains implausibly high at $6810 \pm 1530$\,K. Given this, we conclude that the rise in the measured transmission spectrum toward NUV wavelengths is too steep to be explained by scattering out of the transmission beam. Instead, it would suggest the presence of one or more significant NUV absorbers in the upper atmosphere of WASP-121b, assuming the slope is indeed a feature of the planetary spectrum and not caused by an uncorrected systematic effect in the data.

Second, we tried fitting a gray cloud deck to the G430L and G750L data, with the blue G430L subset excluded. For this scenario (not shown in Figure \ref{fig:trspec_nuvopt}), we obtain a reduced $\chi^2$ of 1.9 for 51 degrees of freedom, which formally rules it out at $3.8\sigma$ confidence. Alternatively, if the $\RpRs$ uncertainties for these optical data have been uniformly underestimated by $\sim 30$\%, this gray cloud scenario would only be excluded at $\sim 1\sigma$ confidence. However, lacking any reason to doubt our inferred $\RpRs$ uncertainties, we propose instead that the optical data exhibit significant spectral variations that cannot be explained by a gray cloud deck.

\subsection{Forward model comparison with optical-NIR data} \label{sec:discussion:forward}

The results of the previous section imply the transmission spectrum of WASP-121b exhibits significant wavelength-dependent opacity variations across the $\sim 0.47$--$1\,\um$ wavelength range. To explore this further, we used the \texttt{ATMO} code \citep{2014A&A...564A..59A,2016A&A...594A..69D,2018MNRAS.474.5158G,2015ApJ...804L..17T,2016ApJ...817L..19T} to generate a small grid of aerosol-free atmosphere models spanning temperature and metallicity, assuming isothermal pressure-temperature (PT) profiles and chemical equilibrium abundances. Specifically, our grid consisted of temperatures ranging from $1000$\,K to $2700$\,K in $100$\,K increments, each evaluated for metallicities of $0.1\times$, $1\times$, $10\times$, $20\times$, $30\times$, $40\times$, and $50\times$ solar. \texttt{ATMO} solves for the gas-phase and condensed-phase chemical equilibrium mole fractions for a given pressure, temperature, and set of elemental abundances \citep{2016A&A...594A..69D}. For the results presented here we consider local condensation, such that the chemistry calculation in each model pressure level is entirely independent of all other pressure levels. We do not account for rainout chemistry, under which condensation deeper within the atmosphere depletes elemental abundances at lower pressures levels \citep{1999ApJ...512..843B,2011ApJ...737...34M,2016ApJ...827..121M}. Rainout could be important in the atmosphere of WASP-121b, but we defer investigation of this effect to future work that includes a more realistic treatment of the PT profile than the isothermal assumption made here. Finally, we applied uniform vertical offsets to $\RpRs$ for each model in order to optimize the match to the data. No further tuning of the models was performed.

None of these equilibrium models are able to explain the absorption at wavelengths shortward of $0.47\,\um$, nor the G141 bump between wavelengths of $1.15$--$1.3\,\um$. We discuss these latter two components of the transmission spectrum further in Sections \ref{sec:discussion:nuv} and \ref{sec:discussion:retrieval}, respectively. For the remaining data -- namely, the STIS data spanning the $0.47$--$1\,\um$ wavelength range and the WFC3 data covering the H$_2$O band centered at $1.4\,\um$ -- we find a good match is obtained for the model with a temperature of $1500$\,K and metallicity of $20 \times$ solar (Figure \ref{fig:trspec_stiswfc3}), which has a reduced $\chi^2$ of 1.0 for 69 degrees of freedom. Similarly good matches to the data are obtained for the $1500$\,K models with metallicities of $10\times$ and $30\times$ solar. These metallicities are broadly consistent with predictions for a $1.18\Mjup$ planet such as WASP-121b \citep{2016ApJ...831...64T}.

\begin{figure}
\centering  % this centres figure in column
\includegraphics[width=\columnwidth]{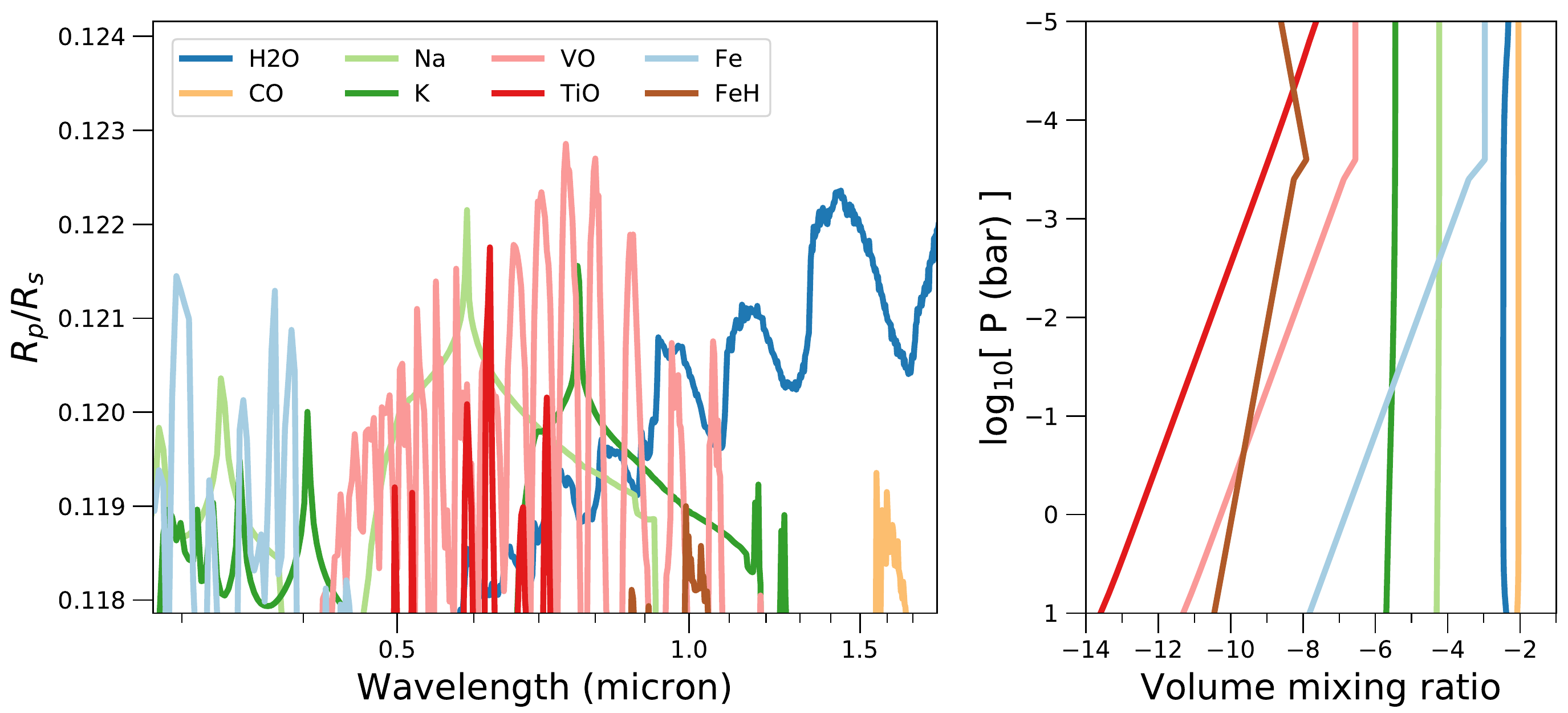}
\caption{(\textit{Left panel}) Individual contributions to the transmission spectrum due to the major radiatively-active species in the best-match forward model shown in Figure \ref{fig:trspec_stiswfc3}, i.e.\ chemical equilibrium for $T=1500$\,K and $20\times$ solar metallicity. Note that continuum opacity due to gas-phase species such as H$_2$ and He is not shown. (\textit{Right panel}) Corresponding pressure-dependent abundances.}
\label{fig:trspec_opacityvmrs}
\end{figure}

Aside from collision-induced asorption and gas-phase Rayleigh scattering, the primary opacity sources of these models are Na and VO at optical wavelengths and H$_2$O at NIR wavelengths. This is illustrated in Figure \ref{fig:trspec_opacityvmrs}, which shows a break-down of the opacity sources in the best-matching chemical equilibrium model. Interestingly, opacity due to TiO is not as significant as VO in the optical, despite Ti being approximately an order of magnitude more abundant than V for solar elemental composition \citep{2009ARA&A..47..481A}. This occurs because for a given pressure, the condensation of Ti species commences at higher temperatures than for V species \citep[e.g.][]{1999ApJ...512..843B,2018A&A...614A...1W}. The isothermal temperature of the best-match model (i.e.\ 1500\,K) is less than the condensation temperature of both Ti$_3$O$_5$(s) and V$_2$O$_3$(s), meaning that these are the dominant forms of Ti and V in the model, respectively. However, since the isothermal temperature is closer to the VO(g)/V$_2$O$_3$(s) condensation temperature than the TiO(g)/Ti$_3$O$_5$(s) condensation temperature, the abundance of VO(g) is larger than for TiO(g). 

In contrast, \cite{2002ApJ...577..974L} found that calcium titanates (e.g.\ CaTiO$_5$) -- which are not currently included in \texttt{ATMO} -- are likely to be the first Ti-bearing condensates to form. Furthermore, arguing from trends in solar system meteorite data and M/L dwarf spectra, Lodders notes that V will likely condense in solid solution with the calcium titanates, resulting in VO gas-phase depletion commencing at the same temperature as TiO gas-phase depletion. However, for hot Jupiters, Ti and V condensation may depend on condensation and mixing timescales, both vertical and horizontal, that are very different to the protostellar nebula and M/L dwarfs. Such details are complex and beyond the scope of the present study. At this stage, we simply note that VO absorption is favored by these HST data for WASP-121b, with no evidence for significant TiO absorption, an interpretation that is corroborated by the free-chemistry retrieval presented in Section \ref{sec:discussion:retrieval} below.

We also note that the best-matching forward model temperature of $1500$\,K is substantially cooler than that of the dayside photosphere, which is inferred to be $\sim 2700$\,K from secondary eclipse measurements \citep{2017Natur.548...58E}. Such a large temperature difference between the dayside photosphere probed during secondary eclipse and the upper atmosphere of the day-night limb probed during primary transit is in fact broadly in line with predictions of 3D general circulation models of ultra-hot Jupiters \citep[e.g.][]{2016ApJ...821....9K}. Furthermore, the best-match temperature of $1500$\,K is likely to be at the lower end of the plausible range, because, as noted above, the forward models we consider here do not include rainout chemistry. Rainout chemistry will likely result in VO condensing at higher temperatures, as the abundance of VO in the upper atmosphere would be determined by the atmospheric temperature profile at higher pressures where the condensation temperature is also higher. Since the appearance or disappearance of VO spectral bands is primarily what determines the ability of our forward models to match the data (Figure \ref{fig:trspec_stiswfc3}), forward models with rainout chemistry would consequently tend to favor higher temperatures. As noted above, we do not consider models with rainout here, as the details will be highly sensitive to the atmospheric PT profile at pressures $>0.1$\,bar, which is unconstrained by the current data.

\subsection{Absorption at NUV wavelengths} \label{sec:discussion:nuv}

We now consider the steep rise in the transmission spectrum at wavelengths shortward of $\sim 0.47\,\um$. As explained in Section \ref{sec:discussion:scattering}, we consider it unlikely that this feature can be explained by Rayleigh scattering due to gas-phase species such as H$_2$ or high-altitude aerosols. In addition, our chemical equilibrium models presented in Section \ref{sec:discussion:forward} do not predict significant absorption above the H$_2$ continuum at these wavelengths. Nonetheless, we find the rise of the transmission spectrum at NUV wavelengths is empirically repeatable. It is recovered by our analysis when the spectroscopic lightcurves for the two G430L visits are fit jointly and also when they are each fit individually (see Section \ref{app:robustness:g430lrepeat}).

One candidate absorber is the mercapto radical, SH, comprised of a sulfur atom and a hydrogen atom. Indeed, SH was predicted by \cite{2009ApJ...701L..20Z} (Z09) to be a strong NUV absorber in hot Jupiter atmospheres. Using a 1D photochemical kinetics code, Z09 found the abundance of SH may peak at pressures around $\sim 1$--$100\,$mbar in typical hot Jupiter atmospheres, with a mixing ratio of $\sim 10$\,ppm (see Figure 2 of Z09). At these pressures, H$_2$S is the most abundant sulfur-bearing phase under chemical equilibrium \citep{2006ApJ...648.1181V}, while atomic H and S are also available due to photodissociation of molecules such as H$_2$ and H$_2$O. The production of SH then proceeds through numerous chemical pathways involving H$_2$S, H, and S \citep[Z09; see also][]{2016ApJ...824..137Z}.

\begin{figure}[t!]
\centering  % this centres figure in column
\includegraphics[width=\columnwidth]{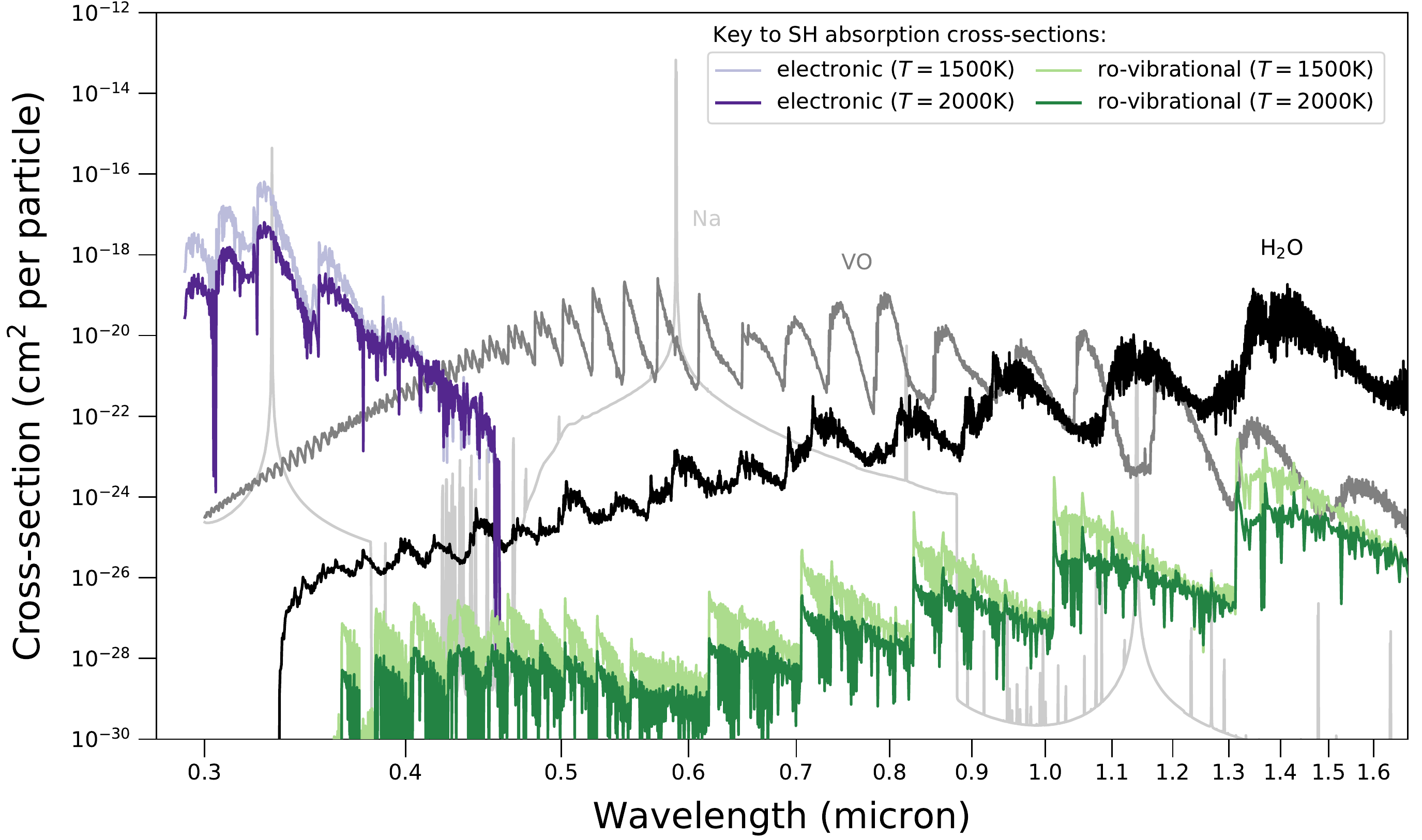}
\caption{Absorption cross-sections for SH. Electronic transitions are from \cite{2009ApJ...701L..20Z} and rotational-vibrational transitions are from ExoMol \citep{2018MNRAS.tmp..913Y}. Cross-sections for H$_2$O, VO, and Na are also shown, weighted by the relative abundances implied by the model shown in Figure \ref{fig:trspec_opacityvmrs}.}
\label{fig:sh_xsects}
\end{figure}

To explore whether or not SH can explain the observed NUV absorption, we performed a simple fit to the 13 shortest-wavelength data points of the transmission spectrum, spanning the $0.3$--$0.47\,\um$ wavelength range (i.e.\ the blue G430L data subset indicated by light blue halos in Figures \ref{fig:trspec_stiswfc3} and \ref{fig:trspec_nuvopt}). As in Section \ref{sec:discussion:scattering}, we followed the methodology outlined in L08. We computed the change in relative planetary radius due to SH absorption, adopting a planetary surface gravity $g = 940$\,cm\,s$^{-2}$ and stellar radius $R_\star = 1.458 \Rsun$ \citep{2016MNRAS.tmp..312D}. We also assumed $\mu = 2.22$ atomic mass units (see Section \ref{sec:speclcs}) and set $\RpRs = 0.120$ as the altitude where H$_2$ becomes optically thick at grazing geometry for a wavelength of $\lambda_0=350$\,nm (Figure \ref{fig:trspec_stiswfc3}), corresponding to a planetary radius of $R_p = 1.702 \,\Rjup$. This in turn translates to an atmospheric pressure of $\sim 20$\,mbar, assuming a temperature of $\sim 1500$--$2000\,$K and an H$_2$ scattering cross-section of $\sigma_0 = 3.51 \times 10^{-27}$\,cm$^2$\,molecule$^{-1}$ for $\lambda_0=350$\,nm \citep[see Section 4.1 of L08; also,][]{2016Natur.529...59S}. Having thus established the pressure scale, we took the temperature-dependent absorption cross-sections for SH and varied the mixing ratio to optimize the match to the NUV transmission spectrum using Equation 1 of L08. For the SH cross-sections, we combined those derived by Z09 with those recently published by the ExoMol project \citep{2018MNRAS.tmp..913Y}. Specifically, the Z09 cross-sections were generated from transitions of the lowest five vibrational levels of the ground electronic state $\text{X}^2\Pi$ to the lowest three vibrational levels of the upper electronic state $\text{A}^2\Sigma^+$ (without predissociation), and exhibit a strong NUV signature. These transitions are not considered in the ExoMol cross-sections, which only account for rotational-vibrational transitions. Both the Z09 and ExoMol cross-sections are shown in Figure \ref{fig:sh_xsects}. 

\begin{figure}[t!]
\centering  % this centres figure in column
\includegraphics[width=0.75\columnwidth]{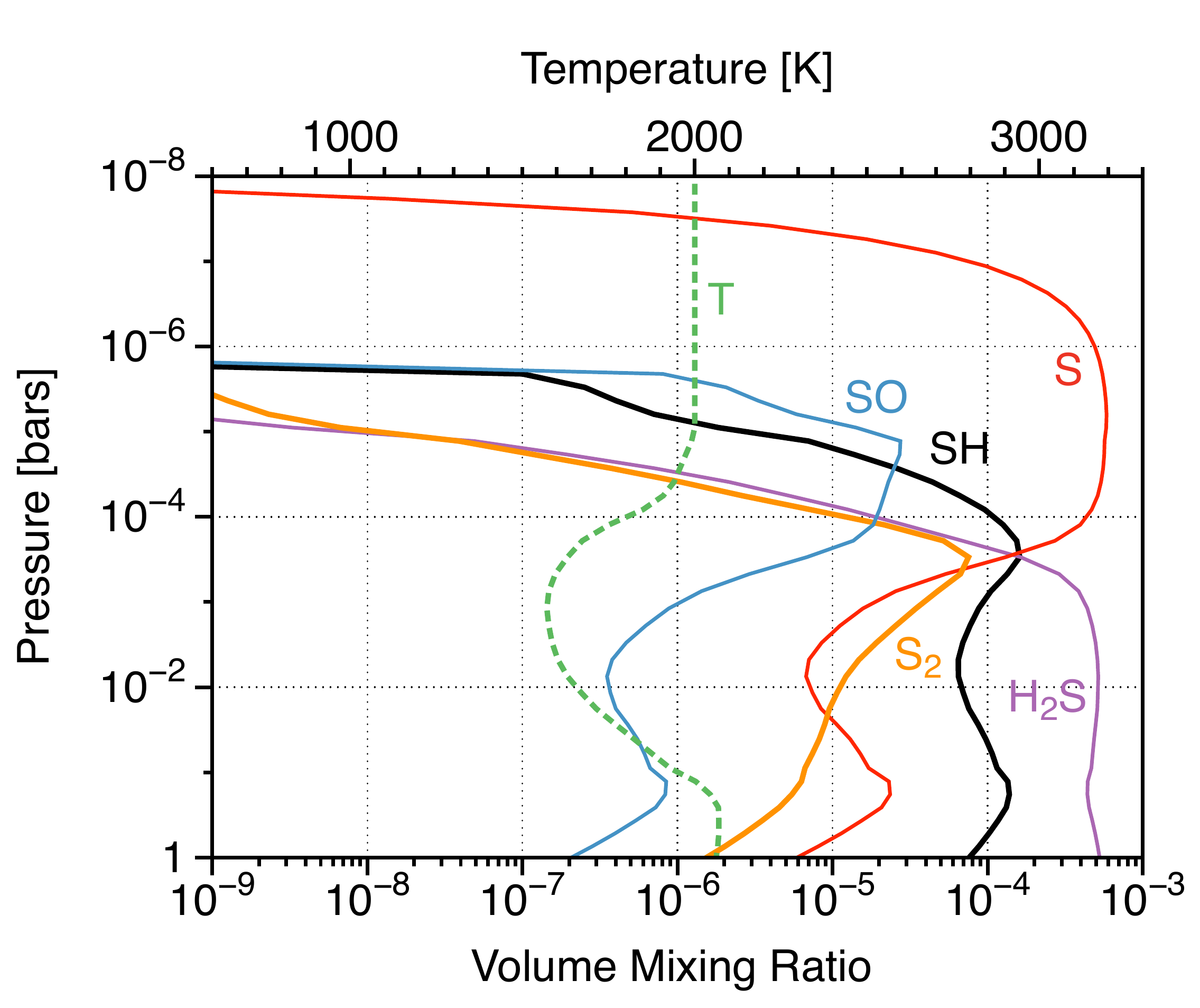}
\caption{Abundance predictions for important sulfur species assuming $20 \times$ solar metallicity and $K_{zz}=10^9$\,cm$^2$\,s$^{-1}$. Dashed green line indicates the adopted PT profile, based on the limb average of a 3D GCM for WASP-121b (Kataria et al., in prep). Calculations were performed using the photochemical kinetics code of \cite{2009ApJ...701L..20Z}, assuming a planet with a hydrogen-dominated atmosphere orbiting an F6V host star at the same distance as WASP-121b.}
\label{fig:sulfur_chem}
\end{figure}

The results of this process are shown in Figure \ref{fig:trspec_nuvopt}. We obtain respectable matches to the data with mixing ratios of $\sim 100$\,ppm and $\sim 20$\,ppm, respectively, for the $T=1500$\,K and $T=2000$\,K absorption cross-sections of Z09. For comparison, Figure \ref{fig:sulfur_chem} shows predicted abundances from the photochemical kinetics code of Z09 for a planet similar to WASP-121b with $20 \times$ solar metallicity and vertical mixing parameter $K_{zz} = 10^9$\,cm$^{2}$\,s$^{-1}$. We find abundances of $\sim 20$--$100$\,ppm are plausible for SH across the bar to mbar pressure range probed by the transmission spectrum, lending some credibility to the hypothesis that it could be the mystery NUV absorber. However, as stressed by Z09, the SH cross-sections remain subject to considerable uncertainty, due to the paucity of available experimental data. This, combined with the low spectral resolution of the G430L data, prevent us from conclusively confirming or ruling out SH at the present time. Other sulfur-bearing compounds that are likely to be abundant, such as SiS, have strong features at NUV wavelengths but remain poorly modeled. \cite{2018ApJ...866...27L} have also flagged gas-phase Fe as an important NUV absorber in ultra-hot Jupiter atmospheres, although we find it is unable to account for the measured signal in the present dataset -- at least under assumptions of equilibrium chemistry for pressures $>10^{-5}$\,bar -- as it was included in the \texttt{ATMO} forward models described in Section \ref{sec:discussion:forward} (see Figure \ref{fig:trspec_opacityvmrs}).

Regardless of the identity of the putative NUV absorber, it likely provides significant heating of the upper atmosphere. For instance, across the $0.3$--$0.47\,\um$ wavelength range, the mean SH absorption cross-section varies from $\sim 10^{-16}$ to $10^{-22}$ \,cm$^{2}$\,molecule$^{-1}$ (Figure \ref{fig:sh_xsects}). Assuming a mixing ratio of $\sim 10$\,ppm, in line with our above estimates, this implies a mean atmospheric opacity (i.e.\ absorption cross-section $\times$ mixing ratio) of $\sim 10^{-24}$\,cm$^{2}$\,molecule$^{-1}$ at the pressures probed in transmission. We find incorporating such an absorber into the 1D radiative-convective atmosphere model of Marley and collaborators \cite[e.g.][]{1999Icar..138..268M,2002ApJ...568..335M,2008ApJ...678.1419F,2008ApJ...689.1327S,2012ApJ...754..135M} would likely heat the atmosphere of WASP-121b by $\sim 500$\,K at mbar pressures. Such heating could, for example, help maintain optical absorbers such as VO and TiO in the gas phase, which in turn would provide further heating of the upper atmosphere. Properly accounting for effects such as these may be important for accurate modeling of planetary circulation and energy budgets.

Absorption of incident UV flux exceeding that predicted by models has also been observed in solar system atmospheres. Two well-known examples are Venus and Jupiter. On Jupiter, a broad reflectivity dip near $0.3\,\um$ has been attributed to a high-altitude dust or haze \citep{1972Icar...16..557O,1972ApJ...173..451A}. The composition of this chromophore is still not known and is generally attributed to some disequilibrium combination of S, N, C, and P species \citep[for a fuller discussion see][]{2004jpsm.book...79W}. Likewise on Venus, dark markings in the atmosphere at UV wavelengths remain poorly understood, well over four decades after their discovery \citep[e.g.,][]{1997veii.conf..415E}. These features have also been attributed to some disequilibrium -- perhaps S-bearing -- absorber \citep[but see][]{1980JGR....85.8141P}.

\subsection{Retrieval analysis of optical-NIR data} \label{sec:discussion:retrieval}

In addition to comparing the data with predictions of forward models that assume chemical equilibrium (Section \ref{sec:discussion:forward}), we performed a free-chemistry retrieval analysis. For these calculations, we treat the abundances of the radiatively-active chemical species as free parameters in the model, rather than solving for the chemical equilibrium abundances at a given temperature. As for the forward models, this was done using \texttt{ATMO}, which can compute transmission spectra for any given atmospheric composition and PT profile. \texttt{ATMO} has previously been used for retrieval analyses of transmission spectra \citep{2017Sci...356..628W,2018AJ....155...29W} and thermal emission spectra \citep{2017Natur.548...58E}.

Since \texttt{ATMO} does not currently include any opacity sources that can explain the steep rise observed at NUV wavelengths (Figure \ref{fig:trspec_stiswfc3}), we restricted the retrieval to optical-NIR wavelengths longward of $0.47\,\um$. We assumed an isothermal PT profile and allowed the limb-averaged temperature ($T_{\textnormal{limb}}$) to vary as a free parameter, as well as the reference planet radius corresponding to the 1\,mbar pressure level ($R_{\textnormal{mbar}}$) effectively providing a floating offset between the model and data. The abundances of H$_2$O, TiO, VO, Na, and FeH were allowed to vary relative to a background atmosphere composition dominated by H$_2$ and He, assuming uniform mixing ratios with pressure. Other gas-phase absorbers such as K and CO were fixed to equilibrium abundances for the final analysis, as these were found to be unconstrained by the current data. Opacity due to aerosol Rayleigh scattering and optically-thick gray cloud was treated using the approach of \cite{2016Natur.529...59S}. Fitting was performed using differential evolution MCMC \citep{2013PASP..125...83E}, as described in our previous work \citep{2017Natur.548...58E,2017Sci...356..628W,2018AJ....155...29W}.

\begin{figure}
\centering  % this centres figure in column
\includegraphics[width=\columnwidth]{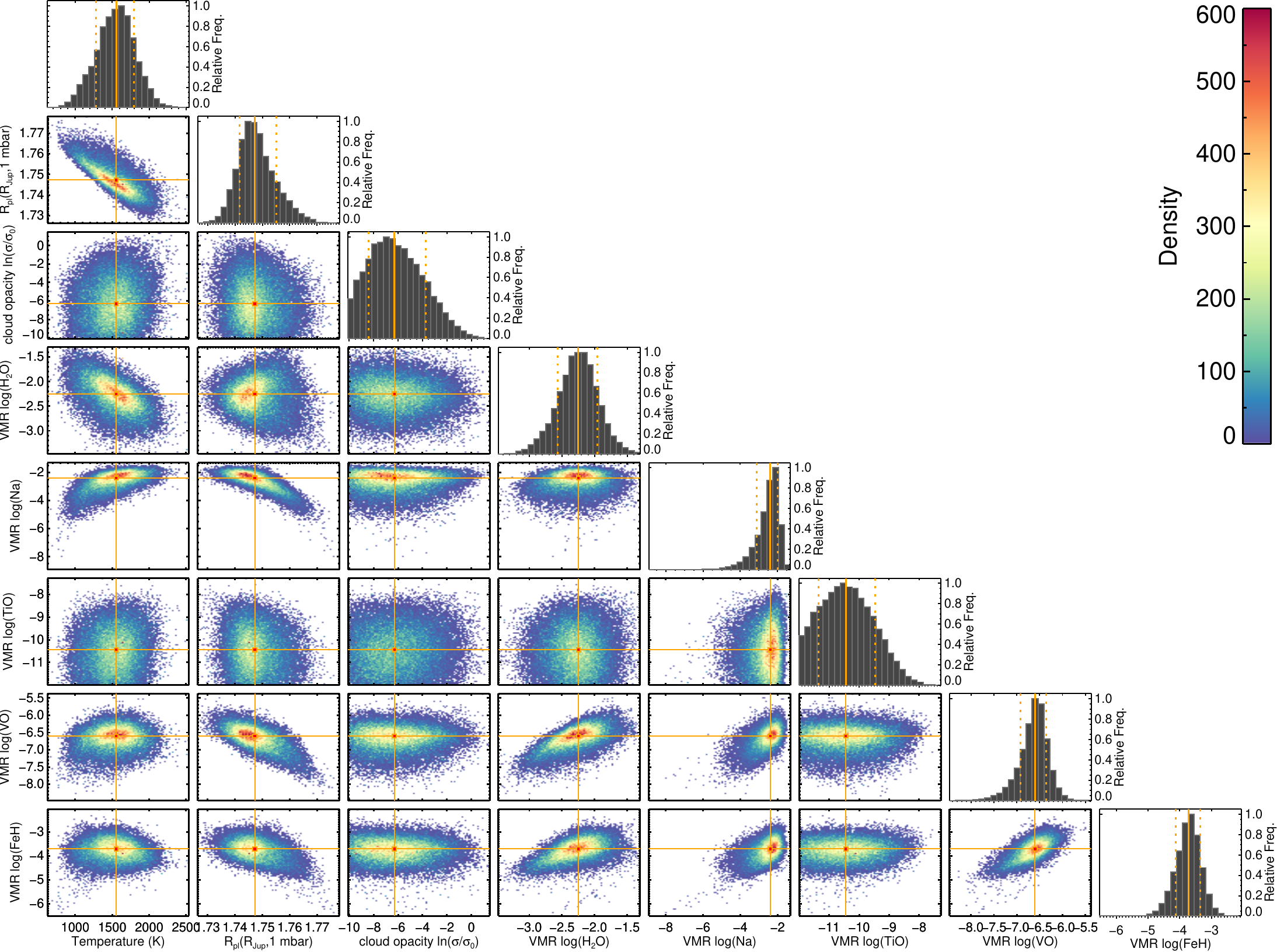}
\caption{Results of the free-chemistry retrieval analysis. \textit{(Off-diagonal panels)} Heat maps showing density of samples drawn from the MCMC analysis for different pairs of parameters. \textit{(Diagonal panels)} Marginalized density distributions for individual parameters. Solid orange lines indicate parameter median values and dashed orange lines indicate ranges spanning 68\% of samples.}
\label{fig:retrieval_density}
\end{figure}

The inferred distributions for each model parameter are summarized in Table \ref{table:retrieval} and shown in Figure \ref{fig:retrieval_density}. We find no evidence for opacity contribution due to aerosols, irrespective of whether they are treated as an enhanced Rayleigh scattering component or an optically-thick gray cloud. For this reason, we only present the results for the case including gray cloud, as the specific treatment of aerosols has negligible impact on the values inferred for the other model parameters.

We obtain a limb-averaged temperature of $T_{\textnormal{limb}} = 1554_{-271}^{+241}$\,K, in agreement with the best-matching chemical equilibrium model presented in Section \ref{sec:discussion:forward}. The inferred abundances for H$_2$O ($-2.2_{-0.3}^{+0.3}$\,dex), VO (${-6.6}_{-0.3}^{+0.2}$\,dex), and TiO (${-10.4}_{-0.9}^{+1.0}$\,dex) are also in good agreement with those predicted by the best-matching equilibrium model (Figure \ref{fig:trspec_opacityvmrs}). The inferred abundance for Na (${-2.4}_{-0.7}^{+0.4}$\,dex; $2\sigma$ lower limit of $-4.22$\,dex) is somewhat higher than the $20 \times$ solar value of $-4.24$\,dex (Figure \ref{fig:trspec_opacityvmrs}). One possibility is that the core of the Na line is probing the planetary thermosphere, where temperatures are higher and the pressure scale height is larger. This would produce a strong Na feature that the retrieval may misinterpret as indicating a high abundance. For instance, \cite{2012MNRAS.422.2477H} detected a strong Na line in the STIS transmission spectrum for HD\,189733b, which high-resolution spectroscopy showed is caused by a thermosphere \citep{2015A&A...577A..62W}.

\begin{figure}
\centering  % this centres figure in column
\includegraphics[width=\columnwidth]{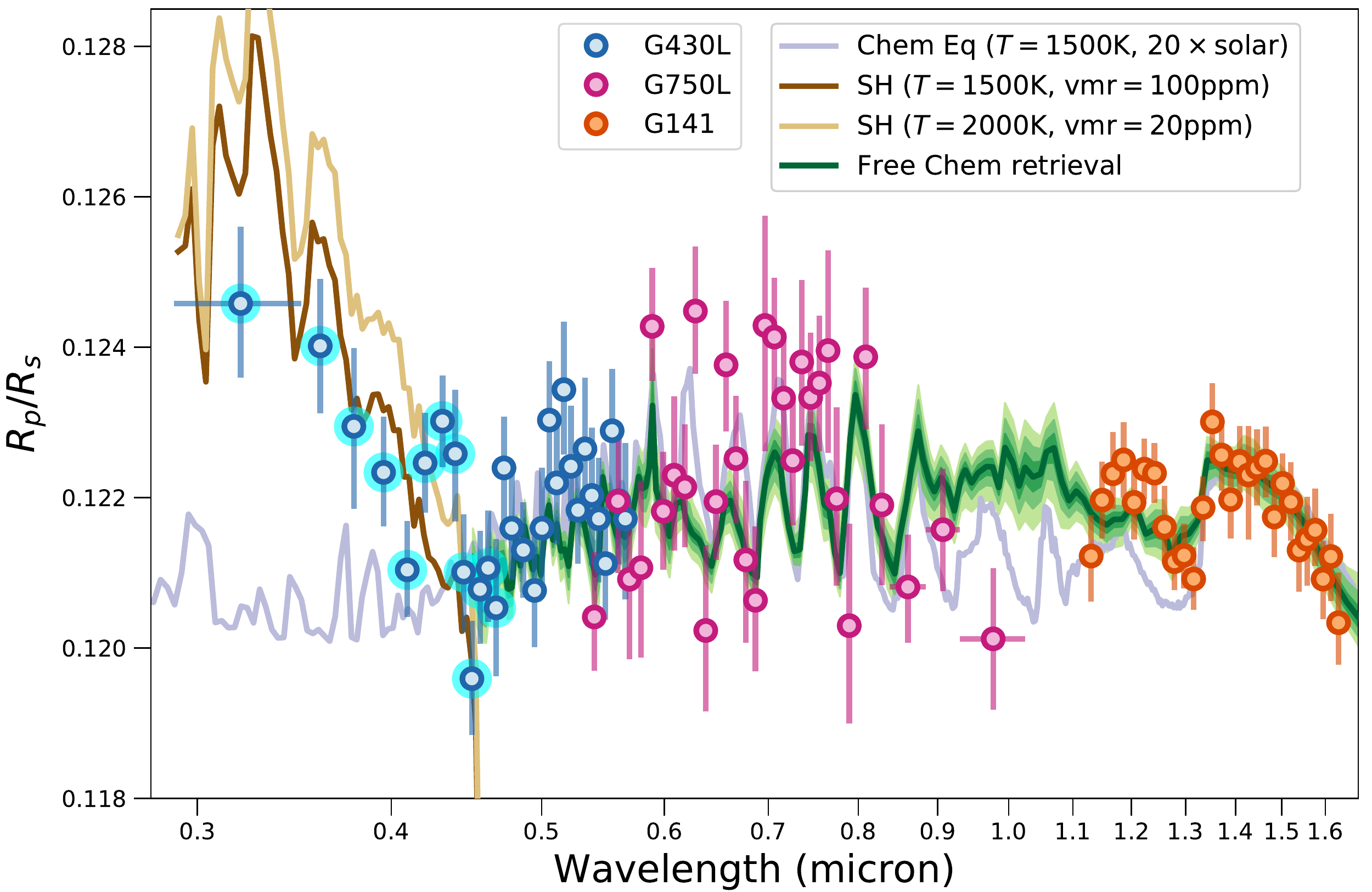}
\caption{Similar to Figure \ref{fig:trspec_stiswfc3}, but showing the distribution of model spectra inferred by the retrieval analysis as well as a hypothetical signal due to SH. The dark green line shows the sample mean at each wavelength and the shaded green areas progressively encompass 68\%, 95.5\%, and 99.7\% of samples about the mean. A significant departure from the chemical equilibrium model occurs between wavelengths of $\sim 0.9$--$1.3\,\um$. This is due to the retrieval inferring a high FeH abundance to explain the bump in the transmission spectrum measured over the short-wavelength half of the G141 bandpass.}
\label{fig:retrieval_specfit}
\end{figure}

In addition, the inferred abundance for FeH (${-3.7}_{-0.4}^{+0.4}$\,dex) is $\sim 5$ orders of magnitude greater than expected for $20 \times$ solar metallicity (Figure \ref{fig:trspec_opacityvmrs}). Such a high FeH abundance -- which we consider implausible -- is driven by the bump in the measured transmission spectrum across the $1.15$--$1.3\,\um$ wavelength range, where FeH has a significant absorption signature \citep[e.g.\ see Figure 7 of][]{2007ApJS..168..140S}. This can be seen clearly in Figure \ref{fig:retrieval_specfit}, which shows the distribution of spectra implied by the retrieval analysis, compared with the best-matching chemical equilibrium model. The inability of our model to simultaneously explain the $1.15$--$1.3\,\um$ bump and the rest of the data results in a moderately-poor overall fit, with a reduced $\chi^2$ of 1.5 for 67 degrees of freedom.

The $1.15$--$1.3\,\um$ bump in the transmission spectrum remains puzzling. It has been recovered by multiple independent analyses of the data performed within our own group, as well as those published by others \citep[e.g.][]{2018AJ....155..156T}. We note that it coincides with a possible spectral feature identified in the dayside thermal spectrum, which was tentatively attributed to VO emission \citep{2017Natur.548...58E}. However, it is difficult to reconcile VO with the feature seen in the transmission spectrum, as it would require increasing the abundance to a level that would be incompatible with the data at optical wavelengths, where VO has a higher opacity. On the other hand, although the host star is photometrically quiet and care has been taken to precisely measure the absolute transit depths for each bandpass (G430L, G750L, G141), it is conceivable that small offsets in $\RpRs$ remain, which, if accounted for, could allow VO to simultaneously explain the transmission spectrum at optical wavelengths along with the $1.15$--$1.3\,\um$ feature. For multi-epoch observations that do not overlap in wavelength such as those considered here, it is impossible to rule out such a scenario with absolute confidence. Upcoming G141 observations should allow a determinination of whether or not the $1.15$--$1.3\,\um$ bump is repeatable. It is also worth noting that a strong thermal gradient over the pressures probed in transmission -- which has not been considered in the present study -- could potentially affect the shape of the transmission spectrum by altering the pressure-dependent scale height $H$ and chemical abundances.

In summary, the retrieval analysis reveals no evidence for aerosols in the optical-NIR transmission spectrum of WASP-121b. The inferred limb-averaged temperature and gas-phase abundances are overall in good agreement with the best-matching forward models of Section \ref{sec:discussion:forward}, which assume chemical equilibrium and $10$--$30\times$ solar metallicity. The primary exception is the inferred FeH abundance, which as described above is far higher than expected for chemical equilibrium and $10$--$30\times$ solar metallicity. We thus conclude that it is unlikely FeH opacity is the true cause of the spectral bump at wavelengths $1.15$--$1.3\,\um$, the provenance of which remains uncertain.

%---------------------------------------------------------------
%\clearpage
\section{Conclusion} \label{sec:conclusion}

We have presented a STIS transmission spectrum for WASP-121b, spanning the $0.3$--$1\,\um$ wavelength range, adding to the $1.15$--$1.65\,\um$ wavelength coverage of published WFC3 data. The new optical data show an increase in atmospheric opacity for wavelengths shortward of $\sim 0.47\,\um$, with a slope that is too steep to be explained by Rayleigh scattering. Instead, assuming the NUV rise is a bona fide signature of the planetary atmosphere, it must be caused by one or more absorbers. We propose SH as a possible candidate, with a mixing ratio of approximately $\sim 20$--$100$\,ppm. Although the identity of the NUV absorber remains uncertain, it should cause substantial heating of the upper atmosphere and therefore could be an important component missing from existing models of highly-irradiated atmospheres. At longer wavelengths between $0.47$--$1\,\um$, we measure significant opacity variations that can be well-explained by VO absorption. Analyzing the STIS and WFC3 data with both free-chemistry retrievals and comparisons to chemical equilibrium forward models, we estimate abundances of H$_2$O and VO approximately $\sim 10$--$30\,\times$ solar. We find no significant evidence for TiO, suggesting it may have condensed from the gas-phase. Our chemical equilibrium forward models are unable to simultaneously reproduce the optical data and the WFC3 bump spanning the $1.15$--$1.35\,\um$ wavelength range. Free-chemistry retrievals are able to do so, but only by invoking an unrealistically high FeH abundance.

Overall, the evidence uncovered here for significant NUV and optical absorption implies a substantial fraction of incident stellar radiation is likely deposited at low pressures in the atmosphere of WASP-121b. Heating via this mechanism could be responsible for the thermal inversion detected on the dayside hemisphere. The broad coherence of this picture is tantalizing, but many unknowns remain. Although we consider the evidence for VO in the existing transmission spectrum to be reasonably strong, further observations are required to confirm or rule it out at high confidence. Similarly, additional observations, along with a more extensive exploration of candidates other than SH, are required to identify the NUV absorber. The possible explanation provided by SH, however, flags the potential importance of non-equilibrium sulfur chemistry in highly-irradiated atmospheres, which until now has received relatively little attention.

%\clearpage 
\bibliographystyle{apj}
\bibliography{wasp121}

\begin{thebibliography}{}
\expandafter\ifx\csname natexlab\endcsname\relax\def\natexlab#1{#1}\fi

\bibitem[{{Allard} {et~al.}(2012){Allard}, {Homeier}, \&
  {Freytag}}]{2012RSPTA.370.2765A}
{Allard}, F., {Homeier}, D., \& {Freytag}, B. 2012, Philosophical Transactions
  of the Royal Society of London Series A, 370, 2765

\bibitem[{{Amundsen} {et~al.}(2014){Amundsen}, {Baraffe}, {Tremblin},
  {Manners}, {Hayek}, {Mayne}, \& {Acreman}}]{2014A&A...564A..59A}
{Amundsen}, D.~S., {Baraffe}, I., {Tremblin}, P., {et~al.} 2014, \aap, 564, A59

\bibitem[{{Arcangeli} {et~al.}(2018){Arcangeli}, {D{\'e}sert}, {Line}, {Bean},
  {Parmentier}, {Stevenson}, {Kreidberg}, {Fortney}, {Mansfield}, \&
  {Showman}}]{2018ApJ...855L..30A}
{Arcangeli}, J., {D{\'e}sert}, J.-M., {Line}, M.~R., {et~al.} 2018, \apjl, 855,
  L30

\bibitem[{{Asplund} {et~al.}(2009){Asplund}, {Grevesse}, {Sauval}, \&
  {Scott}}]{2009ARA&A..47..481A}
{Asplund}, M., {Grevesse}, N., {Sauval}, A.~J., \& {Scott}, P. 2009, \araa, 47,
  481

\bibitem[{{Axel}(1972)}]{1972ApJ...173..451A}
{Axel}, L. 1972, \apj, 173, 451

\bibitem[{{Beatty} {et~al.}(2017){Beatty}, {Madhusudhan}, {Tsiaras}, {Zhao},
  {Gilliland}, {Knutson}, {Shporer}, \& {Wright}}]{2017AJ....154..158B}
{Beatty}, T.~G., {Madhusudhan}, N., {Tsiaras}, A., {et~al.} 2017, \aj, 154, 158

\bibitem[{{Bell} {et~al.}(2017){Bell}, {Nikolov}, {Cowan}, {Barstow}, {Barman},
  {Crossfield}, {Gibson}, {Evans}, {Sing}, {Knutson}, {Kataria}, {Lothringer},
  {Benneke}, \& {Schwartz}}]{2017ApJ...847L...2B}
{Bell}, T.~J., {Nikolov}, N., {Cowan}, N.~B., {et~al.} 2017, \apjl, 847, L2

\bibitem[{{Ben-Jaffel} \& {Ballester}(2013)}]{2013A&A...553A..52B}
{Ben-Jaffel}, L., \& {Ballester}, G.~E. 2013, \aap, 553, A52

\bibitem[{{Berta} {et~al.}(2011){Berta}, {Charbonneau}, {Bean}, {Irwin},
  {Burke}, {D{\'e}sert}, {Nutzman}, \& {Falco}}]{2011ApJ...736...12B}
{Berta}, Z.~K., {Charbonneau}, D., {Bean}, J., {et~al.} 2011, \apj, 736, 12

\bibitem[{{Burrows} \& {Sharp}(1999)}]{1999ApJ...512..843B}
{Burrows}, A., \& {Sharp}, C.~M. 1999, \apj, 512, 843

\bibitem[{{Charbonneau} {et~al.}(2002){Charbonneau}, {Brown}, {Noyes}, \&
  {Gilliland}}]{2002ApJ...568..377C}
{Charbonneau}, D., {Brown}, T.~M., {Noyes}, R.~W., \& {Gilliland}, R.~L. 2002,
  \apj, 568, 377

\bibitem[{{Claret}(2000)}]{2000A&A...363.1081C}
{Claret}, A. 2000, \aap, 363, 1081

\bibitem[{{Delrez} {et~al.}(2016){Delrez}, {Santerne}, {Almenara}, {Anderson},
  {Collier-Cameron}, {D{\'{\i}}az}, {Gillon}, {Hellier}, {Jehin}, {Lendl},
  {Maxted}, {Neveu-VanMalle}, {Pepe}, {Pollacco}, {Queloz}, {S{\'e}gransan},
  {Smalley}, {Smith}, {Triaud}, {Udry}, {Van Grootel}, \&
  {West}}]{2016MNRAS.tmp..312D}
{Delrez}, L., {Santerne}, A., {Almenara}, J.-M., {et~al.} 2016, \mnras,
  doi:10.1093/mnras/stw522

\bibitem[{{Deming} {et~al.}(2013){Deming}, {Wilkins}, {McCullough}, {Burrows},
  {Fortney}, {Agol}, {Dobbs-Dixon}, {Madhusudhan}, {Crouzet}, {Desert},
  {Gilliland}, {Haynes}, {Knutson}, {Line}, {Magic}, {Mandell}, {Ranjan},
  {Charbonneau}, {Clampin}, {Seager}, \& {Showman}}]{2013ApJ...774...95D}
{Deming}, D., {Wilkins}, A., {McCullough}, P., {et~al.} 2013, \apj, 774, 95

\bibitem[{{Demory} {et~al.}(2015){Demory}, {Ehrenreich}, {Queloz}, {Seager},
  {Gilliland}, {Chaplin}, {Proffitt}, {Gillon}, {G{\"u}nther}, {Benneke},
  {Dumusque}, {Lovis}, {Pepe}, {S{\'e}gransan}, {Triaud}, \&
  {Udry}}]{2015MNRAS.450.2043D}
{Demory}, B.-O., {Ehrenreich}, D., {Queloz}, D., {et~al.} 2015, \mnras, 450,
  2043

\bibitem[{{Drummond} {et~al.}(2016){Drummond}, {Tremblin}, {Baraffe},
  {Amundsen}, {Mayne}, {Venot}, \& {Goyal}}]{2016A&A...594A..69D}
{Drummond}, B., {Tremblin}, P., {Baraffe}, I., {et~al.} 2016, \aap, 594, A69

\bibitem[{{Eastman} {et~al.}(2013){Eastman}, {Gaudi}, \&
  {Agol}}]{2013PASP..125...83E}
{Eastman}, J., {Gaudi}, B.~S., \& {Agol}, E. 2013, \pasp, 125, 83

\bibitem[{{Eaton} {et~al.}(2003){Eaton}, {Henry}, \&
  {Fekel}}]{2003ASSL..288..189E}
{Eaton}, J.~A., {Henry}, G.~W., \& {Fekel}, F.~C. 2003, in Astrophysics and
  Space Science Library, Vol. 288, Astrophysics and Space Science Library, ed.
  T.~D. {Oswalt}, 189

\bibitem[{{Ehrenreich} {et~al.}(2015){Ehrenreich}, {Bourrier}, {Wheatley},
  {Lecavelier des Etangs}, {H{\'e}brard}, {Udry}, {Bonfils}, {Delfosse},
  {D{\'e}sert}, {Sing}, \& {Vidal-Madjar}}]{2015Natur.522..459E}
{Ehrenreich}, D., {Bourrier}, V., {Wheatley}, P.~J., {et~al.} 2015, \nat, 522,
  459

\bibitem[{{Espinoza} {et~al.}(2018){Espinoza}, {Rackham}, {Jord{\'a}n}, {Apai},
  {L{\'o}pez-Morales}, {Osip}, {Grimm}, {Hoeijmakers}, {Wilson}, {Bixel},
  {McGruder}, {Rodler}, {Weaver}, {Lewis}, {Fortney}, \&
  {Fraine}}]{2018MNRAS.tmp.2569E}
{Espinoza}, N., {Rackham}, B.~V., {Jord{\'a}n}, A., {et~al.} 2018, \mnras,
  arXiv:1807.10652

\bibitem[{{Esposito} {et~al.}(1997){Esposito}, {Bertaux}, {Krasnopolsky},
  {Moroz}, \& {Zasova}}]{1997veii.conf..415E}
{Esposito}, L.~W., {Bertaux}, J.-L., {Krasnopolsky}, V., {Moroz}, V.~I., \&
  {Zasova}, L.~V. 1997, in Venus II: Geology, Geophysics, Atmosphere, and Solar
  Wind Environment, ed. S.~W. {Bougher}, D.~M. {Hunten}, \& R.~J. {Phillips},
  415

\bibitem[{{Evans} {et~al.}(2013){Evans}, {Pont}, {Sing}, {Aigrain}, {Barstow},
  {D{\'e}sert}, {Gibson}, {Heng}, {Knutson}, \& {Lecavelier des
  Etangs}}]{2013ApJ...772L..16E}
{Evans}, T.~M., {Pont}, F., {Sing}, D.~K., {et~al.} 2013, \apjl, 772, L16

\bibitem[{{Evans} {et~al.}(2016){Evans}, {Sing}, {Wakeford}, {Nikolov},
  {Ballester}, {Drummond}, {Kataria}, {Gibson}, {Amundsen}, \&
  {Spake}}]{2016ApJ...822L...4E}
{Evans}, T.~M., {Sing}, D.~K., {Wakeford}, H.~R., {et~al.} 2016, \apjl, 822, L4

\bibitem[{{Evans} {et~al.}(2017){Evans}, {Sing}, {Kataria}, {Goyal}, {Nikolov},
  {Wakeford}, {Deming}, {Marley}, {Amundsen}, {Ballester}, {Barstow},
  {Ben-Jaffel}, {Bourrier}, {Buchhave}, {Cohen}, {Ehrenreich}, {Garc{\'{\i}}a
  Mu{\~n}oz}, {Henry}, {Knutson}, {Lavvas}, {Lecavelier Des Etangs}, {Lewis},
  {L{\'o}pez-Morales}, {Mandell}, {Sanz-Forcada}, {Tremblin}, \&
  {Lupu}}]{2017Natur.548...58E}
{Evans}, T.~M., {Sing}, D.~K., {Kataria}, T., {et~al.} 2017, \nat, 548, 58

\bibitem[{{Fischer} {et~al.}(2016){Fischer}, {Knutson}, {Sing}, {Henry},
  {Williamson}, {Fortney}, {Burrows}, {Kataria}, {Nikolov}, {Showman},
  {Ballester}, {D{\'e}sert}, {Aigrain}, {Deming}, {Lecavelier des Etangs}, \&
  {Vidal-Madjar}}]{2016ApJ...827...19F}
{Fischer}, P.~D., {Knutson}, H.~A., {Sing}, D.~K., {et~al.} 2016, \apj, 827, 19

\bibitem[{{Flower}(1996)}]{1996ApJ...469..355F}
{Flower}, P.~J. 1996, \apj, 469, 355

\bibitem[{{Foreman-Mackey} {et~al.}(2013){Foreman-Mackey}, {Hogg}, {Lang}, \&
  {Goodman}}]{2013PASP..125..306F}
{Foreman-Mackey}, D., {Hogg}, D.~W., {Lang}, D., \& {Goodman}, J. 2013, \pasp,
  125, 306

\bibitem[{{Fortney}(2005)}]{2005MNRAS.364..649F}
{Fortney}, J.~J. 2005, \mnras, 364, 649

\bibitem[{{Fortney} {et~al.}(2008){Fortney}, {Lodders}, {Marley}, \&
  {Freedman}}]{2008ApJ...678.1419F}
{Fortney}, J.~J., {Lodders}, K., {Marley}, M.~S., \& {Freedman}, R.~S. 2008,
  \apj, 678, 1419

\bibitem[{{Fossati} {et~al.}(2010){Fossati}, {Haswell}, {Froning}, {Hebb},
  {Holmes}, {Kolb}, {Helling}, {Carter}, {Wheatley}, {Collier Cameron},
  {Loeillet}, {Pollacco}, {Street}, {Stempels}, {Simpson}, {Udry}, {Joshi},
  {West}, {Skillen}, \& {Wilson}}]{2010ApJ...714L.222F}
{Fossati}, L., {Haswell}, C.~A., {Froning}, C.~S., {et~al.} 2010, \apjl, 714,
  L222

\bibitem[{{Fraine} {et~al.}(2014){Fraine}, {Deming}, {Benneke}, {Knutson},
  {Jord{\'a}n}, {Espinoza}, {Madhusudhan}, {Wilkins}, \&
  {Todorov}}]{2014Natur.513..526F}
{Fraine}, J., {Deming}, D., {Benneke}, B., {et~al.} 2014, \nat, 513, 526

\bibitem[{{Gaia Collaboration} {et~al.}(2018){Gaia Collaboration}, {Brown},
  {Vallenari}, {Prusti}, {de Bruijne}, {Babusiaux}, {Bailer-Jones}, {Biermann},
  {Evans}, {Eyer}, \& et~al.}]{2018A&A...616A...1G}
{Gaia Collaboration}, {Brown}, A.~G.~A., {Vallenari}, A., {et~al.} 2018, \aap,
  616, A1

\bibitem[{Gelman \& Rubin(1992)}]{GelmanRubin92}
Gelman, A., \& Rubin, D.~B. 1992, Stat. Sci., 7, 457

\bibitem[{{Gibson}(2014)}]{2014MNRAS.445.3401G}
{Gibson}, N.~P. 2014, \mnras, 445, 3401

\bibitem[{{Gibson} {et~al.}(2012){Gibson}, {Aigrain}, {Roberts}, {Evans},
  {Osborne}, \& {Pont}}]{2012MNRAS.419.2683G}
{Gibson}, N.~P., {Aigrain}, S., {Roberts}, S., {et~al.} 2012, \mnras, 419, 2683

\bibitem[{{Gibson} {et~al.}(2017){Gibson}, {Nikolov}, {Sing}, {Barstow},
  {Evans}, {Kataria}, \& {Wilson}}]{2017MNRAS.467.4591G}
{Gibson}, N.~P., {Nikolov}, N., {Sing}, D.~K., {et~al.} 2017, \mnras, 467, 4591

\bibitem[{{Goyal} {et~al.}(2018){Goyal}, {Mayne}, {Sing}, {Drummond},
  {Tremblin}, {Amundsen}, {Evans}, {Carter}, {Spake}, {Baraffe}, {Nikolov},
  {Manners}, {Chabrier}, \& {Hebrard}}]{2018MNRAS.474.5158G}
{Goyal}, J.~M., {Mayne}, N., {Sing}, D.~K., {et~al.} 2018, \mnras, 474, 5158

\bibitem[{{Haynes} {et~al.}(2015){Haynes}, {Mandell}, {Madhusudhan}, {Deming},
  \& {Knutson}}]{2015ApJ...806..146H}
{Haynes}, K., {Mandell}, A.~M., {Madhusudhan}, N., {Deming}, D., \& {Knutson},
  H. 2015, \apj, 806, 146

\bibitem[{{Henry}(1999)}]{1999PASP..111..845H}
{Henry}, G.~W. 1999, \pasp, 111, 845

\bibitem[{{Hubeny} {et~al.}(2003){Hubeny}, {Burrows}, \&
  {Sudarsky}}]{2003ApJ...594.1011H}
{Hubeny}, I., {Burrows}, A., \& {Sudarsky}, D. 2003, \apj, 594, 1011

\bibitem[{{Huitson} {et~al.}(2012){Huitson}, {Sing}, {Vidal-Madjar},
  {Ballester}, {Lecavelier des Etangs}, {D{\'e}sert}, \&
  {Pont}}]{2012MNRAS.422.2477H}
{Huitson}, C.~M., {Sing}, D.~K., {Vidal-Madjar}, A., {et~al.} 2012, \mnras,
  422, 2477

\bibitem[{{Huitson} {et~al.}(2013){Huitson}, {Sing}, {Pont}, {Fortney},
  {Burrows}, {Wilson}, {Ballester}, {Nikolov}, {Gibson}, {Deming}, {Aigrain},
  {Evans}, {Henry}, {Lecavelier des Etangs}, {Showman}, {Vidal-Madjar}, \&
  {Zahnle}}]{2013MNRAS.434.3252H}
{Huitson}, C.~M., {Sing}, D.~K., {Pont}, F., {et~al.} 2013, \mnras, 434, 3252

\bibitem[{{Kataria} {et~al.}(2016){Kataria}, {Sing}, {Lewis}, {Visscher},
  {Showman}, {Fortney}, \& {Marley}}]{2016ApJ...821....9K}
{Kataria}, T., {Sing}, D.~K., {Lewis}, N.~K., {et~al.} 2016, \apj, 821, 9

\bibitem[{{Kempton} {et~al.}(2017){Kempton}, {Bean}, \&
  {Parmentier}}]{2017ApJ...845L..20K}
{Kempton}, E.~M.-R., {Bean}, J.~L., \& {Parmentier}, V. 2017, \apjl, 845, L20

\bibitem[{{Kreidberg} {et~al.}(2014){Kreidberg}, {Bean}, {D{\'e}sert},
  {Benneke}, {Deming}, {Stevenson}, {Seager}, {Berta-Thompson}, {Seifahrt}, \&
  {Homeier}}]{2014Natur.505...69K}
{Kreidberg}, L., {Bean}, J.~L., {D{\'e}sert}, J.-M., {et~al.} 2014, \nat, 505,
  69

\bibitem[{{Kreidberg} {et~al.}(2015){Kreidberg}, {Line}, {Bean}, {Stevenson},
  {D{\'e}sert}, {Madhusudhan}, {Fortney}, {Barstow}, {Henry}, {Williamson}, \&
  {Showman}}]{2015ApJ...814...66K}
{Kreidberg}, L., {Line}, M.~R., {Bean}, J.~L., {et~al.} 2015, \apj, 814, 66

\bibitem[{{Kreidberg} {et~al.}(2018){Kreidberg}, {Line}, {Parmentier},
  {Stevenson}, {Louden}, {Bonnefoy}, {Faherty}, {Henry}, {Williamson},
  {Stassun}, {Beatty}, {Bean}, {Fortney}, {Showman}, {D{\'e}sert}, \&
  {Arcangeli}}]{2018AJ....156...17K}
{Kreidberg}, L., {Line}, M.~R., {Parmentier}, V., {et~al.} 2018, \aj, 156, 17

\bibitem[{{Lecavelier Des Etangs} {et~al.}(2008){Lecavelier Des Etangs},
  {Pont}, {Vidal-Madjar}, \& {Sing}}]{2008A&A...481L..83L}
{Lecavelier Des Etangs}, A., {Pont}, F., {Vidal-Madjar}, A., \& {Sing}, D.
  2008, \aap, 481, L83

\bibitem[{{Lodders}(2002)}]{2002ApJ...577..974L}
{Lodders}, K. 2002, \apj, 577, 974

\bibitem[{{Lothringer} {et~al.}(2018){Lothringer}, {Barman}, \&
  {Koskinen}}]{2018ApJ...866...27L}
{Lothringer}, J.~D., {Barman}, T., \& {Koskinen}, T. 2018, \apj, 866, 27

\bibitem[{{Madhusudhan} {et~al.}(2011){Madhusudhan}, {Burrows}, \&
  {Currie}}]{2011ApJ...737...34M}
{Madhusudhan}, N., {Burrows}, A., \& {Currie}, T. 2011, \apj, 737, 34

\bibitem[{{Magic} {et~al.}(2013){Magic}, {Collet}, {Asplund}, {Trampedach},
  {Hayek}, {Chiavassa}, {Stein}, \& {Nordlund}}]{2013A&A...557A..26M}
{Magic}, Z., {Collet}, R., {Asplund}, M., {et~al.} 2013, \aap, 557, A26

\bibitem[{{Mandel} \& {Agol}(2002)}]{2002ApJ...580L.171M}
{Mandel}, K., \& {Agol}, E. 2002, \apjl, 580, L171

\bibitem[{{Mansfield} {et~al.}(2018){Mansfield}, {Bean}, {Line}, {Parmentier},
  {Kreidberg}, {D{\'e}sert}, {Fortney}, {Stevenson}, {Arcangeli}, \&
  {Dragomir}}]{2018AJ....156...10M}
{Mansfield}, M., {Bean}, J.~L., {Line}, M.~R., {et~al.} 2018, \aj, 156, 10

\bibitem[{{Marley} \& {McKay}(1999)}]{1999Icar..138..268M}
{Marley}, M.~S., \& {McKay}, C.~P. 1999, \icarus, 138, 268

\bibitem[{{Marley} {et~al.}(2012){Marley}, {Saumon}, {Cushing}, {Ackerman},
  {Fortney}, \& {Freedman}}]{2012ApJ...754..135M}
{Marley}, M.~S., {Saumon}, D., {Cushing}, M., {et~al.} 2012, \apj, 754, 135

\bibitem[{{Marley} {et~al.}(2002){Marley}, {Seager}, {Saumon}, {Lodders},
  {Ackerman}, {Freedman}, \& {Fan}}]{2002ApJ...568..335M}
{Marley}, M.~S., {Seager}, S., {Saumon}, D., {et~al.} 2002, \apj, 568, 335

\bibitem[{{Mbarek} \& {Kempton}(2016)}]{2016ApJ...827..121M}
{Mbarek}, R., \& {Kempton}, E.~M.-R. 2016, \apj, 827, 121

\bibitem[{{Nikolov} {et~al.}(2016){Nikolov}, {Sing}, {Gibson}, {Fortney},
  {Evans}, {Barstow}, {Kataria}, \& {Wilson}}]{2016ApJ...832..191N}
{Nikolov}, N., {Sing}, D.~K., {Gibson}, N.~P., {et~al.} 2016, \apj, 832, 191

\bibitem[{{Nikolov} {et~al.}(2014){Nikolov}, {Sing}, {Pont}, {Burrows},
  {Fortney}, {Ballester}, {Evans}, {Huitson}, {Wakeford}, {Wilson}, {Aigrain},
  {Deming}, {Gibson}, {Henry}, {Knutson}, {Lecavelier des Etangs}, {Showman},
  {Vidal-Madjar}, \& {Zahnle}}]{2014MNRAS.437...46N}
{Nikolov}, N., {Sing}, D.~K., {Pont}, F., {et~al.} 2014, \mnras, 437, 46

\bibitem[{{Nikolov} {et~al.}(2015){Nikolov}, {Sing}, {Burrows}, {Fortney},
  {Henry}, {Pont}, {Ballester}, {Aigrain}, {Wilson}, {Huitson}, {Gibson},
  {D{\'e}sert}, {Etangs}, {Showman}, {Vidal-Madjar}, {Wakeford}, \&
  {Zahnle}}]{2015MNRAS.447..463N}
{Nikolov}, N., {Sing}, D.~K., {Burrows}, A.~S., {et~al.} 2015, \mnras, 447, 463

\bibitem[{{Nikolov} {et~al.}(2018){Nikolov}, {Sing}, {Goyal}, {Henry},
  {Wakeford}, {Evans}, {L{\'o}pez-Morales}, {Garc{\'{\i}}a Mu{\~n}oz},
  {Ben-Jaffel}, {Sanz-Forcada}, {Ballester}, {Kataria}, {Barstow}, {Bourrier},
  {Buchhave}, {Cohen}, {Deming}, {Ehrenreich}, {Knutson}, {Lavvas}, {Lecavelier
  des Etangs}, {Lewis}, {Mandell}, \& {Williamson}}]{2018MNRAS.474.1705N}
{Nikolov}, N., {Sing}, D.~K., {Goyal}, J., {et~al.} 2018, \mnras, 474, 1705

\bibitem[{{Nugroho} {et~al.}(2017){Nugroho}, {Kawahara}, {Masuda}, {Hirano},
  {Kotani}, \& {Tajitsu}}]{2017AJ....154..221N}
{Nugroho}, S.~K., {Kawahara}, H., {Masuda}, K., {et~al.} 2017, \aj, 154, 221

\bibitem[{{Owen} \& {Sagan}(1972)}]{1972Icar...16..557O}
{Owen}, T., \& {Sagan}, C. 1972, \icarus, 16, 557

\bibitem[{{Parmentier} {et~al.}(2013){Parmentier}, {Showman}, \&
  {Lian}}]{2013A&A...558A..91P}
{Parmentier}, V., {Showman}, A.~P., \& {Lian}, Y. 2013, \aap, 558, A91

\bibitem[{{Parmentier} {et~al.}(2018){Parmentier}, {Line}, {Bean}, {Mansfield},
  {Kreidberg}, {Lupu}, {Visscher}, {D{\'e}sert}, {Fortney}, {Deleuil},
  {Arcangeli}, {Showman}, \& {Marley}}]{2018A&A...617A.110P}
{Parmentier}, V., {Line}, M.~R., {Bean}, J.~L., {et~al.} 2018, \aap, 617, A110

\bibitem[{{Pollack} {et~al.}(1980){Pollack}, {Toon}, {Whitten}, {Boese},
  {Ragent}, {Tomasko}, {Eposito}, {Travis}, \& {Wiedman}}]{1980JGR....85.8141P}
{Pollack}, J.~B., {Toon}, O.~B., {Whitten}, R.~C., {et~al.} 1980, \jgr, 85,
  8141

\bibitem[{{Pont} {et~al.}(2008){Pont}, {Knutson}, {Gilliland}, {Moutou}, \&
  {Charbonneau}}]{2008MNRAS.385..109P}
{Pont}, F., {Knutson}, H., {Gilliland}, R.~L., {Moutou}, C., \& {Charbonneau},
  D. 2008, \mnras, 385, 109

\bibitem[{{Rackham} {et~al.}(2018){Rackham}, {Apai}, \&
  {Giampapa}}]{2018ApJ...853..122R}
{Rackham}, B.~V., {Apai}, D., \& {Giampapa}, M.~S. 2018, \apj, 853, 122

\bibitem[{{Saumon} \& {Marley}(2008)}]{2008ApJ...689.1327S}
{Saumon}, D., \& {Marley}, M.~S. 2008, \apj, 689, 1327

\bibitem[{{Seager} \& {Sasselov}(1998)}]{1998ApJ...502L.157S}
{Seager}, S., \& {Sasselov}, D.~D. 1998, \apjl, 502, L157

\bibitem[{{Seager} \& {Sasselov}(2000)}]{2000ApJ...537..916S}
---. 2000, \apj, 537, 916

\bibitem[{{Sedaghati} {et~al.}(2017){Sedaghati}, {Boffin}, {MacDonald},
  {Gandhi}, {Madhusudhan}, {Gibson}, {Oshagh}, {Claret}, \&
  {Rauer}}]{2017Natur.549..238S}
{Sedaghati}, E., {Boffin}, H.~M.~J., {MacDonald}, R.~J., {et~al.} 2017, \nat,
  549, 238

\bibitem[{{Sharp} \& {Burrows}(2007)}]{2007ApJS..168..140S}
{Sharp}, C.~M., \& {Burrows}, A. 2007, \apjs, 168, 140

\bibitem[{{Showman} {et~al.}(2009){Showman}, {Fortney}, {Lian}, {Marley},
  {Freedman}, {Knutson}, \& {Charbonneau}}]{2009ApJ...699..564S}
{Showman}, A.~P., {Fortney}, J.~J., {Lian}, Y., {et~al.} 2009, \apj, 699, 564

\bibitem[{{Sing} {et~al.}(2011){Sing}, {Pont}, {Aigrain}, {Charbonneau},
  {D{\'e}sert}, {Gibson}, {Gilliland}, {Hayek}, {Henry}, {Knutson}, {Lecavelier
  Des Etangs}, {Mazeh}, \& {Shporer}}]{2011MNRAS.416.1443S}
{Sing}, D.~K., {Pont}, F., {Aigrain}, S., {et~al.} 2011, \mnras, 416, 1443

\bibitem[{{Sing} {et~al.}(2015){Sing}, {Wakeford}, {Showman}, {Nikolov},
  {Fortney}, {Burrows}, {Ballester}, {Deming}, {Aigrain}, {D{\'e}sert},
  {Gibson}, {Henry}, {Knutson}, {Lecavelier des Etangs}, {Pont},
  {Vidal-Madjar}, {Williamson}, \& {Wilson}}]{2015MNRAS.446.2428S}
{Sing}, D.~K., {Wakeford}, H.~R., {Showman}, A.~P., {et~al.} 2015, \mnras, 446,
  2428

\bibitem[{{Sing} {et~al.}(2016){Sing}, {Fortney}, {Nikolov}, {Wakeford},
  {Kataria}, {Evans}, {Aigrain}, {Ballester}, {Burrows}, {Deming},
  {D{\'e}sert}, {Gibson}, {Henry}, {Huitson}, {Knutson}, {Etangs}, {Pont},
  {Showman}, {Vidal-Madjar}, {Williamson}, \& {Wilson}}]{2016Natur.529...59S}
{Sing}, D.~K., {Fortney}, J.~J., {Nikolov}, N., {et~al.} 2016, \nat, 529, 59

\bibitem[{{Spake} {et~al.}(2018){Spake}, {Sing}, {Evans}, {Oklop{\v c}i{\'c}},
  {Bourrier}, {Kreidberg}, {Rackham}, {Irwin}, {Ehrenreich}, {Wyttenbach},
  {Wakeford}, {Zhou}, {Chubb}, {Nikolov}, {Goyal}, {Henry}, {Williamson},
  {Blumenthal}, {Anderson}, {Hellier}, {Charbonneau}, {Udry}, \&
  {Madhusudhan}}]{2018Natur.557...68S}
{Spake}, J.~J., {Sing}, D.~K., {Evans}, T.~M., {et~al.} 2018, \nat, 557, 68

\bibitem[{{Spiegel} {et~al.}(2009){Spiegel}, {Silverio}, \&
  {Burrows}}]{2009ApJ...699.1487S}
{Spiegel}, D.~S., {Silverio}, K., \& {Burrows}, A. 2009, \apj, 699, 1487

\bibitem[{{Stevenson} {et~al.}(2014){Stevenson}, {D{\'e}sert}, {Line}, {Bean},
  {Fortney}, {Showman}, {Kataria}, {Kreidberg}, {McCullough}, {Henry},
  {Charbonneau}, {Burrows}, {Seager}, {Madhusudhan}, {Williamson}, \&
  {Homeier}}]{2014Sci...346..838S}
{Stevenson}, K.~B., {D{\'e}sert}, J.-M., {Line}, M.~R., {et~al.} 2014, Science,
  346, 838

\bibitem[{{Thorngren} {et~al.}(2016){Thorngren}, {Fortney}, {Murray-Clay}, \&
  {Lopez}}]{2016ApJ...831...64T}
{Thorngren}, D.~P., {Fortney}, J.~J., {Murray-Clay}, R.~A., \& {Lopez}, E.~D.
  2016, \apj, 831, 64

\bibitem[{{Tremblin} {et~al.}(2016){Tremblin}, {Amundsen}, {Chabrier},
  {Baraffe}, {Drummond}, {Hinkley}, {Mourier}, \&
  {Venot}}]{2016ApJ...817L..19T}
{Tremblin}, P., {Amundsen}, D.~S., {Chabrier}, G., {et~al.} 2016, \apjl, 817,
  L19

\bibitem[{{Tremblin} {et~al.}(2015){Tremblin}, {Amundsen}, {Mourier},
  {Baraffe}, {Chabrier}, {Drummond}, {Homeier}, \&
  {Venot}}]{2015ApJ...804L..17T}
{Tremblin}, P., {Amundsen}, D.~S., {Mourier}, P., {et~al.} 2015, \apjl, 804,
  L17

\bibitem[{{Tsiaras} {et~al.}(2018){Tsiaras}, {Waldmann}, {Zingales},
  {Rocchetto}, {Morello}, {Damiano}, {Karpouzas}, {Tinetti}, {McKemmish},
  {Tennyson}, \& {Yurchenko}}]{2018AJ....155..156T}
{Tsiaras}, A., {Waldmann}, I.~P., {Zingales}, T., {et~al.} 2018, \aj, 155, 156

\bibitem[{{Vidal-Madjar} {et~al.}(2003){Vidal-Madjar}, {Lecavelier des Etangs},
  {D{\'e}sert}, {Ballester}, {Ferlet}, {H{\'e}brard}, \&
  {Mayor}}]{2003Natur.422..143V}
{Vidal-Madjar}, A., {Lecavelier des Etangs}, A., {D{\'e}sert}, J.-M., {et~al.}
  2003, \nat, 422, 143

\bibitem[{{Visscher} {et~al.}(2006){Visscher}, {Lodders}, \&
  {Fegley}}]{2006ApJ...648.1181V}
{Visscher}, C., {Lodders}, K., \& {Fegley}, Jr., B. 2006, \apj, 648, 1181

\bibitem[{{Wakeford} {et~al.}(2017){Wakeford}, {Sing}, {Kataria}, {Deming},
  {Nikolov}, {Lopez}, {Tremblin}, {Amundsen}, {Lewis}, {Mandell}, {Fortney},
  {Knutson}, {Benneke}, \& {Evans}}]{2017Sci...356..628W}
{Wakeford}, H.~R., {Sing}, D.~K., {Kataria}, T., {et~al.} 2017, Science, 356,
  628

\bibitem[{{Wakeford} {et~al.}(2018){Wakeford}, {Sing}, {Deming}, {Lewis},
  {Goyal}, {Wilson}, {Barstow}, {Kataria}, {Drummond}, {Evans}, {Carter},
  {Nikolov}, {Knutson}, {Ballester}, \& {Mandell}}]{2018AJ....155...29W}
{Wakeford}, H.~R., {Sing}, D.~K., {Deming}, D., {et~al.} 2018, \aj, 155, 29

\bibitem[{{West} {et~al.}(2004){West}, {Baines}, {Friedson}, {Banfield},
  {Ragent}, \& {Taylor}}]{2004jpsm.book...79W}
{West}, R.~A., {Baines}, K.~H., {Friedson}, A.~J., {et~al.} 2004, {Jovian
  clouds and haze}, ed. F.~{Bagenal}, T.~E. {Dowling}, \& W.~B. {McKinnon},
  79--104

\bibitem[{{Woitke} {et~al.}(2018){Woitke}, {Helling}, {Hunter}, {Millard},
  {Turner}, {Worters}, {Blecic}, \& {Stock}}]{2018A&A...614A...1W}
{Woitke}, P., {Helling}, C., {Hunter}, G.~H., {et~al.} 2018, \aap, 614, A1

\bibitem[{{Wyttenbach} {et~al.}(2015){Wyttenbach}, {Ehrenreich}, {Lovis},
  {Udry}, \& {Pepe}}]{2015A&A...577A..62W}
{Wyttenbach}, A., {Ehrenreich}, D., {Lovis}, C., {Udry}, S., \& {Pepe}, F.
  2015, \aap, 577, A62

\bibitem[{{Yurchenko} {et~al.}(2018){Yurchenko}, {Bond}, {Gorman}, {Lodi},
  {McKemmish}, {Nunn}, {Shah}, \& {Tennyson}}]{2018MNRAS.tmp..913Y}
{Yurchenko}, S.~N., {Bond}, W., {Gorman}, M.~N., {et~al.} 2018, \mnras,
  arXiv:1803.09724

\bibitem[{{Zahnle} {et~al.}(2009){Zahnle}, {Marley}, {Freedman}, {Lodders}, \&
  {Fortney}}]{2009ApJ...701L..20Z}
{Zahnle}, K., {Marley}, M.~S., {Freedman}, R.~S., {Lodders}, K., \& {Fortney},
  J.~J. 2009, \apjl, 701, L20

\bibitem[{{Zahnle} {et~al.}(2016){Zahnle}, {Marley}, {Morley}, \&
  {Moses}}]{2016ApJ...824..137Z}
{Zahnle}, K., {Marley}, M.~S., {Morley}, C.~V., \& {Moses}, J.~I. 2016, \apj,
  824, 137

\end{thebibliography}

\begin{table}
\begin{minipage}{\columnwidth}
  \centering
\scriptsize
\caption{Results of white lightcurve MCMC analyses. Quoted values give sample medians with uncertainties corresponding to ranges encompassing $68\%$ of samples about the median. Note that $\sigma$ values were not fit directly as part of the MCMC analysis but obtained by multiplying the $\beta$ values by the formal photon noise values for each lightcurve. \label{table:whitefit}}
\begin{tabular}{cccccc}

\hline \\ 
&& \multicolumn{4}{c}{$\aRs$ and $b$ allowed to vary} \\ \cline{3-6} && G430Lv1 & G430Lv2 & G750L & G141 \medskip \\ \cline{3-6} 
 \\ 
\smallskip 
$\RpRs$        && \multicolumn{2}{c}{\hrulefill { $0.1226_{-0.0006}^{+0.0006}$} \hrulefill} & $0.1223_{-0.0005}^{+0.0004}$ & $0.1216_{-0.0004}^{+0.0004}$ \\ 
$u_1$          && \multicolumn{2}{c}{\hrulefill { $0.50_{-0.06}^{+0.06}$} \hrulefill} & $0.21_{-0.09}^{+0.08}$ & $0.16_{-0.05}^{+0.06}$ \\ 
$u_2$          && \multicolumn{2}{c}{\hrulefill { $0.11_{-0.11}^{+0.11}$} \hrulefill} & $0.25_{-0.14}^{+0.15}$ & $0.10_{-0.10}^{+0.09}$ \\ 
$\aRs$         && \multicolumn{2}{c}{\hrulefill { $3.87_{-0.04}^{+0.04}$} \hrulefill} & $3.88_{-0.05}^{+0.03}$ & $3.83_{-0.04}^{+0.02}$ \\ 
$b$            && \multicolumn{2}{c}{\hrulefill { $0.05_{-0.04}^{+0.05}$} \hrulefill} & $0.08_{-0.06}^{+0.08}$ & $0.06_{-0.04}^{+0.06}$ \\ 
$i$ ($^\circ$) && \multicolumn{2}{c}{\hrulefill { $89.3_{-0.8}^{+0.5}$} \hrulefill} & $88.8_{-1.2}^{+0.8}$ & $89.1_{-1.0}^{+0.6}$ \\ 
$\Tmid$ (MJD)  &&  $57685.74504_{-0.00057}^{+0.00061}$ & $57698.49467_{-0.00061}^{+0.00058}$ & $57704.86884_{-0.00023}^{+0.00018}$ & $57424.38323_{-0.00039}^{+0.00019}$ \\ 
$\beta$        &&  $1.29_{-0.17}^{+0.18}$ & $1.16_{-0.17}^{+0.19}$ & $1.36_{-0.11}^{+0.12}$ & $1.08_{-0.11}^{+0.13}$ \\ 
$\sigma$ (ppm) &&  $116_{-15}^{+17}$ & $105_{-15}^{+17}$ & $141_{-12}^{+12}$ & $67_{-7}^{+8}$ \\ 
$c_0$          &&  $1.0003_{-0.0006}^{+0.0005}$ & $0.9991_{-0.0003}^{+0.0004}$ & $1.0018_{-0.0013}^{+0.0022}$ & $1.0000_{-0.0004}^{+0.0003}$ \\ 
$c_1$          &&  $0.000159_{-0.000001}^{+0.000001}$ & $0.000895_{-0.000007}^{+0.000007}$ & $0.000113_{-0.000006}^{+0.000005}$ & $-0.000199_{-0.000006}^{+0.000006}$ \\ 
$A$ (ppm)      &&  $963_{-251}^{+473}$ & $593_{-159}^{+263}$ & $1708_{-727}^{+1177}$ & $486_{-141}^{+261}$ \\ 
$\ln \iLphi$   &&  $-1.12_{-0.92}^{+0.75}$ & $0.15_{-0.70}^{+0.54}$ & $-3.57_{-1.09}^{+1.20}$ & $-0.68_{-0.48}^{+0.42}$ \\ 
$\ln \iLx$     &&  $-0.96_{-0.58}^{+0.43}$ & $-1.14_{-1.14}^{+0.93}$ & $-4.23_{-1.06}^{+1.14}$ & $-1.03_{-0.82}^{+0.68}$ \\ 
$\ln \iLy$     &&  $-1.18_{-1.56}^{+1.30}$ & $-0.89_{-1.27}^{+0.85}$ & $-4.75_{-0.93}^{+1.02}$ & $-3.24_{-1.72}^{+1.55}$ \\ \\ \hline \\ 
&& \multicolumn{4}{c}{$\aRs$ and $b$ held fixed} \\ \cline{3-6}&& G430Lv1 & G430Lv2 & G750L & G141 \medskip \\ \cline{3-6} 
 \\ 
\smallskip 
$\RpRs$        && \multicolumn{2}{c}{\hrulefill { $0.1223_{-0.0006}^{+0.0006}$} \hrulefill} & $0.1219_{-0.0005}^{+0.0004}$ & $0.1218_{-0.0004}^{+0.0004}$ \\ 
$u_1$          && \multicolumn{2}{c}{\hrulefill { $0.51_{-0.06}^{+0.06}$} \hrulefill} & $0.18_{-0.09}^{+0.08}$ & $0.16_{-0.05}^{+0.06}$ \\ 
$u_2$          && \multicolumn{2}{c}{\hrulefill { $0.11_{-0.11}^{+0.11}$} \hrulefill} & $0.33_{-0.14}^{+0.15}$ & $0.08_{-0.10}^{+0.09}$ \\ 
$\aRs$         && \multicolumn{4}{c}{\hrulefill { $3.86$ (fixed)} \hrulefill} \\ 
$b$            && \multicolumn{4}{c}{\hrulefill { $0.06$ (fixed)} \hrulefill} \\ 
$i$ ($^\circ$) && \multicolumn{4}{c}{\hrulefill { $89.1$ (fixed)} \hrulefill} \\ 
$\Tmid$ (MJD)  &&  $57685.74516_{-0.00021}^{+0.00021}$ & $57698.49459_{-0.00022}^{+0.00022}$ & $57704.86860_{-0.00018}^{+0.00017}$ & $57424.38341_{-0.00010}^{+0.00009}$ \\ 
$\beta$        &&  $1.33_{-0.16}^{+0.17}$ & $1.21_{-0.19}^{+0.19}$ & $1.36_{-0.11}^{+0.12}$ & $1.10_{-0.11}^{+0.13}$ \\ 
$\sigma$ (ppm) &&  $120_{-14}^{+15}$ & $109_{-17}^{+17}$ & $141_{-11}^{+13}$ & $68_{-7}^{+8}$ \\ 
$c_0$          &&  $1.0002_{-0.0009}^{+0.0006}$ & $0.9992_{-0.0004}^{+0.0006}$ & $1.0012_{-0.0010}^{+0.0020}$ & $0.9999_{-0.0005}^{+0.0003}$ \\ 
$c_1$          &&  $0.000128_{-0.000002}^{+0.000002}$ & $0.000897_{-0.000010}^{+0.000011}$ & $0.000225_{-0.000045}^{+0.000040}$ & $-0.000218_{-0.000042}^{+0.000043}$ \\ 
$A$ (ppm)      &&  $1175_{-371}^{+767}$ & $720_{-238}^{+445}$ & $1323_{-611}^{+1252}$ & $565_{-181}^{+362}$ \\ 
$\ln \iLphi$   &&  $-1.30_{-0.75}^{+0.59}$ & $-0.08_{-1.14}^{+0.64}$ & $-2.67_{-1.39}^{+1.61}$ & $-0.79_{-0.50}^{+0.43}$ \\ 
$\ln \iLx$     &&  $-1.66_{-1.48}^{+1.02}$ & $-1.29_{-1.70}^{+1.18}$ & $-4.55_{-1.00}^{+1.19}$ & $-1.55_{-0.93}^{+0.76}$ \\ 
$\ln \iLy$     &&  $-1.89_{-1.22}^{+1.30}$ & $-1.27_{-1.28}^{+0.97}$ & $-4.67_{-0.96}^{+1.06}$ & $-4.07_{-1.28}^{+1.74}$ \\ \\ \hline

\end{tabular}
\end{minipage}
\end{table}

\begin{table}
\begin{minipage}{\columnwidth}
  \centering
\scriptsize
\caption{Results of G430L spectroscopic lightcurve fits for selected parameters. \label{table:specfit_g430l}}
\begin{tabular}{cccccccccccc}
  \hline \\
  $\lambda$ (\AA) & $\RpRs$ & $u_1$ & $u_2$ & $\beta_{\rm{v1}}$ & $\sigma_{\rm{v1}}$ (ppm) & $\beta_{\rm{v2}}$ & $\sigma_{\rm{v2}}$ (ppm) \medskip \\ \hline
  \\

2898-3499 & $0.1246_{-0.0011}^{+0.0010}$ & $0.55_{-0.08}^{+0.08}$ & $0.21_{-0.13}^{+0.13}$ & $1.16_{-0.10}^{+0.10}$ & $565_{-49}^{+47}$ & $1.15_{-0.11}^{+0.10}$ & $557_{-52}^{+50}$ \\
3499-3700 & $0.1238_{-0.0010}^{+0.0009}$ & $0.41_{-0.09}^{+0.09}$ & $0.30_{-0.14}^{+0.14}$ & $1.03_{-0.09}^{+0.11}$ & $593_{-54}^{+61}$ & $1.10_{-0.09}^{+0.10}$ & $635_{-51}^{+55}$ \\
3700-3868 & $0.1235_{-0.0011}^{+0.0012}$ & $0.41_{-0.10}^{+0.10}$ & $0.39_{-0.15}^{+0.14}$ & $1.11_{-0.11}^{+0.11}$ & $550_{-54}^{+56}$ & $1.13_{-0.11}^{+0.11}$ & $559_{-53}^{+52}$ \\
3868-4041 & $0.1223_{-0.0008}^{+0.0008}$ & $0.58_{-0.08}^{+0.08}$ & $0.23_{-0.12}^{+0.12}$ & $1.13_{-0.09}^{+0.09}$ & $439_{-36}^{+36}$ & $1.13_{-0.10}^{+0.10}$ & $440_{-37}^{+39}$ \\
4041-4151 & $0.1211_{-0.0007}^{+0.0006}$ & $0.60_{-0.07}^{+0.07}$ & $0.15_{-0.11}^{+0.12}$ & $1.06_{-0.08}^{+0.09}$ & $438_{-34}^{+38}$ & $1.02_{-0.09}^{+0.10}$ & $422_{-36}^{+41}$ \\
4151-4261 & $0.1227_{-0.0007}^{+0.0007}$ & $0.62_{-0.07}^{+0.07}$ & $0.09_{-0.12}^{+0.12}$ & $1.02_{-0.09}^{+0.10}$ & $399_{-36}^{+37}$ & $1.06_{-0.12}^{+0.12}$ & $413_{-47}^{+46}$ \\
4261-4371 & $0.1230_{-0.0006}^{+0.0006}$ & $0.50_{-0.08}^{+0.08}$ & $0.14_{-0.12}^{+0.12}$ & $1.15_{-0.09}^{+0.10}$ & $452_{-35}^{+38}$ & $0.93_{-0.10}^{+0.11}$ & $365_{-40}^{+42}$ \\
4371-4426 & $0.1225_{-0.0009}^{+0.0009}$ & $0.54_{-0.09}^{+0.08}$ & $0.21_{-0.13}^{+0.13}$ & $0.98_{-0.10}^{+0.11}$ & $504_{-52}^{+56}$ & $1.09_{-0.09}^{+0.09}$ & $560_{-46}^{+48}$ \\
4426-4481 & $0.1209_{-0.0007}^{+0.0008}$ & $0.62_{-0.08}^{+0.08}$ & $0.08_{-0.13}^{+0.12}$ & $0.92_{-0.10}^{+0.10}$ & $463_{-51}^{+50}$ & $1.09_{-0.09}^{+0.09}$ & $551_{-47}^{+46}$ \\
4481-4536 & $0.1196_{-0.0007}^{+0.0007}$ & $0.58_{-0.08}^{+0.07}$ & $0.18_{-0.12}^{+0.12}$ & $1.09_{-0.08}^{+0.09}$ & $539_{-41}^{+45}$ & $0.96_{-0.09}^{+0.10}$ & $473_{-45}^{+47}$ \\
4536-4591 & $0.1208_{-0.0007}^{+0.0008}$ & $0.48_{-0.08}^{+0.08}$ & $0.21_{-0.13}^{+0.13}$ & $1.11_{-0.09}^{+0.09}$ & $554_{-44}^{+45}$ & $1.07_{-0.09}^{+0.09}$ & $531_{-44}^{+45}$ \\
4591-4646 & $0.1211_{-0.0008}^{+0.0007}$ & $0.43_{-0.08}^{+0.07}$ & $0.29_{-0.12}^{+0.12}$ & $1.03_{-0.10}^{+0.10}$ & $507_{-50}^{+50}$ & $1.00_{-0.09}^{+0.10}$ & $493_{-44}^{+48}$ \\
4646-4701 & $0.1205_{-0.0009}^{+0.0010}$ & $0.55_{-0.09}^{+0.09}$ & $0.14_{-0.14}^{+0.14}$ & $1.09_{-0.10}^{+0.10}$ & $546_{-48}^{+52}$ & $1.03_{-0.11}^{+0.10}$ & $517_{-53}^{+52}$ \\
4701-4756 & $0.1224_{-0.0006}^{+0.0007}$ & $0.52_{-0.08}^{+0.08}$ & $0.11_{-0.12}^{+0.12}$ & $0.98_{-0.11}^{+0.10}$ & $492_{-54}^{+50}$ & $0.86_{-0.10}^{+0.10}$ & $434_{-50}^{+53}$ \\
4756-4811 & $0.1216_{-0.0007}^{+0.0007}$ & $0.45_{-0.08}^{+0.08}$ & $0.24_{-0.12}^{+0.12}$ & $0.99_{-0.09}^{+0.09}$ & $498_{-46}^{+44}$ & $0.94_{-0.09}^{+0.09}$ & $471_{-45}^{+47}$ \\
4811-4921 & $0.1214_{-0.0006}^{+0.0006}$ & $0.45_{-0.08}^{+0.08}$ & $0.10_{-0.12}^{+0.12}$ & $1.00_{-0.10}^{+0.10}$ & $373_{-36}^{+36}$ & $1.06_{-0.10}^{+0.10}$ & $395_{-38}^{+38}$ \\
4921-4976 & $0.1208_{-0.0008}^{+0.0008}$ & $0.43_{-0.08}^{+0.08}$ & $0.21_{-0.13}^{+0.12}$ & $1.10_{-0.10}^{+0.10}$ & $557_{-48}^{+48}$ & $1.04_{-0.09}^{+0.10}$ & $522_{-46}^{+50}$ \\
4976-5030 & $0.1216_{-0.0009}^{+0.0008}$ & $0.45_{-0.09}^{+0.09}$ & $0.20_{-0.14}^{+0.14}$ & $1.06_{-0.09}^{+0.09}$ & $539_{-46}^{+48}$ & $1.11_{-0.09}^{+0.09}$ & $562_{-46}^{+46}$ \\
5030-5085 & $0.1230_{-0.0008}^{+0.0008}$ & $0.40_{-0.08}^{+0.08}$ & $0.16_{-0.13}^{+0.13}$ & $1.08_{-0.08}^{+0.09}$ & $542_{-41}^{+48}$ & $1.03_{-0.09}^{+0.10}$ & $516_{-47}^{+50}$ \\
5085-5140 & $0.1222_{-0.0008}^{+0.0007}$ & $0.50_{-0.09}^{+0.08}$ & $0.07_{-0.13}^{+0.14}$ & $1.06_{-0.09}^{+0.09}$ & $531_{-45}^{+47}$ & $1.01_{-0.10}^{+0.10}$ & $507_{-52}^{+51}$ \\
5140-5195 & $0.1234_{-0.0008}^{+0.0008}$ & $0.36_{-0.09}^{+0.09}$ & $0.20_{-0.13}^{+0.13}$ & $0.99_{-0.10}^{+0.10}$ & $511_{-51}^{+54}$ & $1.09_{-0.08}^{+0.09}$ & $561_{-43}^{+49}$ \\
5195-5250 & $0.1224_{-0.0008}^{+0.0008}$ & $0.36_{-0.08}^{+0.08}$ & $0.18_{-0.13}^{+0.13}$ & $0.95_{-0.10}^{+0.10}$ & $486_{-49}^{+54}$ & $1.15_{-0.09}^{+0.10}$ & $592_{-46}^{+49}$ \\
5250-5305 & $0.1215_{-0.0007}^{+0.0007}$ & $0.48_{-0.09}^{+0.08}$ & $0.17_{-0.13}^{+0.13}$ & $1.02_{-0.09}^{+0.09}$ & $529_{-46}^{+45}$ & $0.89_{-0.10}^{+0.10}$ & $462_{-52}^{+51}$ \\
5305-5360 & $0.1223_{-0.0009}^{+0.0009}$ & $0.38_{-0.09}^{+0.09}$ & $0.16_{-0.14}^{+0.13}$ & $1.10_{-0.09}^{+0.10}$ & $572_{-46}^{+54}$ & $1.05_{-0.09}^{+0.10}$ & $545_{-48}^{+52}$ \\
5360-5415 & $0.1221_{-0.0010}^{+0.0009}$ & $0.27_{-0.09}^{+0.09}$ & $0.27_{-0.13}^{+0.14}$ & $1.06_{-0.10}^{+0.11}$ & $559_{-53}^{+59}$ & $0.99_{-0.10}^{+0.10}$ & $523_{-52}^{+53}$ \\
5415-5469 & $0.1218_{-0.0008}^{+0.0009}$ & $0.34_{-0.09}^{+0.08}$ & $0.30_{-0.13}^{+0.13}$ & $1.07_{-0.09}^{+0.10}$ & $573_{-48}^{+52}$ & $1.04_{-0.10}^{+0.10}$ & $553_{-52}^{+52}$ \\
5469-5524 & $0.1210_{-0.0008}^{+0.0007}$ & $0.38_{-0.08}^{+0.08}$ & $0.21_{-0.13}^{+0.13}$ & $0.93_{-0.09}^{+0.11}$ & $502_{-51}^{+57}$ & $1.06_{-0.09}^{+0.10}$ & $571_{-48}^{+54}$ \\
5524-5579 & $0.1229_{-0.0008}^{+0.0008}$ & $0.40_{-0.09}^{+0.08}$ & $0.14_{-0.13}^{+0.14}$ & $1.06_{-0.09}^{+0.09}$ & $573_{-47}^{+48}$ & $1.13_{-0.09}^{+0.09}$ & $613_{-48}^{+50}$ \\
5579-5634 & $0.1219_{-0.0009}^{+0.0008}$ & $0.28_{-0.09}^{+0.09}$ & $0.36_{-0.14}^{+0.13}$ & $0.93_{-0.10}^{+0.10}$ & $514_{-53}^{+54}$ & $1.06_{-0.10}^{+0.10}$ & $582_{-54}^{+53}$ \\
5634-5688 & $0.1218_{-0.0010}^{+0.0010}$ & $0.43_{-0.08}^{+0.09}$ & $0.16_{-0.13}^{+0.13}$ & $1.02_{-0.10}^{+0.10}$ & $565_{-53}^{+57}$ & $1.23_{-0.08}^{+0.08}$ & $682_{-43}^{+47}$ \\
  
\\ \hline
\end{tabular}
\end{minipage}
\end{table}

\begin{table}
\begin{minipage}{\columnwidth}
  \centering
\scriptsize
\caption{Similar to Table \ref{table:specfit_g430l}, but for the G750L spectroscopic lightcurve fits. \label{table:specfit_g750l}}
\begin{tabular}{cccccccccc}
\hline \\ 
$\lambda$ (\AA) & $\RpRs$ & $u_1$ & $u_2$ & $\beta$ & $\sigma$ (ppm) \medskip \\ \hline
\\

5263-5550 & $0.1205_{-0.0007}^{+0.0008}$ & $0.36_{-0.08}^{+0.07}$ & $0.30_{-0.12}^{+0.12}$ & $1.02_{-0.07}^{+0.08}$ & $403_{-29}^{+30}$ \\
5550-5648 & $0.1220_{-0.0010}^{+0.0010}$ & $0.39_{-0.08}^{+0.08}$ & $0.22_{-0.13}^{+0.14}$ & $1.04_{-0.07}^{+0.08}$ & $609_{-43}^{+45}$ \\
5648-5745 & $0.1209_{-0.0011}^{+0.0010}$ & $0.36_{-0.09}^{+0.08}$ & $0.30_{-0.13}^{+0.13}$ & $1.05_{-0.08}^{+0.08}$ & $597_{-45}^{+47}$ \\
5745-5843 & $0.1211_{-0.0012}^{+0.0012}$ & $0.31_{-0.09}^{+0.09}$ & $0.31_{-0.13}^{+0.13}$ & $1.03_{-0.10}^{+0.09}$ & $569_{-52}^{+50}$ \\
5843-5940 & $0.1243_{-0.0007}^{+0.0008}$ & $0.25_{-0.08}^{+0.08}$ & $0.28_{-0.13}^{+0.12}$ & $1.03_{-0.07}^{+0.08}$ & $563_{-40}^{+43}$ \\
5940-6038 & $0.1218_{-0.0008}^{+0.0008}$ & $0.26_{-0.08}^{+0.08}$ & $0.26_{-0.13}^{+0.13}$ & $1.04_{-0.07}^{+0.07}$ & $576_{-39}^{+41}$ \\
6038-6135 & $0.1223_{-0.0010}^{+0.0010}$ & $0.26_{-0.09}^{+0.08}$ & $0.28_{-0.14}^{+0.14}$ & $1.15_{-0.08}^{+0.08}$ & $627_{-43}^{+44}$ \\
6135-6233 & $0.1221_{-0.0008}^{+0.0008}$ & $0.21_{-0.08}^{+0.08}$ & $0.34_{-0.13}^{+0.13}$ & $1.03_{-0.08}^{+0.08}$ & $561_{-42}^{+43}$ \\
6233-6330 & $0.1245_{-0.0008}^{+0.0009}$ & $0.30_{-0.08}^{+0.09}$ & $0.20_{-0.13}^{+0.13}$ & $1.05_{-0.08}^{+0.08}$ & $575_{-42}^{+45}$ \\
6330-6428 & $0.1202_{-0.0011}^{+0.0011}$ & $0.30_{-0.09}^{+0.09}$ & $0.23_{-0.13}^{+0.14}$ & $1.08_{-0.09}^{+0.08}$ & $583_{-48}^{+45}$ \\
6428-6526 & $0.1219_{-0.0008}^{+0.0008}$ & $0.22_{-0.09}^{+0.09}$ & $0.22_{-0.13}^{+0.13}$ & $1.01_{-0.07}^{+0.08}$ & $547_{-40}^{+44}$ \\
6526-6623 & $0.1238_{-0.0009}^{+0.0008}$ & $0.21_{-0.09}^{+0.09}$ & $0.18_{-0.13}^{+0.13}$ & $1.13_{-0.07}^{+0.08}$ & $634_{-39}^{+44}$ \\
6623-6721 & $0.1225_{-0.0009}^{+0.0008}$ & $0.28_{-0.08}^{+0.08}$ & $0.21_{-0.13}^{+0.13}$ & $0.96_{-0.08}^{+0.09}$ & $532_{-45}^{+48}$ \\
6721-6818 & $0.1212_{-0.0012}^{+0.0010}$ & $0.19_{-0.09}^{+0.09}$ & $0.25_{-0.13}^{+0.13}$ & $1.03_{-0.10}^{+0.10}$ & $572_{-57}^{+54}$ \\
6818-6916 & $0.1206_{-0.0009}^{+0.0010}$ & $0.21_{-0.09}^{+0.08}$ & $0.38_{-0.14}^{+0.14}$ & $1.03_{-0.09}^{+0.09}$ & $579_{-52}^{+50}$ \\
6916-7014 & $0.1243_{-0.0017}^{+0.0015}$ & $0.16_{-0.10}^{+0.10}$ & $0.25_{-0.14}^{+0.14}$ & $1.09_{-0.09}^{+0.09}$ & $620_{-53}^{+53}$ \\
7014-7111 & $0.1241_{-0.0008}^{+0.0008}$ & $0.22_{-0.09}^{+0.08}$ & $0.19_{-0.13}^{+0.13}$ & $1.03_{-0.07}^{+0.08}$ & $594_{-42}^{+46}$ \\
7111-7209 & $0.1233_{-0.0009}^{+0.0009}$ & $0.15_{-0.09}^{+0.09}$ & $0.26_{-0.13}^{+0.13}$ & $1.13_{-0.07}^{+0.08}$ & $664_{-42}^{+45}$ \\
7209-7307 & $0.1225_{-0.0009}^{+0.0008}$ & $0.22_{-0.09}^{+0.09}$ & $0.23_{-0.13}^{+0.13}$ & $1.04_{-0.07}^{+0.08}$ & $627_{-44}^{+47}$ \\
7307-7404 & $0.1238_{-0.0011}^{+0.0011}$ & $0.18_{-0.09}^{+0.09}$ & $0.22_{-0.13}^{+0.13}$ & $1.05_{-0.09}^{+0.09}$ & $655_{-54}^{+55}$ \\
7404-7502 & $0.1233_{-0.0008}^{+0.0009}$ & $0.07_{-0.09}^{+0.09}$ & $0.29_{-0.14}^{+0.13}$ & $1.06_{-0.07}^{+0.07}$ & $684_{-46}^{+46}$ \\
7502-7600 & $0.1235_{-0.0009}^{+0.0009}$ & $0.16_{-0.09}^{+0.09}$ & $0.16_{-0.13}^{+0.13}$ & $0.96_{-0.08}^{+0.08}$ & $631_{-50}^{+51}$ \\
7600-7698 & $0.1240_{-0.0014}^{+0.0013}$ & $0.21_{-0.09}^{+0.09}$ & $0.30_{-0.14}^{+0.14}$ & $1.07_{-0.08}^{+0.08}$ & $722_{-55}^{+55}$ \\
7698-7795 & $0.1220_{-0.0012}^{+0.0012}$ & $0.12_{-0.09}^{+0.09}$ & $0.32_{-0.13}^{+0.14}$ & $1.14_{-0.08}^{+0.08}$ & $803_{-54}^{+57}$ \\
7795-7991 & $0.1203_{-0.0013}^{+0.0014}$ & $0.16_{-0.09}^{+0.10}$ & $0.26_{-0.13}^{+0.13}$ & $1.10_{-0.10}^{+0.09}$ & $595_{-52}^{+50}$ \\
7991-8186 & $0.1239_{-0.0010}^{+0.0009}$ & $0.16_{-0.09}^{+0.09}$ & $0.27_{-0.13}^{+0.13}$ & $1.01_{-0.07}^{+0.08}$ & $609_{-44}^{+47}$ \\
8186-8381 & $0.1219_{-0.0011}^{+0.0011}$ & $0.16_{-0.09}^{+0.09}$ & $0.25_{-0.14}^{+0.13}$ & $1.06_{-0.08}^{+0.08}$ & $699_{-50}^{+51}$ \\
8381-8840 & $0.1208_{-0.0007}^{+0.0007}$ & $0.08_{-0.08}^{+0.08}$ & $0.32_{-0.13}^{+0.12}$ & $0.93_{-0.08}^{+0.08}$ & $432_{-35}^{+37}$ \\
8840-9299 & $0.1216_{-0.0008}^{+0.0008}$ & $0.10_{-0.09}^{+0.08}$ & $0.20_{-0.13}^{+0.13}$ & $1.11_{-0.07}^{+0.08}$ & $548_{-36}^{+40}$ \\
9299-10245 & $0.1201_{-0.0009}^{+0.0009}$ & $0.16_{-0.08}^{+0.09}$ & $0.22_{-0.13}^{+0.13}$ & $1.14_{-0.08}^{+0.08}$ & $562_{-37}^{+38}$ \\
\\ \hline
\end{tabular}
\end{minipage}
\end{table}

\begin{table}
\caption{Results of free-chemistry retrieval analysis \label{table:retrieval}}
\begin{center}
\begin{tabular}{cc} 
\hline \\ 
Parameter & Value \medskip \\ 
\hline \\ 
\smallskip $R_{\textnormal{mbar}}$ ($\Rjup$) & ${1.747}_{-0.006}^{+0.008}$ \\ \smallskip 
 $T_{\textnormal{limb}}$ (K) & ${1554}_{-271}^{+241}$ \\ \smallskip 
 $\log_{10}$[\,H$_2$O\,] & ${-2.2}_{-0.3}^{+0.3}$ \\ \smallskip 
 $\log_{10}$[\,VO\,] & ${-6.6}_{-0.3}^{+0.2}$ \\ \smallskip 
 $\log_{10}$[\,TiO\,]$^{\textnormal{a}}$ & $<-7.9$ \\ \smallskip 
 $\log_{10}$[\,Na\,] & ${-2.4}_{-0.7}^{+0.4}$ \\ \smallskip 
 $\log_{10}$[\,FeH\,] & ${-3.7}_{-0.4}^{+0.4}$ \\ \smallskip 
 $\ln$[\,$\sigma_{\textnormal{cloud}}/\sigma_0$\,] & ${-6.3}_{-2.1}^{+2.5}$ \\ \hline
\end{tabular}
\vspace{5pt} \\
$^{\textnormal{a}}$\, \footnotesize $3\sigma$ upper limit
\end{center}
\end{table}

%%%%%%%%%%%%%%%%%%%%%%%%%%%%%%%%%%%%%%%%%%%%%
\clearpage
\appendix

\counterwithin{figure}{section}

\section{Raw Spectroscopic Lightcurves} \label{app:rawspeclcs}

\begin{figure}[b!]
\centering  % this centres figure in column
\includegraphics[width=\columnwidth]{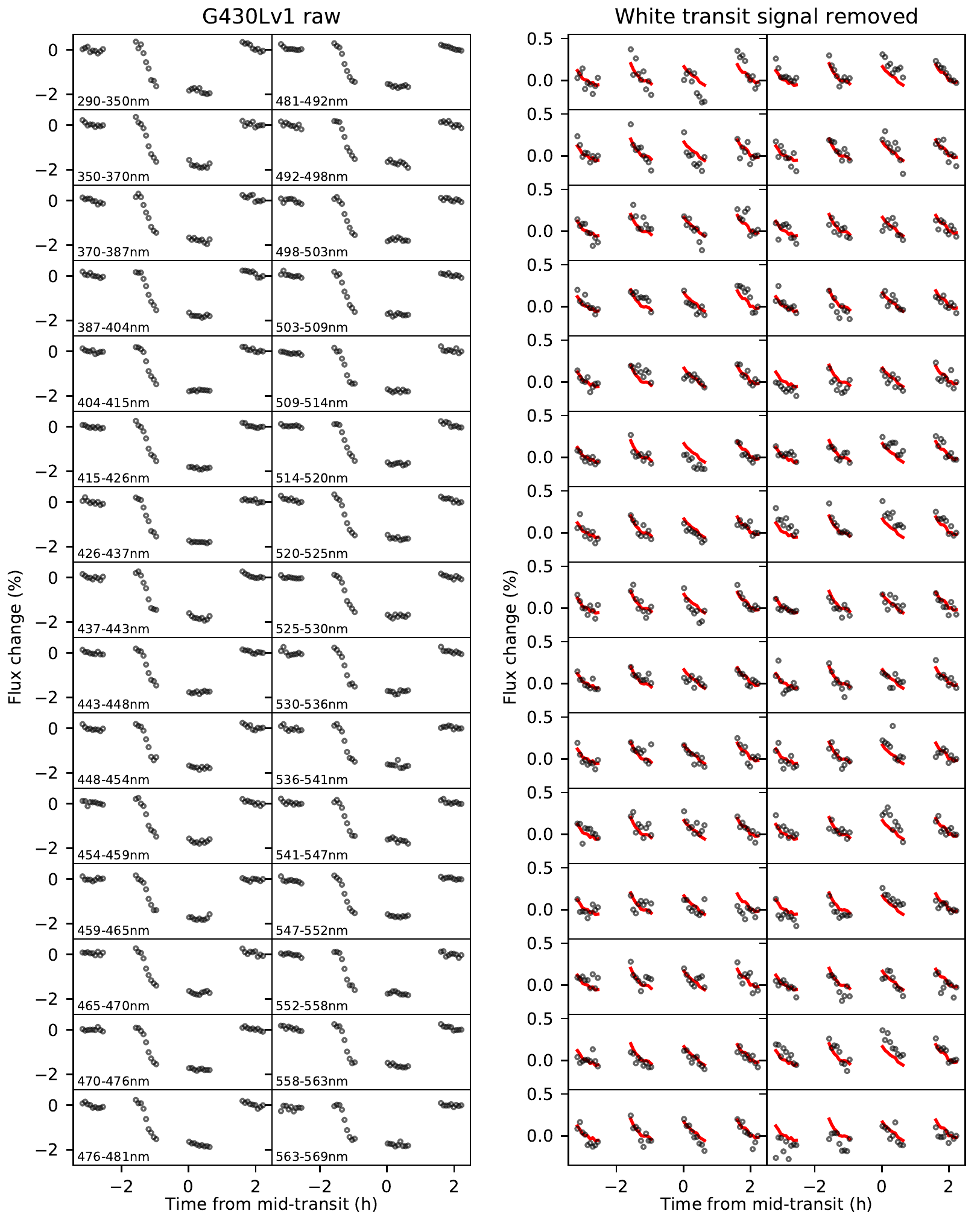}
\caption{\textit{(Left two columns)} Raw lightcurves for each spectroscopic channel of the G430Lv1 dataset. \textit{(Right two columns)} Black circles show residuals after subtracting the best-fit white lightcurve transit signal from each of the raw spectroscopic lightcurves, to highlight the systematics component. Red lines show the best-fit white lightcurve systematics model, to emphasize variation in systematics across spectroscopic channels.}
\label{fig:speclcs_raw_g430lv1}
\end{figure}

\begin{figure}
\centering  % this centres figure in column
\includegraphics[width=\columnwidth]{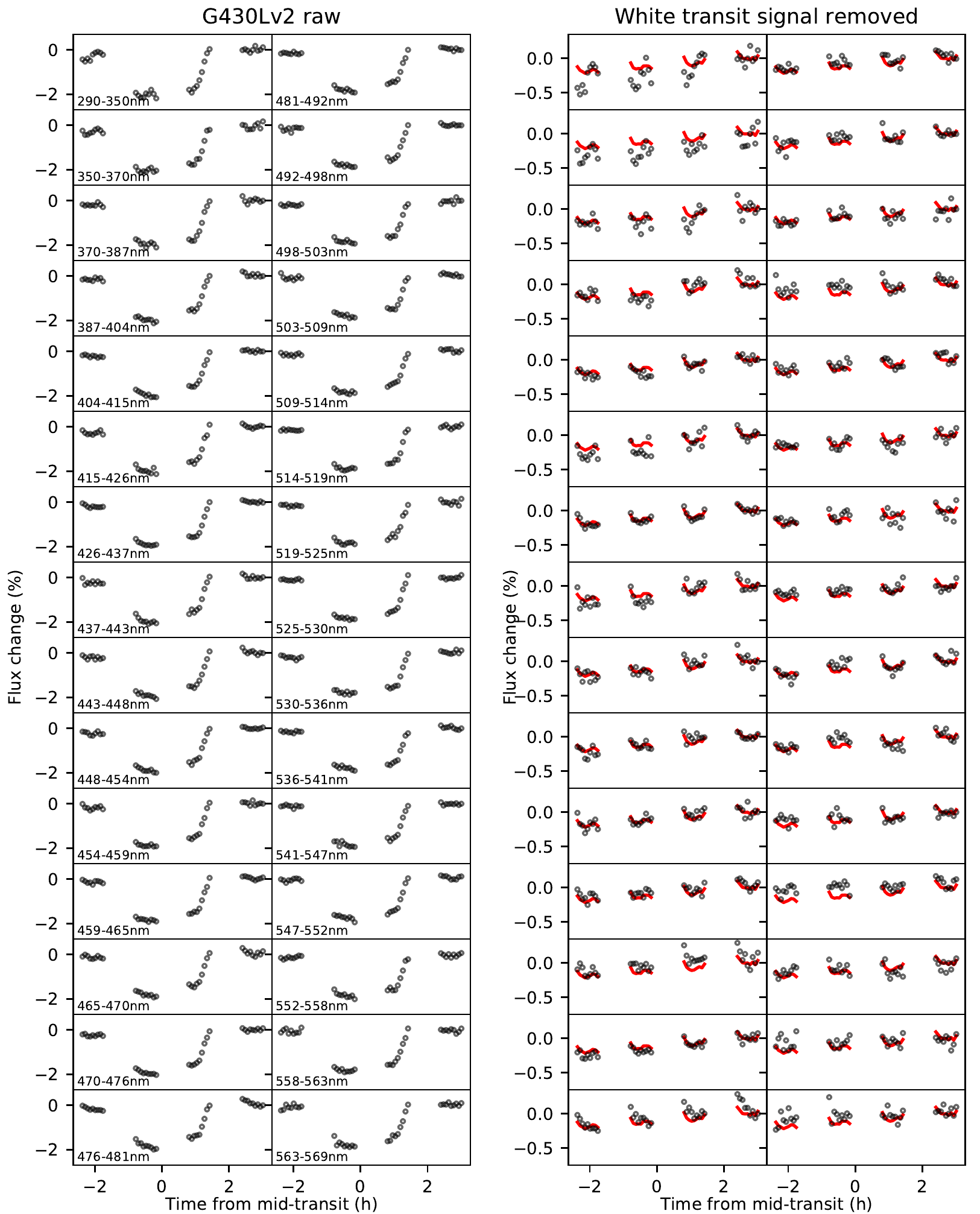}
\caption{The same as Figure \ref{fig:speclcs_raw_g430lv1}, but for the G430Lv2 dataset.}
\label{fig:speclcs_raw_g430lv2}
\end{figure}

\begin{figure}
\centering  % this centres figure in column
\includegraphics[width=\columnwidth]{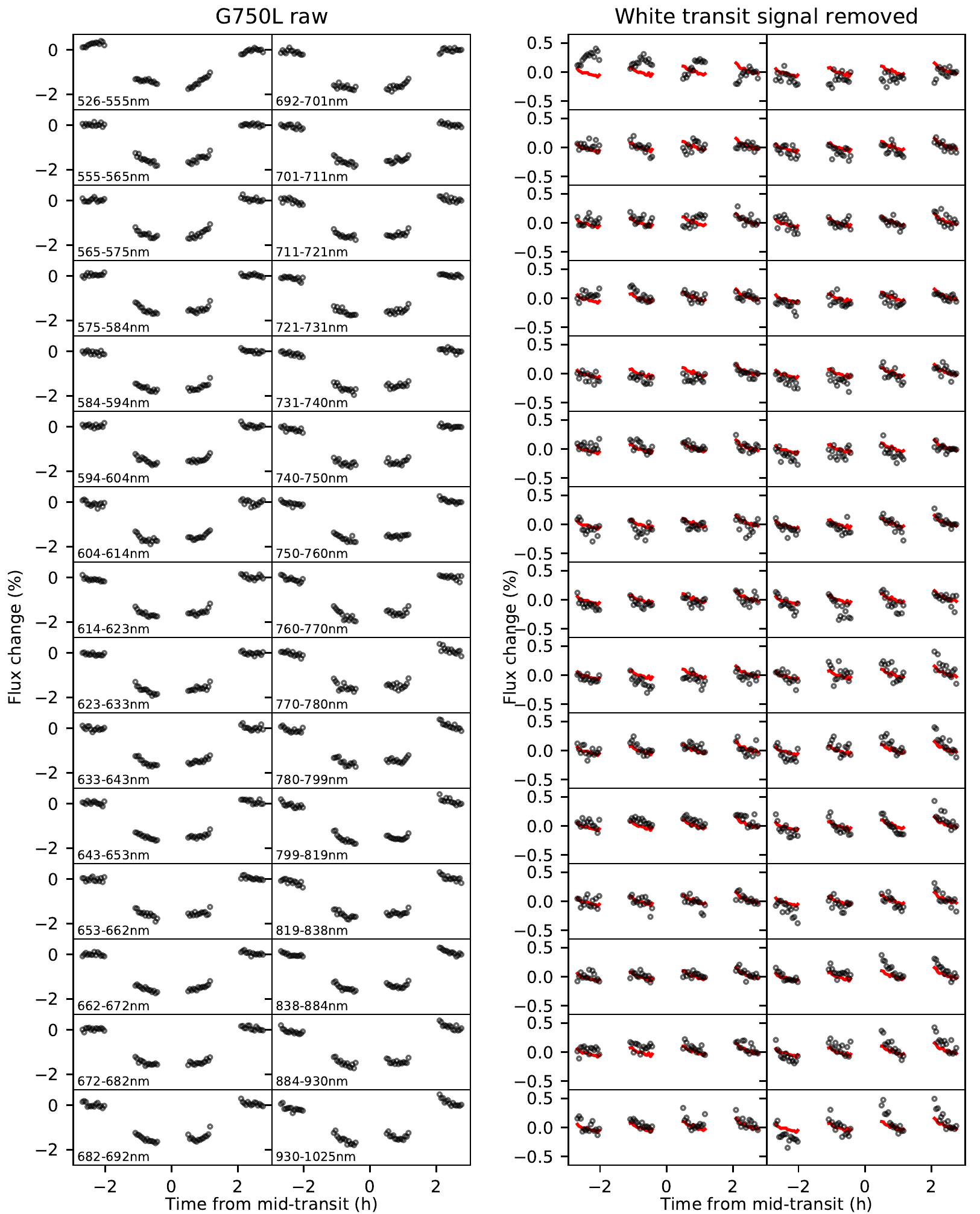}
\caption{The same as Figure \ref{fig:speclcs_raw_g430lv1}, but for the G750L dataset.}
\label{fig:speclcs_raw_g750l}
\end{figure}

HST lightcurves are strongly affected by instrumental systematics that must be accounted for as part of the lightcurve fitting process. For this reason, we present the raw spectroscopic lightcurves for the G430Lv1, G430Lv2, and G750L datasets in Figures \ref{fig:speclcs_raw_g430lv1}, \ref{fig:speclcs_raw_g430lv2}, and \ref{fig:speclcs_raw_g750l}, respectively. These figures also show the residuals after dividing the raw spectroscopic lightcurves by the corresponding white lightcurve best-fit transit signal, making it easier to inspect the systematics. Red lines indicate the best-fit systematics of the corresponding white lightcurve in order to highlight the wavelength-dependent nature of the systematics, which must be modeled individually for each spectroscopic channel.

\begin{figure}
\centering  % this centres figure in column
\includegraphics[width=0.93\columnwidth]{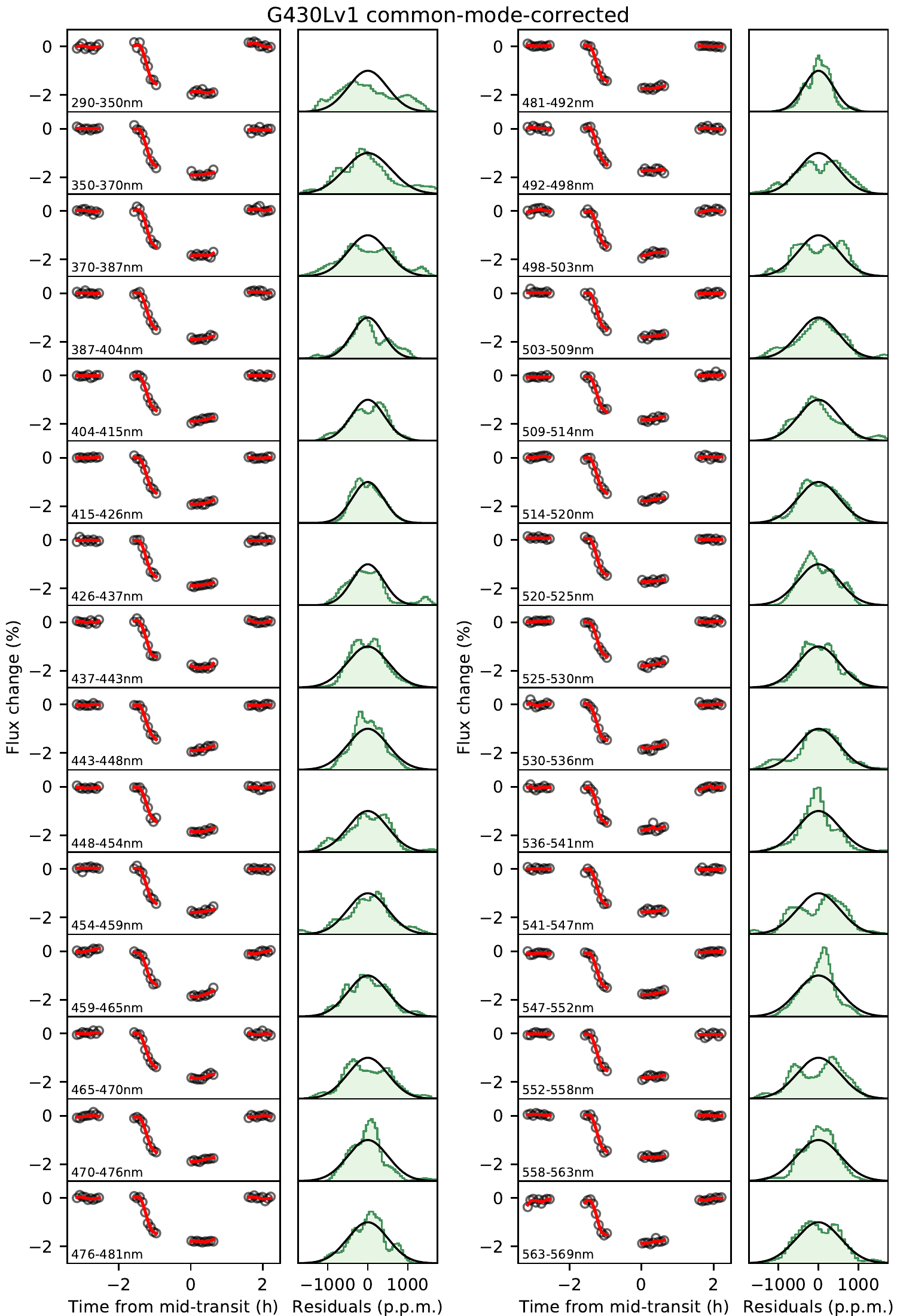}
\caption{\textit{(First and third columns)} Black circles show spectroscopic lightcurves for the G430Lv1 dataset after applying a common-mode correction. Red lines show best-fit GP models that simultaneously account for the transit signal and systematics. \textit{(Second and fourth columns)} Histograms of residuals between the data and best-fit GP model for each spectroscopic lightcurve, generated the same way as those shown in Figure \ref{fig:whitefit_lcs}. Black lines show normalized normal distributions with standard deviation equal to photon noise.}
\label{fig:speclcs_cmcorr_g430lv1}
\end{figure}

\begin{figure}
\centering  % this centres figure in column
\includegraphics[width=0.93\columnwidth]{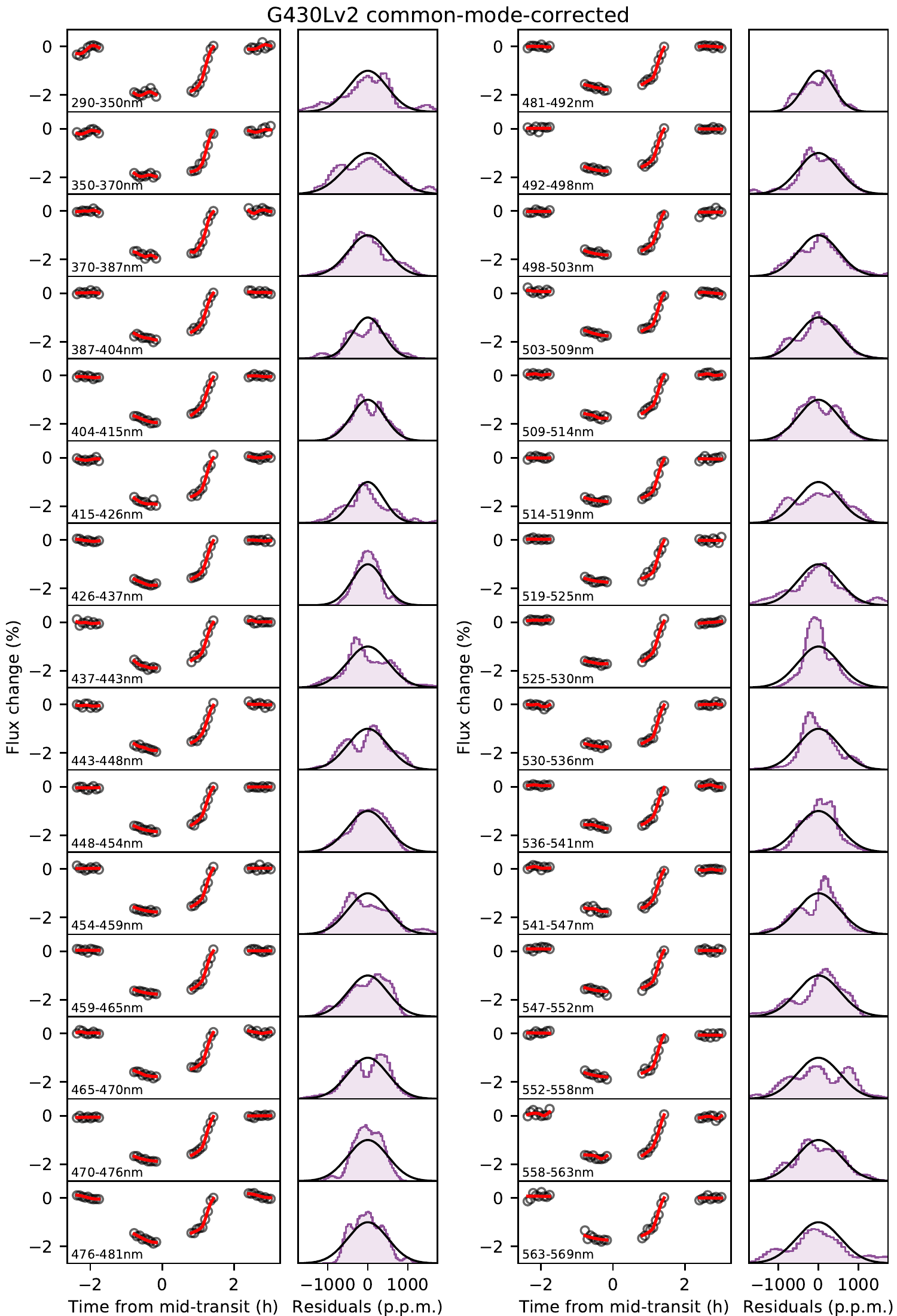}
\caption{The same as Figure \ref{fig:speclcs_cmcorr_g430lv1}, but for the G430Lv2 dataset.}
\label{fig:speclcs_cmcorr_g430lv2}
\end{figure}

\begin{figure}
\centering  % this centres figure in column
\includegraphics[width=0.93\columnwidth]{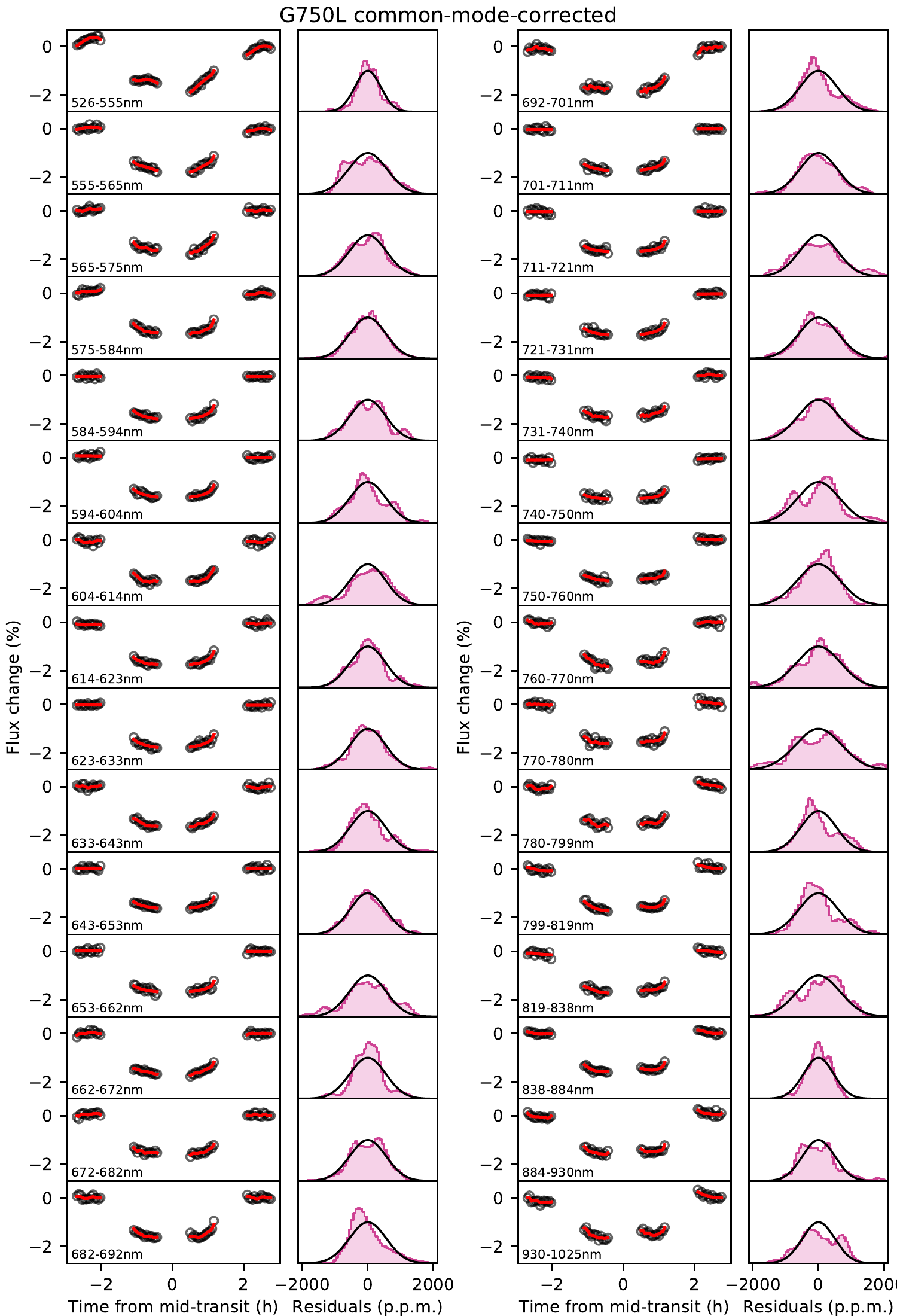}
\caption{The same as Figure \ref{fig:speclcs_cmcorr_g430lv1}, but for the G750L dataset.}
\label{fig:speclcs_cmcorr_g750l}
\end{figure}

As described in Section \ref{sec:speclcs}, however, we do apply a common-mode correction before fitting the spectroscopic lightcurves. The common-mode corrections are constructed for each dataset from the residuals of the corresponding white lightcurve with the best-fit transit signal removed. Figures \ref{fig:speclcs_cmcorr_g430lv1}, \ref{fig:speclcs_cmcorr_g430lv2}, and \ref{fig:speclcs_cmcorr_g750l} show the spectroscopic lightcurves after applying common-mode corrections for the G430Lv1, G430Lv2, and G750L datasets, respectively. Red lines indicate the best-fit GP model for each spectroscopic channel, which includes both the systematics and transit signals. Histograms of residuals are also shown for each spectroscopic channel. These were generated by taking 1000 random draws from the best-fit GP model, subtracting each of these from the data, then binning the resulting residuals. 

\clearpage

\section{Investigating the robustness of the measured transmission spectrum} \label{app:robustness}

In this section, we consider a number of affects unrelated to the planet itself that could potentially introduce biases to the inferred transmission spectrum.

\subsection{Sensitivity to limb darkening treatment} \label{app:robustness:ld}

As described in Sections \ref{sec:whitelc} and \ref{sec:speclcs}, for our main lightcurve analyses we adopted quadratic limb darkening profiles and allowed both coefficients ($u_1$, $u_2$) to vary as free parameters in the fitting. However, we also repeated the analyses using the four-parameter nonlinear law of \cite{2000A&A...363.1081C}, with coefficients fixed to values obtained by fitting to the limb darkened profile of the STAGGER 3D stellar model described in Section \ref{sec:whitelc}. For the white lightcurve analyses, we found the planet parameters inferred using the two limb darkening treatments (i.e.\ `free quadratic' and `fixed nonlinear') were consistent to within $1\sigma$, with only a single exception. Namely, for the fixed nonlinear analysis of the G750L lightcurve, we obtain $\aRs = 3.59_{-0.13}^{+0.12}$ and $b = 0.36_{-0.08}^{+0.07}$, which are both somewhat discrepant relative to the values inferred for the other lightcurves (Table \ref{table:whitefit}).

\begin{figure}
\centering  % this centres figure in column
\includegraphics[width=0.7\columnwidth]{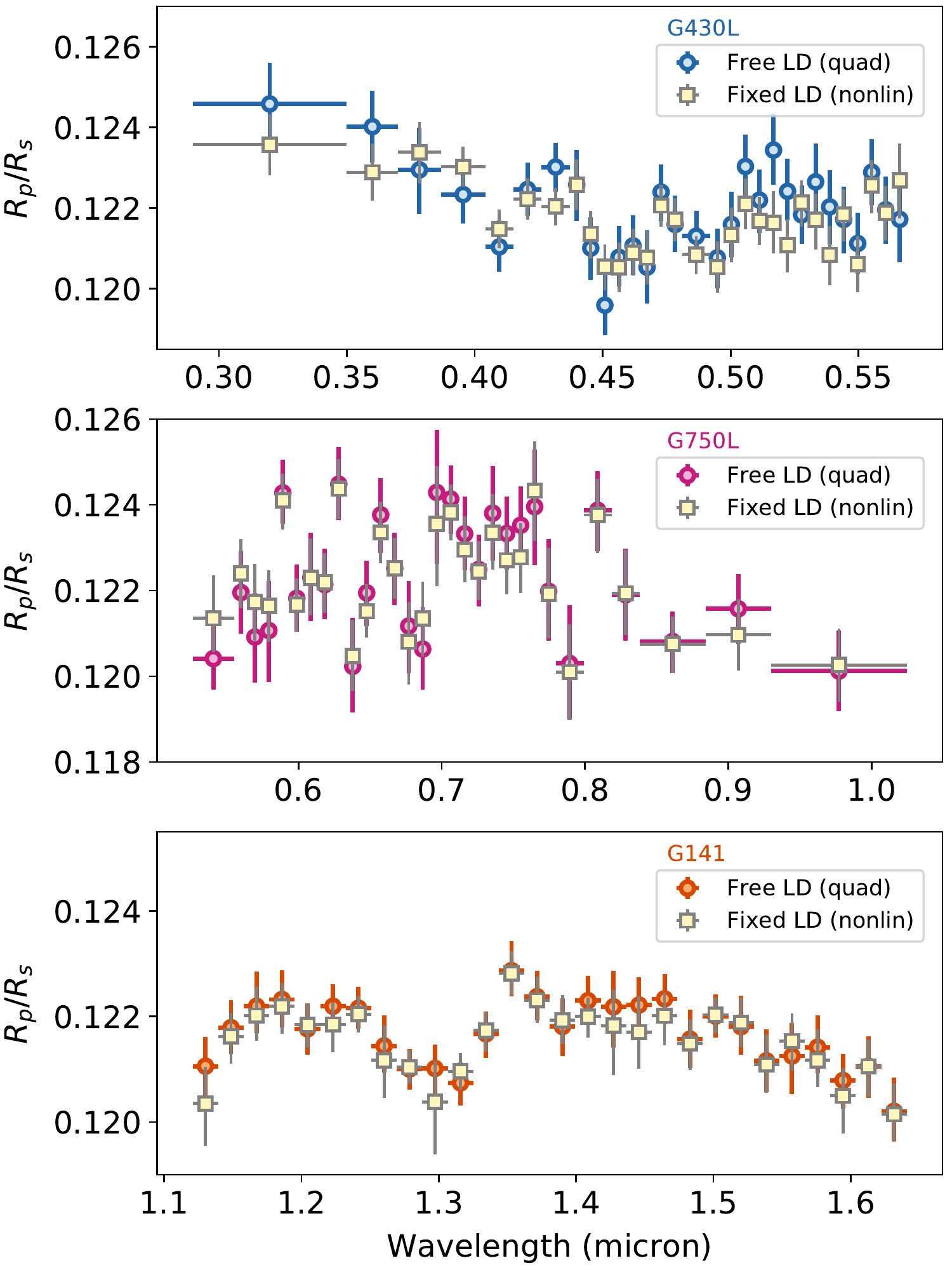}
\caption{Sensitivity of inferred transmission spectrum to limb darkening treatment for the G430L \textit{(top row)}, G750L \textit{(middle row)}, and G141 \textit{(bottom row)} datasets. Colored circles show results obtained assuming a quadratic law with coefficients allowed to vary in the lightcurve fits and pale yellow squares show results obtained assuming a four-parameter nonlinear law with coefficients fixed to values estimated from the stellar model described in the text.}
\label{fig:trspec_ldsensitivity}
\end{figure}

For the spectroscopic lightcurve fits, the effect of the two limb darkening treatments on the recovered transmission spectrum is illustrated in Figure \ref{fig:trspec_ldsensitivity}. The differences are negligible for the G750L and G141 datasets. For the G430L dataset, which at bluer wavelengths is more strongly affected by limb darkening, the transmission spectrum is systematically shifted to lower values for the fixed nonlinear analysis. Even so, the offset is less than $1\sigma$ for almost all of the spectroscopic channels and does not affect the interpretation of the transmission spectrum. We therefore conclude that our results are insensitive to the choice of limb darkening treatment.

\subsection{Including time $t$ as a GP input variable} \label{sec:discussion''gpt}

We repeated the GP fits to the white lightcurves (as described in Section \ref{sec:whitelc}) and spectroscopic lightcurves (as described in Section \ref{sec:speclcs}) with time $t$ provided as a fourth input variable in addition to $\{ \phi, x, y \}$. This was done to allow for possible departures from the linear function of $t$ that we assumed for the baseline trend. For instance, \cite{2015MNRAS.450.2043D} report a ramp-like baseline trend for observations of Alpha Cen A spanning 16 and 9 consecutive HST orbits. We note, however, that Alpha Cen A has a brightness of $V=0$\,mag, compared to $V=10.5$\,mag for WASP-121, which may result in especially pronounced systematics. We also note that analyses of STIS lightcurves often assume linear time baselines, and in a number of instances have been verified by independent observations using different instruments \citep[e.g.][]{2013MNRAS.434.3252H,2016ApJ...827...19F,2016ApJ...832..191N,2018MNRAS.tmp.2569E}. Furthermore, in our experience of STIS lightcurves, baseline trend departures from a linear function of $t$ often correlate with $x$ and $y$, and would therefore be accounted for by the GP fits adopting only $\{ \phi, x, y \}$ as input variables.

Given this, it is unsurprising that for the white lightcurve fits with $t$ included as an additional GP input, we obtain results consistent at the $1\sigma$ level with those reported in Table \ref{table:whitefit} for all but one free parameter. The single exception is the $\RpRs$ value inferred from the joint analysis of the two G430L datasets, which was $0.1233_{-0.0006}^{+0.0006}$, compared with the value of $0.1223_{-0.0006}^{0.0006}$ obtained without $t$ as a GP input. However, this is not a statistically significant difference.

\begin{figure}
\centering  % this centres figure in column
\includegraphics[width=0.7\columnwidth]{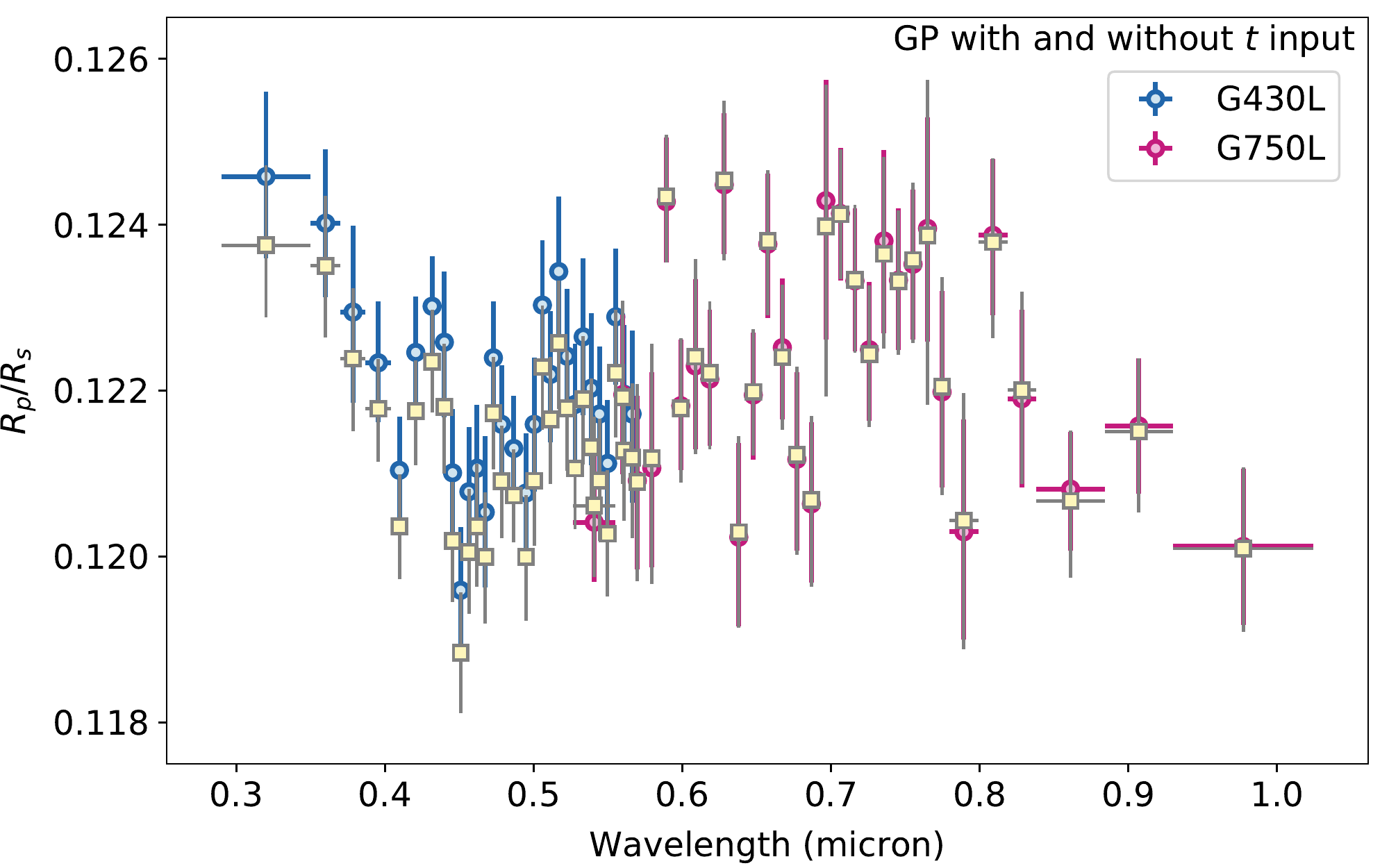}
\caption{Comparison of the STIS transmission spectrum obtained with (pale yellow squares) and without (colored circles) time $t$ as a GP input variable.}
\label{fig:trspec_gpt}
\end{figure}
  
  Similarly, for the spectroscopic lightcurve GP analyses including $t$ as an input, we obtain estimates for $\RpRs$ that are within $0.1\sigma$ of those listed in Tables \ref{table:specfit_g430l} and \ref{table:specfit_g750l} for the majority of channels. Specifically, this was the case for 14 of the G430L channels and 26 of the G750L channels. For all remaining channels, the $\RpRs$ estimates were within $1\sigma$ of those obtained without $t$ as a GP input. However, the uncertainties for $\RpRs$ increased by a median of $\sim 10\%$ for both the G430L and G750L datasets when $t$ was included as a GP input. We plot the resulting transmission spectrum in Figure \ref{fig:trspec_gpt} and report the results in Table \ref{table:specfit_gpt}.

We suspect the inclusion of $t$ as an additional GP input results in over-estimated uncertainties for $\RpRs$. For most channels, we found the inferred correlation length scale $L_t$ is large compared to $L_\phi$, $L_x$, and $L_y$, implying $t$ is a relatively unimportant input variable. In practice, $t$ likely plays a very similar role to $x$ and $y$ as a GP input (consider the second and third rows of Figure \ref{fig:stis_timeseries}, after accounting for the repeatable orbit-to-orbit variations in $x$ and $y$). Thus, by including $t$ as an input variable, we may be introducing an extra source of degeneracy to the systematics model which is not justified by the data. This in turn could artificially broaden the posterior distribution for parameters such as $\RpRs$. For these reasons, we present the transmission spectrum reported in Section \ref{sec:discussion} as our nominal measurement, and include this slightly more conservative analysis here for completeness. Under both analyses, our basic interpretation of the spectrum remains the same.
  
\begin{table}
\begin{minipage}{\columnwidth}
  \centering
\scriptsize
\caption{Results of spectroscopic lightcurve fits with $t$ as an additional GP input for selected parameters. \label{table:specfit_gpt}}
\begin{tabular}{ccccccccc} 
\hline \\ 
\multicolumn{4}{c}{G430L} && \multicolumn{4}{c}{G750L} \medskip \\ \cline{1-4} \cline{6-9} 
$\lambda$ (\AA) & $\RpRs$ & $u_1$ & $u_2$ & \quad & $\lambda$ (\AA) & $\RpRs$ & $u_1$ & $u_2$ \medskip \\ \cline{1-4} \cline{6-9}
$2898$-$3499$ & $0.1246_{-0.0011}^{+0.0011}$ & $0.55_{-0.09}^{+0.10}$ & $0.21_{-0.15}^{+0.14}$ && $5263$-$5550$ & $0.1206_{-0.0009}^{+0.0011}$ & $0.37_{-0.08}^{+0.08}$ & $0.29_{-0.13}^{+0.13}$ \\ 
$3499$-$3700$ & $0.1236_{-0.0010}^{+0.0010}$ & $0.42_{-0.09}^{+0.09}$ & $0.31_{-0.13}^{+0.15}$ && $5550$-$5648$ & $0.1219_{-0.0010}^{+0.0012}$ & $0.39_{-0.08}^{+0.09}$ & $0.23_{-0.14}^{+0.13}$ \\ 
$3700$-$3868$ & $0.1242_{-0.0013}^{+0.0012}$ & $0.40_{-0.10}^{+0.10}$ & $0.39_{-0.16}^{+0.14}$ && $5648$-$5745$ & $0.1209_{-0.0012}^{+0.0012}$ & $0.36_{-0.09}^{+0.09}$ & $0.30_{-0.13}^{+0.14}$ \\ 
$3868$-$4041$ & $0.1227_{-0.0009}^{+0.0009}$ & $0.56_{-0.09}^{+0.08}$ & $0.27_{-0.13}^{+0.14}$ && $5745$-$5843$ & $0.1212_{-0.0015}^{+0.0014}$ & $0.30_{-0.09}^{+0.09}$ & $0.32_{-0.14}^{+0.14}$ \\ 
$4041$-$4151$ & $0.1215_{-0.0010}^{+0.0010}$ & $0.59_{-0.08}^{+0.09}$ & $0.16_{-0.14}^{+0.14}$ && $5843$-$5940$ & $0.1243_{-0.0008}^{+0.0007}$ & $0.25_{-0.09}^{+0.08}$ & $0.28_{-0.13}^{+0.13}$ \\ 
$4151$-$4261$ & $0.1223_{-0.0007}^{+0.0007}$ & $0.63_{-0.07}^{+0.07}$ & $0.10_{-0.12}^{+0.12}$ && $5940$-$6038$ & $0.1218_{-0.0009}^{+0.0008}$ & $0.27_{-0.09}^{+0.09}$ & $0.25_{-0.13}^{+0.14}$ \\ 
$4261$-$4371$ & $0.1230_{-0.0007}^{+0.0007}$ & $0.51_{-0.08}^{+0.08}$ & $0.14_{-0.12}^{+0.13}$ && $6038$-$6135$ & $0.1224_{-0.0012}^{+0.0012}$ & $0.25_{-0.09}^{+0.09}$ & $0.26_{-0.13}^{+0.14}$ \\ 
$4371$-$4426$ & $0.1224_{-0.0009}^{+0.0010}$ & $0.53_{-0.09}^{+0.09}$ & $0.23_{-0.13}^{+0.15}$ && $6135$-$6233$ & $0.1222_{-0.0009}^{+0.0009}$ & $0.20_{-0.08}^{+0.09}$ & $0.34_{-0.14}^{+0.13}$ \\ 
$4426$-$4481$ & $0.1206_{-0.0007}^{+0.0008}$ & $0.64_{-0.09}^{+0.08}$ & $0.07_{-0.12}^{+0.13}$ && $6233$-$6330$ & $0.1245_{-0.0010}^{+0.0010}$ & $0.29_{-0.09}^{+0.08}$ & $0.20_{-0.13}^{+0.13}$ \\ 
$4481$-$4536$ & $0.1196_{-0.0008}^{+0.0007}$ & $0.59_{-0.08}^{+0.07}$ & $0.18_{-0.12}^{+0.13}$ && $6330$-$6428$ & $0.1203_{-0.0012}^{+0.0012}$ & $0.31_{-0.09}^{+0.09}$ & $0.23_{-0.13}^{+0.14}$ \\ 
$4536$-$4591$ & $0.1208_{-0.0008}^{+0.0008}$ & $0.49_{-0.08}^{+0.08}$ & $0.22_{-0.13}^{+0.13}$ && $6428$-$6526$ & $0.1220_{-0.0008}^{+0.0008}$ & $0.21_{-0.08}^{+0.09}$ & $0.22_{-0.13}^{+0.13}$ \\ 
$4591$-$4646$ & $0.1211_{-0.0008}^{+0.0008}$ & $0.43_{-0.08}^{+0.08}$ & $0.30_{-0.13}^{+0.12}$ && $6526$-$6623$ & $0.1238_{-0.0009}^{+0.0009}$ & $0.21_{-0.09}^{+0.09}$ & $0.18_{-0.14}^{+0.13}$ \\ 
$4646$-$4701$ & $0.1204_{-0.0010}^{+0.0010}$ & $0.54_{-0.10}^{+0.10}$ & $0.17_{-0.15}^{+0.15}$ && $6623$-$6721$ & $0.1224_{-0.0009}^{+0.0009}$ & $0.28_{-0.08}^{+0.09}$ & $0.21_{-0.13}^{+0.14}$ \\ 
$4701$-$4756$ & $0.1224_{-0.0007}^{+0.0007}$ & $0.51_{-0.09}^{+0.08}$ & $0.12_{-0.13}^{+0.14}$ && $6721$-$6818$ & $0.1212_{-0.0012}^{+0.0011}$ & $0.20_{-0.09}^{+0.10}$ & $0.25_{-0.14}^{+0.14}$ \\ 
$4756$-$4811$ & $0.1217_{-0.0007}^{+0.0008}$ & $0.45_{-0.08}^{+0.08}$ & $0.25_{-0.13}^{+0.13}$ && $6818$-$6916$ & $0.1207_{-0.0011}^{+0.0010}$ & $0.21_{-0.09}^{+0.09}$ & $0.37_{-0.14}^{+0.14}$ \\ 
$4811$-$4921$ & $0.1213_{-0.0006}^{+0.0006}$ & $0.45_{-0.08}^{+0.07}$ & $0.10_{-0.12}^{+0.13}$ && $6916$-$7014$ & $0.1240_{-0.0021}^{+0.0017}$ & $0.17_{-0.10}^{+0.11}$ & $0.26_{-0.14}^{+0.14}$ \\ 
$4921$-$4976$ & $0.1208_{-0.0009}^{+0.0009}$ & $0.43_{-0.09}^{+0.09}$ & $0.22_{-0.14}^{+0.13}$ && $7014$-$7111$ & $0.1241_{-0.0008}^{+0.0008}$ & $0.22_{-0.08}^{+0.08}$ & $0.20_{-0.13}^{+0.13}$ \\ 
$4976$-$5030$ & $0.1214_{-0.0010}^{+0.0009}$ & $0.45_{-0.09}^{+0.10}$ & $0.21_{-0.14}^{+0.15}$ && $7111$-$7209$ & $0.1233_{-0.0009}^{+0.0009}$ & $0.15_{-0.09}^{+0.08}$ & $0.26_{-0.13}^{+0.13}$ \\ 
$5030$-$5085$ & $0.1230_{-0.0008}^{+0.0008}$ & $0.40_{-0.08}^{+0.09}$ & $0.18_{-0.13}^{+0.14}$ && $7209$-$7307$ & $0.1224_{-0.0009}^{+0.0008}$ & $0.22_{-0.08}^{+0.08}$ & $0.24_{-0.13}^{+0.13}$ \\ 
$5085$-$5140$ & $0.1221_{-0.0008}^{+0.0009}$ & $0.49_{-0.09}^{+0.09}$ & $0.10_{-0.14}^{+0.14}$ && $7307$-$7404$ & $0.1236_{-0.0011}^{+0.0012}$ & $0.19_{-0.10}^{+0.09}$ & $0.22_{-0.14}^{+0.14}$ \\ 
$5140$-$5195$ & $0.1226_{-0.0013}^{+0.0012}$ & $0.38_{-0.09}^{+0.10}$ & $0.20_{-0.14}^{+0.14}$ && $7404$-$7502$ & $0.1233_{-0.0009}^{+0.0009}$ & $0.07_{-0.09}^{+0.09}$ & $0.29_{-0.13}^{+0.13}$ \\ 
$5195$-$5250$ & $0.1223_{-0.0007}^{+0.0008}$ & $0.35_{-0.08}^{+0.08}$ & $0.21_{-0.13}^{+0.13}$ && $7502$-$7600$ & $0.1236_{-0.0010}^{+0.0009}$ & $0.16_{-0.09}^{+0.09}$ & $0.16_{-0.13}^{+0.13}$ \\ 
$5250$-$5305$ & $0.1217_{-0.0008}^{+0.0007}$ & $0.45_{-0.09}^{+0.09}$ & $0.21_{-0.14}^{+0.14}$ && $7600$-$7698$ & $0.1239_{-0.0020}^{+0.0019}$ & $0.21_{-0.10}^{+0.09}$ & $0.28_{-0.13}^{+0.14}$ \\ 
$5305$-$5360$ & $0.1223_{-0.0010}^{+0.0010}$ & $0.39_{-0.10}^{+0.10}$ & $0.16_{-0.15}^{+0.14}$ && $7698$-$7795$ & $0.1220_{-0.0013}^{+0.0013}$ & $0.13_{-0.09}^{+0.09}$ & $0.31_{-0.13}^{+0.13}$ \\ 
$5360$-$5415$ & $0.1221_{-0.0010}^{+0.0011}$ & $0.27_{-0.10}^{+0.10}$ & $0.27_{-0.15}^{+0.14}$ && $7795$-$7991$ & $0.1204_{-0.0015}^{+0.0015}$ & $0.16_{-0.10}^{+0.10}$ & $0.26_{-0.15}^{+0.14}$ \\ 
$5415$-$5469$ & $0.1214_{-0.0010}^{+0.0010}$ & $0.33_{-0.09}^{+0.09}$ & $0.32_{-0.14}^{+0.15}$ && $7991$-$8186$ & $0.1238_{-0.0012}^{+0.0010}$ & $0.16_{-0.09}^{+0.09}$ & $0.28_{-0.14}^{+0.13}$ \\ 
$5469$-$5524$ & $0.1206_{-0.0009}^{+0.0009}$ & $0.36_{-0.09}^{+0.09}$ & $0.26_{-0.14}^{+0.14}$ && $8186$-$8381$ & $0.1220_{-0.0011}^{+0.0012}$ & $0.16_{-0.09}^{+0.09}$ & $0.24_{-0.13}^{+0.13}$ \\ 
$5524$-$5579$ & $0.1228_{-0.0008}^{+0.0009}$ & $0.40_{-0.09}^{+0.09}$ & $0.16_{-0.13}^{+0.13}$ && $8381$-$8840$ & $0.1207_{-0.0009}^{+0.0008}$ & $0.09_{-0.09}^{+0.08}$ & $0.32_{-0.13}^{+0.13}$ \\ 
$5579$-$5634$ & $0.1217_{-0.0009}^{+0.0009}$ & $0.28_{-0.09}^{+0.10}$ & $0.39_{-0.15}^{+0.15}$ && $8840$-$9299$ & $0.1215_{-0.0010}^{+0.0009}$ & $0.10_{-0.09}^{+0.09}$ & $0.20_{-0.13}^{+0.13}$ \\ 
$5634$-$5688$ & $0.1216_{-0.0011}^{+0.0011}$ & $0.44_{-0.09}^{+0.10}$ & $0.17_{-0.16}^{+0.14}$ && $9299$-$10245$ & $0.1201_{-0.0010}^{+0.0010}$ & $0.16_{-0.09}^{+0.09}$ & $0.23_{-0.13}^{+0.14}$ \\ 
\\ \hline
\end{tabular}
\end{minipage}
\end{table}

\subsection{Host star activity} \label{app:robustness:variability}

Host star activity in the form of dark and bright spots has the potential to introduce transit depth offsets between datasets, as well as chromatic biases within individual datasets. We have been monitoring WASP-121 with the Celestron 14-inch (C14) Automated Imaging Telescope (AIT) at Fairborn Observatory in southern Arizona \citep{1999PASP..111..845H,2003ASSL..288..189E}. Observations were conducted over two campaigns using the Cousins $R$ photometric bandpass. The first campaign spanned 2017 Jan 27 to Apr 23 and the second campaign spanned 2018 Feb 22 to Apr 8. The CCD images were used to compute differential magnitudes with respect to the mean brightness of ten of the most constant comparison stars in the same field. Further details of our data acquisition, reduction procedures, and analysis of the data can be found in \cite{2015MNRAS.446.2428S}, which describes a similar monitoring program for the planetary host star WASP-31. Although our observations for WASP-121 were made after the HST transit observations, they allow us to constrain the photometric variability of the F6V host star over timescales spanning multiple stellar rotation periods.

\begin{figure}
\centering  % this centres figure in column
\includegraphics[width=\columnwidth]{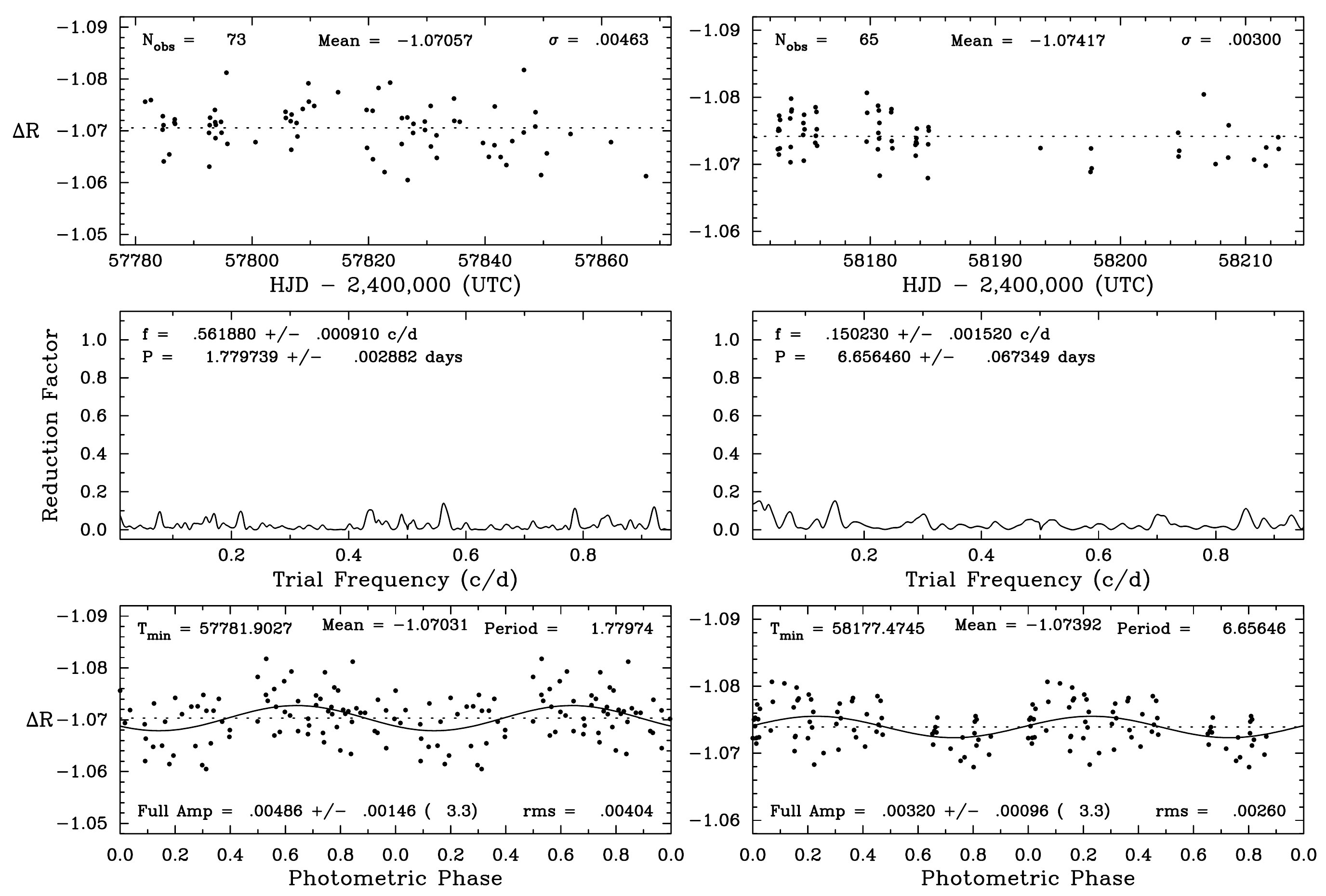}
\caption{Ground-based photometric monitoring data obtained in the Cousins $R$ bandpass using the C14-AIT at Fairborn Observatory. First and second columns show data from the 2017 and 2018 observing campaigns, respectively. \textit{(Top row)} Differential photometry. \textit{(Middle row)} Photometry periodogram. \textit{(Bottom row)} Photometry phase-folded to the period corresponding to the peak of the periodogram. Note that data points coinciding with transits of WASP-121b have been removed from these plots.}
\label{fig:spot_groundphot}
\end{figure}

Due to the southern declination of WASP-121 ($-39^\circ\,05^\prime\,51^{\prime\prime}$) and the northern latitude of Fairborn Observatory ($+31^\circ\,41^\prime\,18^{\prime\prime}$), the C14-AIT observations were made at large zenith angles between $70$--$80$ degrees, corresponding to airmass values of $3$--$5$. The panels in the first row of Figure \ref{fig:spot_groundphot} show the resulting photometry, after removing a small number of points that coincided with transits of WASP-121b. We measure a mean differential brightness for WASP-121 relative to the comparison stars of $-1.07057$\,mag for the 2017 campaign and $-1.07417$\,mag for the 2018 campaign. The standard deviation about the yearly mean was found to be $4.6$\,mmag for the 2017 campaign and $3.0$\,mmag for the 2018 campaign. The telescope CCD was replaced between the 2017 and 2018 campaigns, which may explain the lower scatter in the 2018 campaign, as well as the $3.6$\,mmag change in differential brightness. For comparison, \cite{2016MNRAS.tmp..312D} monitored WASP-121 over approximately six weeks using the TRAPPIST 60\,cm telescope and reported standard deviations in the night-to-night photometry of $1.6$\,mmag in the $B$ band, $1.3$\,mmag in the $V$ band, and $1.1$\,mmag in the $z^{\prime}$ band.

The second row of Figure \ref{fig:spot_groundphot} shows the frequency spectra for each C14-AIT campaign. The horizontal axis covers frequencies between $0.005$--$0.95$\,day$^{-1}$, corresponding to a period range of $1.05$--$200$\,day. No significant periodicity is detected for either campaign. In the third row of Figure \ref{fig:spot_groundphot} we plot the photometry phase-folded at a period of $1.78$\,days for the 2017 campaign and $6.66$\,days for the 2018 campaign, corresponding to the (insignificant) peaks of the respective periodograms. We obtain peak-to-peak amplitudes of $0.00486 \pm 0.00146$\,mag for the 2017 campaign and $0.00320 \pm 0.00096$\,mag for the 2018 campaign. In both cases, these amplitudes are comparable to the scatter in the residuals. A similar search for periodic signals in the WASP and TRAPPIST photometry performed by \cite{2016MNRAS.tmp..312D} also failed to uncover any evidence for periodic signals above the $\sim 1$\,mmag level. We therefore conclude WASP-121 is photometrically stable over multi-week periods to at least the $5$\,mmag level and likely to the $1$\,mmag level or better. No significant periodicity has yet been detected and our ability to constrain the variability is currently limited by the available photometric precision.

The lack of detected photometric variability for WASP-121 implies transit depth measurements should not vary significantly from epoch-to-epoch due to intrinsic stellar activity. This is consistent with the good agreement we obtain for the two G430L visits (Section \ref{app:robustness:g430lrepeat}) and also across the overlapping wavelength range covered by the G430L and G750L bandpasses (Section \ref{app:robustness:overlap}). In addition, we observe no strong evidence for spot-crossing events in the transit lightcurves. However, a persistent, unocculted spot coverage could conceivably introduce chromatic biases to the measured transit depth while remaining undetected in the photometric monitoring data, due to the lack of time-varying signal. Given the apparently near-polar orientation of the planetary orbit \citep{2016MNRAS.tmp..312D}, the persistent spot coverage would not necessarily need to be uniform in longitude. 

To quantify possible chromatic effects due to persistent unocculted spots, we follow a similar approach to that of \cite{2011ApJ...736...12B}. First, for a star without spots, the out-of-transit flux $F_{\textnormal{o.o.t.}}$ will be:
\begin{eqnarray}
F_{\textnormal{o.o.t.}} &=& A_{\star} f_{\star}
\end{eqnarray}
where $A_\star$ is the area of the stellar disc and $f_\star$ is the stellar flux per unit area. Assuming a non-luminous nightside hemisphere for the planet and ignoring limb darkening, the measured in-transit flux $F_{\textnormal{i.t.}}$ will be:
\begin{eqnarray}
F_{\textnormal{i.t.}} &=& ( A_{\star} - A_p ) f_{\star}
\end{eqnarray}
where $A_p$ is the area of the planet disc. This gives a relative transit depth $D=1-F_{\textnormal{i.t.}}/F_{\textnormal{o.o.t.}}$ of:
\begin{eqnarray}
D &=& A_p/A_\star
\end{eqnarray}
For a star with unocculted spots, the measured out-of-transit flux $\hat{F}_{\textnormal{o.o.t.}}$ will be:
\begin{eqnarray}
\hat{F}_{\textnormal{o.o.t.}} &=& ( A_{\star}-A_{\bullet} ) f_{\star} + A_{\bullet} f_{\bullet}
\end{eqnarray}
where $A_{\bullet}$ is the area covered by spots and $f_{\bullet}$ is the spot flux per unit area. The corresponding in-transit flux $\hat{F}_{\textnormal{i.t.}}$ will be measured as:
\begin{eqnarray}
\hat{F}_{\textnormal{i.t.}} &=& ( A_{\star}-A_{\bullet}-A_p ) f_{\star} + A_{\bullet} f_{\bullet}
\end{eqnarray}
It follows that the measured relative transit depth $\hat{D}=1-\hat{F}_{\textnormal{i.t.}}/\hat{F}_{\textnormal{o.o.t.}}$ will be:
\begin{eqnarray}
\hat{D} &=& \frac{D}{1-\alpha\left[1-\beta(\lambda)\right]} \label{eq:spotmodel}
\end{eqnarray}
where $\alpha \equiv A_\bullet/A_\star$ is the fractional area of the stellar disc covered by unocculted spots and $\beta(\lambda) \equiv f_\bullet(\lambda)/f_\star(\lambda)$ is the wavelength-dependent flux ratio of the spots and stellar photosphere. The chromatic bias $\kappa(\lambda) = \hat{D}-D$ due to unocculted spots will therefore be:
\begin{eqnarray}
 \kappa(\lambda) &=& D \left( \frac{\eta(\lambda)}{1-\eta(\lambda)} \right) \label{eq:chrombias}
\end{eqnarray}
where $\eta(\lambda) \equiv \alpha\left[ 1-\beta(\lambda)\right]$. This is equivalent to the spot corrections applied in previous studies such as \cite{2011MNRAS.416.1443S} and \cite{2013MNRAS.434.3252H}, and the transit light source effect described by \cite{2018ApJ...853..122R}.

\begin{figure}
\centering  % this centres figure in column
\includegraphics[width=\columnwidth]{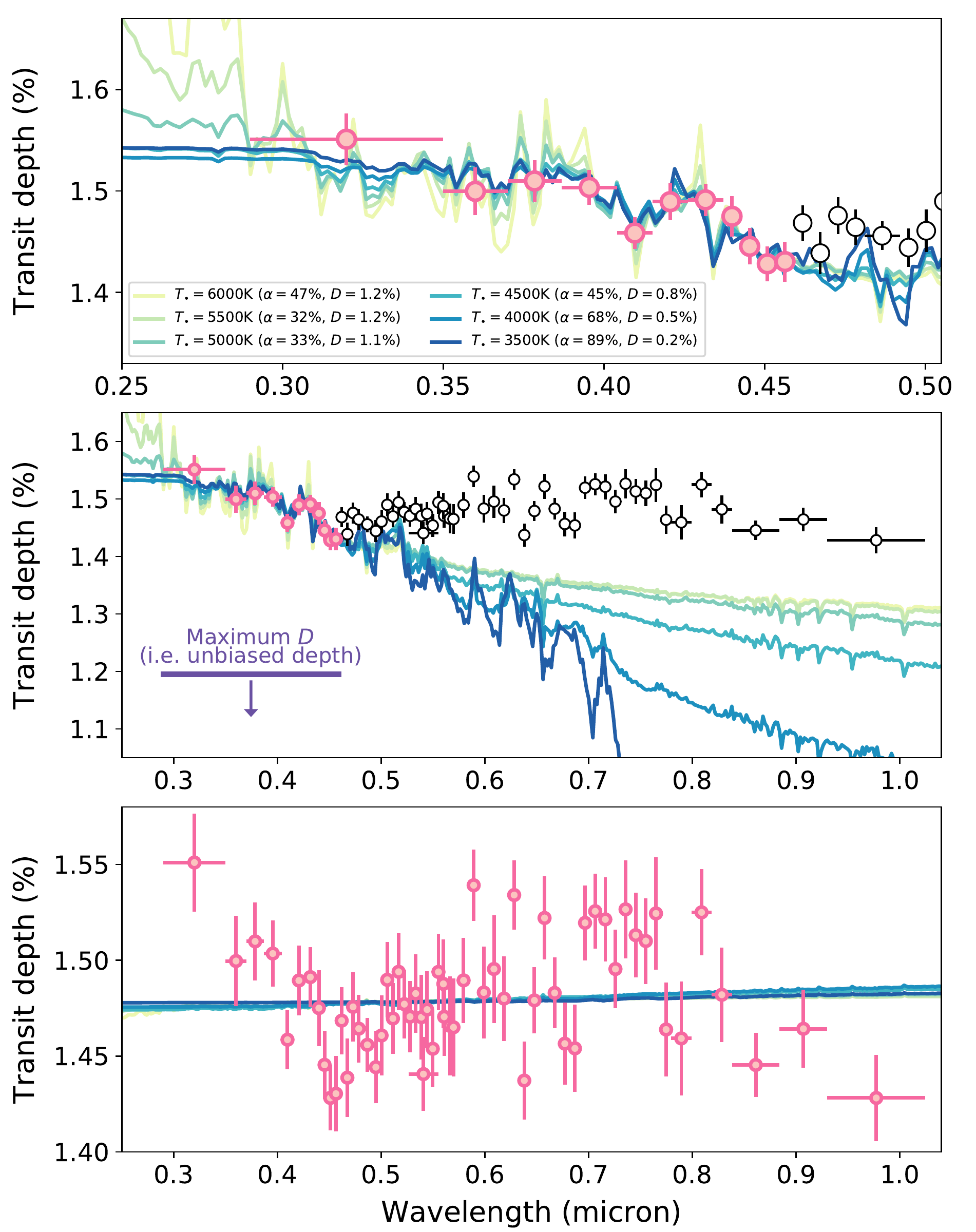}
\caption{Fits to the data assuming transit depth variations are caused by chromatic biases due to unocculted star spots. In all panels, data used in the fit are indicated by pink circles and other data points are indicated by unfilled black circles. \textit{(Top panel)} Fits to only the blue G430L data with inferred fractional spot coverages $\alpha$ and unbiased transit depths $D$ listed in the legend for different assumed spot temperatures. \textit{(Middle panel)} The same, but over an extended wavelength range to illustrate the predicted chromatic biases at longer wavelengths. Horizontal purple line indicates the implied range for the unbiased transit depths. \textit{(Bottom panel)} Fits to the full STIS dataset.}
\label{fig:spot_tests}
\end{figure}

Under the assumption that the true transit depth $D$ does not vary across the wavelength of interest, we fit the model given by Equation \ref{eq:spotmodel} to the transit depths derived from the measured $\RpRs$ values given in Table \ref{table:specfit_g430l}. In these fits, we allowed $D$ and $\alpha$ to vary as free parameters, while for $f_\star(\lambda)$ we adopted a PHOENIX stellar model from the BT-Settl grid \citep{2012RSPTA.370.2765A} with properties similar to WASP-121 ($T_\star=6500$, $\log g = 4.0$\,cgs, $\textnormal{[Fe/H]=0}$\,dex). For $f_\bullet(\lambda)$, we used the same BT-Settl stellar model and repeated the fitting process for a range of assumed spot temperatures $T_\bullet$ ranging from $6000$\,K down to $3500$\,K in increments of $500$\,K. The results are shown in Figure \ref{fig:spot_tests}. 

If we restrict attention to the blue G430L data only, we find unocculted spots can reproduce the shape of the measured spectrum. However, this requires invoking fractional coverages $\alpha$ ranging from 32\% for $T_\bullet = 5500$\,K to well over 50\% for other $T_\bullet$ values. Such large spot coverages would likely have a significant effect on the spectral typing of the star and be at odds with the modest X-ray flux we measure for WASP-121 (see below). Furthermore, the true transit depth $D$ would be no deeper than $1.2$\%, which is significantly lower than the measured transit depths $\hat{D}$ of $>1.4$\% (Figure \ref{fig:spot_tests}). We also find the chromatic bias $\kappa(\lambda)$ given by Equation \ref{eq:chrombias} would vary by $\Delta \kappa = 2200$\,ppm from the G430L to G141 bandpasses for $T_\bullet = 6000$\,K, and by more for lower $T_\bullet$ values. Instead, we measure consistent transit depths in G430L and G141 bandpasses, to a precision of $\sim 200$\,ppm (Table \ref{table:whitefit}). Thus, under the unocculted spot scenario, the unbiased transit depth would be $>2200$\,ppm deeper in the G141 bandpass relative to the G430L bandpass. This is equivalent to a change in the transmission spectrum of $>10\,H$, where $H$ is a pressure scale height, since the change in measured transit depth due to $1H$ is $\sim 150$--$200$\,ppm for WASP-121b. Even larger differences would be expected for $T_\bullet <6000$\,K. Meanwhile, we are unable to reproduce the data if we attempt to fit the full STIS wavelength range using the same unocculted spot model (bottom panel of Figure \ref{fig:spot_tests}). For these reasons, we consider it unlikely that unocculted spots can explain the measured spectrum.

Finally, we provide a brief report on measurements of WASP-121 made over approximately two hours on 2017 Apr 6 using the XMM-Newton space observatory (Obs ID 0804790601, P.I.\ Sanz-Forcada). Data were collected simultaneously at X-ray wavelengths ($0.12$--$2.48$\,keV; $5$--$100$\,\AA) with the XMM-EPIC instrument and UV wavelengths ($1685$--$2480$\,\AA) with the XMM-OM instrument. The X-ray fluxes, combined with the distance to the system (i.e.\ 272\,pc; see Section \ref{sec:intro}), imply an X-ray luminosity $\log_{10}L_{\rm X}=29.02$ (cgs). Using the multicolor brightness of WASP-121 ($V=10.51$\,mag, $B=11.0$\,mag) with the bolometric corrections of \cite{1996ApJ...469..355F}, we calculate a bolometric luminosity $\log_{10} L_{\rm bol}=34.14$ (cgs). This implies $\log_{10} L_{\rm X}/L_{\rm bol}=-5.12$, which is consistent with a low activity star. The XMM-OM UV timeseries does not show evidence for significant variability, while the XMM-EPIC X-ray timeseries may show some variability, although the statistics are poor and currently it is not possible to give a firm assessment. Further details will be provided in Sanz-Forcada et al.\ (in prep.).

\subsection{Repeatability of the G430L observations} \label{app:robustness:g430lrepeat}

In Section \ref{sec:speclcs}, we presented the results of our primary G430L spectroscopic lightcurve analysis, for which $\RpRs$ was treated as a shared parameter fit jointly across both visits. However, we also analyzed each visit individually to check the measurement repeatability. The resulting transmission spectra are shown in the top panel of Figure \ref{fig:trspec_g430l} and exhibit good agreement. If we consider the median $\RpRs$ values inferred from the joint analysis to be the `ground truth', we can quantify the likelihood of measuring the transmission spectra for each individual visit using using $\chi^2_{\nu} = \left( \sum{(\rho_{i}-\rho_{0,i})^2/\sigma_{i}^2} \right)/\nu$, where: $\rho_{i}$ is the median $\RpRs$ value inferred for the $i$th channel of the individual visit, with corresponding uncertainty $\sigma_{i}$; $\rho_{0,i}$ is the corresponding $\RpRs$ value inferred from the joint analysis; and $\nu$ is the number of degrees of freedom, which in this case is equal to the number of spectroscopic channels. We obtain $\chi^2_{\nu}=1.0$ for G430Lv1 and $\chi^2_{\nu}=0.7$ for G430Lv2, implying the transmission spectra inferred for each visit individually are consistent with being random draws of an underlying distribution centered at the $\RpRs$ values obtained from the joint analysis.

\subsection{Consistency of G430L-G750L overlap} \label{app:robustness:overlap}

There is some overlap between the G430L and G750L gratings, spanning approximately $0.55$--$0.57\,\um$ in wavelength (Figure \ref{fig:example_spectra}). The transmission spectra recovered for both gratings across this overlap region are consistent with each other, to within the measurement uncertainties (e.g.\ Figure \ref{fig:retrieval_specfit}). This gives some further reassurance that stellar variability or instrumental systematics have not introduced significant biases in the measured transit depth level from one observation to the next. 

To test this more explicitly, we generated lightcurves spanning the full $0.55$--$0.57\,\um$ overlapping wavelength range for the G430Lv1, G430Lv2, and G750L datasets and fit them using the approach described in Section \ref{sec:speclcs}. A joint fit to the two G430L lightcurves gave $\RpRs = 0.1225 \pm 0.0006$, while a fit to the G750L lightcurve gave $\RpRs = 0.1216 \pm 0.0007$. Combining the uncertainties in quadrature, these results are consistent at the $1\,\sigma$ level. 

\begin{figure}
\centering  % this centres figure in column
\includegraphics[width=0.7\columnwidth]{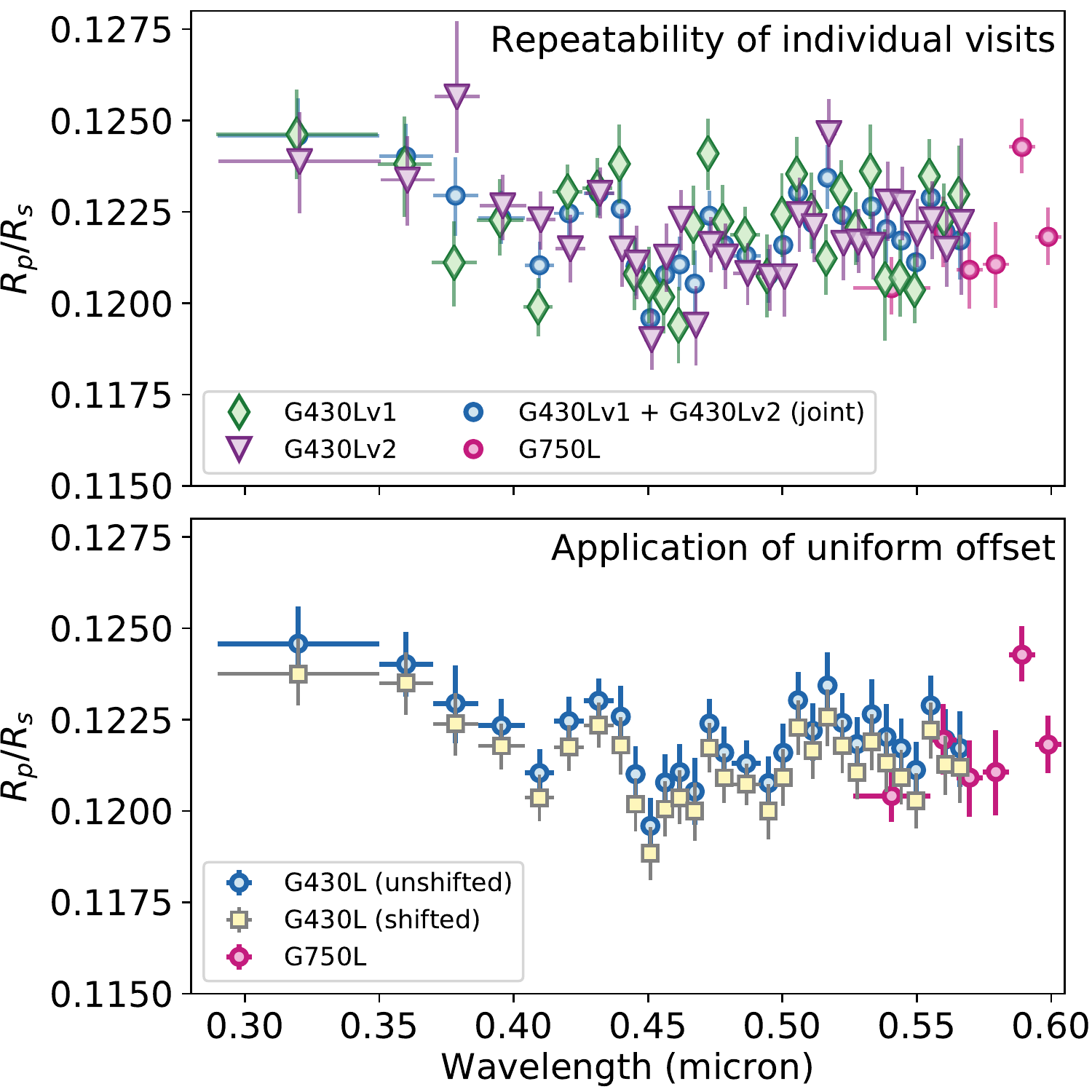}
\caption{Measured transmission spectrum over the $0.3$--$0.6\,\um$ wavelength range, focusing on the G430L dataset. \textit{(Top panel)} Analyses of each visit individually (green triangles and purple diamonds) as well as the joint analysis of both visits simultaneously (filled blue circles). \textit{(Bottom panel)} Application of a uniform vertical offset to align the overlapping range of the G430L and G750L gratings.}
\label{fig:trspec_g430l}
\end{figure}

One possibility would be to apply a wavelength-uniform offset to either the G430L or G750L transmission spectrum, commensurate with the difference of $\sim 0.0009$ measured for $\RpRs$ across this overlapping wavelength range. The application of such an offset  to the G430L spectrum is shown in the bottom panel of Figure \ref{fig:trspec_g430l}. However, given the small amplitude of this offset relative to the measurement uncertainties, we found it did not affect our physical interpretation of the transmission spectrum. For example, the forward model described in Section \ref{sec:discussion:forward}, which assumes chemical equilibrium with $20 \times$ solar elemental abundances and a temperature of $T=1500$\,K, is still compatible with the data (excluding the NUV rise and $1.15-1.3\,\um$ bump) with a reduced $\chi^2$ of 0.9. Nonetheless, it is still worth emphasizing that the overall levels of the transmission spectrum subsets (i.e.\ G430L, G750L, G141) are each subject to some uncertainty, on the order of the corresponding white lightcurve $\RpRs$ uncertainty (Table \ref{table:whitefit}).

\acknowledgements The authors are grateful to the anonymous referee for constructive feedback that improved the quality of this manuscript. We thank Richard Freedman for providing the SH absorption cross-sections. Support for program GO-14767 was provided by NASA through a grant from the Space Telescope Science Institute, which is operated by the Association of Universities for Research in Astronomy, Inc., under NASA contract NAS 5-26555. T.M.E., D.K.S., and N.N.\ acknowledge funding from the European Research Council under the European Unions Seventh Framework Programme (FP7/2007-2013)/ERC grant agreement no.\ 336792. G.W.H.\ and M.H.W.\ acknowledge support from Tennessee State University and the State of Tennessee through its Centers of Excellence program. J.S.F.\ acknowledges funding by the Spanish MINECO grant AYA2016-79425-C3-2-P. J.K.B.\ is supported by a Royal Astronomical Society Research Fellowship. This work has been carried out in the frame of the National Centre for Competence in Research PlanetS supported by the Swiss National Science Foundation (SNSF). V.B.\ and D.E.\ have received funding from the European Research Council (ERC) under the European Union’s Horizon 2020 research and innovation programme (project {\sc Four Aces}; grant agreement no.\ 724427).

\vspace{5mm}
\facilities{ HST(STIS and WFC3) }

%\software{}

%\bibliography
\end{document}